\PassOptionsToPackage{sort&compress}{natbib}
\PassOptionsToPackage{table}{xcolor}
\documentclass{thirdintel}
\usepackage{newtxtext,newtxmath}

\ifPDFTeX
  \renewcommand{\sfdefault}{qhv}
\else
\fi
\usepackage[utf8]{inputenc}

\usepackage{array,tabularx}
\usepackage[section]{placeins}
\usepackage{xurl}

\setcitestyle{numbers,square,comma}
\hypersetup{
  pdftitle={RubatoCodec: Budgeted Adaptive Speech Tokenization with Streaming Decoding},
  pdfauthor={Chihiro Arata; Shohei Taniguchi; Gouki Minegishi; Kan Hatakeyama-Sato},
  pdfsubject={Speech tokenization and streaming synthesis}}
\makeatletter
\newcommand{\bstctlcite}[1]{\@bsphack\if@filesw
  \immediate\write\@auxout{\string\citation{#1}}\fi\@esphack}
\makeatother
\makeatletter
\renewcommand{\AB@authnote}[1]{{\normalfont\textsuperscript{#1}}}
\renewcommand{\AB@affilnote}[1]{{\normalfont\textsuperscript{#1}}}
\makeatother

\newcolumntype{L}{>{\raggedright\arraybackslash}X}
\newcolumntype{Y}{>{\raggedleft\arraybackslash}X}
\newcolumntype{P}[1]{>{\raggedright\arraybackslash}p{#1}}
\newcolumntype{R}[1]{>{\raggedleft\arraybackslash}p{#1}}
\renewcommand{\arraystretch}{1.15}
\newcommand{\rowrule}{\arrayrulecolor{black!25}\midrule[0.4pt]\arrayrulecolor{black}}

\title{\textbf{RubatoCodec: Budgeted Adaptive Speech\protect\\
Tokenization with Streaming Decoding}}
\author[1]{Chihiro Arata}
\author[1]{Shohei Taniguchi}
\author[1]{Gouki Minegishi}
\author[1]{Kan Hatakeyama-Sato}
\affil[1]{Third Intelligence, Inc.}
\tishorttitle{RubatoCodec}
\tidate{}
\ticontact{\href{mailto:chihiro.arata@third-intelligence.com}{chihiro.arata@third-intelligence.com}}

\begin{document}
\bstctlcite{BSTcontrol}
\maketitle
\begin{abstract}
\setlength{\parindent}{0pt}
\noindent Neural speech codecs provide discrete tokens for speech generation. Lower frame rates shorten token sequences, but each token must then carry more linguistic and acoustic information. Variable-frame-rate codecs address this trade-off by adapting token allocation to speech content, but threshold-based allocation lets content determine the token count, so the sequence length a speech generator must produce varies across utterances and datasets at a fixed setting. We present RubatoCodec, a variable-frame-rate speech codec built on X-Codec-2 with an exact utterance-level event budget and streaming decoding. Its Adaptive Firing mechanism uses changes in codec features and waveform spectra to divide speech into variable-duration events without a learned boundary predictor. Discrete codes and an explicit duration represent each event, allowing a speech generator to predict and decode one event at a time. Controlled experiments show that adaptive allocation outperforms fixed-rate coding in reconstruction fidelity and intelligibility at low event rates. A model trained on large-scale multilingual speech achieves competitive reconstruction quality across English, Japanese, and Mandarin, with the best intelligibility preservation among the compared codecs on the evaluated English and Japanese benchmarks at comparable bitrates. In a matched text-to-speech comparison, the adaptive representation yields lower WER and higher UTMOS and UTMOSv2 with comparable speaker similarity. The speech generator also learns the codec's utterance-level event budget without an explicit generation-time rate constraint. These results demonstrate that budgeted adaptive speech representations support both high-quality reconstruction and streaming text-to-speech generation. Code and weights are available at \url{https://github.com/third-intelligence/RubatoCodec} and \url{https://huggingface.co/third-intelligence/RubatoCodec}.

\end{abstract}

\section{Introduction}\label{sec:introduction}

Neural speech codecs encode waveforms as discrete tokens for reconstruction \cite{zeghidour2022soundstream,defossez2023encodec,kumar2023dac} and language-model-based generation, including AudioLM \cite{borsos2023audiolm}, VALL-E \cite{wang2023valle}, and recent synthesis and spoken dialogue systems \cite{defossez2024moshi,ye2025llasa,hu2026qwen3tts,gong2026mossaudiotokenizer}. For generation, the codec's frame rate determines the temporal sequence length: ten seconds of speech requires 500 positions at 50 Hz. Lower rates shorten this sequence but require each position to represent a longer acoustic interval. Recent tokenizers operate at 12.5 Hz \cite{defossez2024moshi,hu2026qwen3tts,gong2026mossaudiotokenizer}, where each token spans 80 ms. At this resolution, where boundaries fall relative to speech content matters: a fixed grid can merge a rapid phonetic transition into a single token while spending as many tokens on a pause as on the transition.

Speech is temporally nonstationary: pauses and steady vowels change slowly, whereas transitions change rapidly. Variable-rate representations and variable-frame-rate codecs exploit this by assigning different durations to frames \cite{dieleman2021variablerate,zhang2025tfc,wang2026codecslime,zheng2026varstok,li2026flexicodec,sohn2026dtmcodec}. Threshold-based allocation, however, lets content determine the frame count, so rates vary across utterances and datasets at a fixed setting \cite{li2026flexicodec,sohn2026dtmcodec}. For example, one released threshold-based codec ranges from 46.1 to 81.5 events/s under a single threshold across our five evaluation datasets (Section~\ref{sec:event-rate-control-across-datasets}). Calibrating the threshold on representative data can match an average rate, but it neither fixes the count for an individual utterance nor preserves that average when the data distribution shifts. Sequence length, and with it downstream modeling cost, then cannot be determined from audio duration alone. An explicit event budget removes this dependence: placement adapts to content, while utterance duration determines the sequence length.

Prescribed-count segmentation has been explored in codecs and speech representation methods \cite{zhang2025tfc,wang2026codecslime,zhang2026wertrap}. We build on this principle with a codec that uses a deterministic firing signal to allocate events, learns duration-aware representations of those events, and decodes them causally.

RubatoCodec separates the number of events from their placement. A retention ratio \(R\) fixes the number of events for \(T\) latent frames at \(K=\operatorname{round}(RT)\) (Section~\ref{sec:budget-constrained-event-allocation}), and Adaptive Firing places them by cumulative allocation of a signal derived from latent dissimilarity and waveform spectral flux, without a learned boundary predictor. The default \(R=0.25\) yields 12.5 events/s on average from a 50-Hz latent grid, with variable event durations. The name reflects the musical term rubato: local timing flexes while the overall tempo is preserved.

RubatoCodec builds on X-Codec-2 \cite{ye2025llasa}, retaining its acoustic/semantic encoder and spectral synthesis backbone while introducing adaptive aggregation and causal decoding. Each event carries discrete codes and an explicit duration, allowing a downstream model to predict and decode completed events incrementally. We evaluate the design through controlled codec comparisons, multilingual reconstruction, and downstream TTS training.

Our contributions are:

\begin{enumerate}
\item \textbf{Budgeted content-adaptive speech tokenization.} Adaptive Firing combines an exact utterance-level event budget with cumulative allocation driven by latent dissimilarity and waveform spectral flux. An onset-anchored aggregator combines frame features and duration within each event to construct its quantizer input. Event boundaries require neither a learned boundary predictor nor rate-penalty optimization.
\item \textbf{Controlled analysis of adaptive allocation.} Matched comparisons show reconstruction gains over fixed-rate downsampling at low event rates, and a placement-only comparison indicates that content-adaptive boundaries contribute to these gains. We also characterize how event rate and quantizer depth trade off at matched bitrates.
\item \textbf{A code--duration interface for streaming generation.} Each event carries semantically supervised codes and an explicit duration, which a causal decoder renders as they become available. Large-scale multilingual evaluation shows competitive reconstruction quality, including the best intelligibility preservation among the compared codecs on the evaluated English and Japanese benchmarks at comparable bitrates. In matched TTS training, the representation yields lower WER and higher UTMOS and UTMOSv2 scores with comparable speaker similarity, and the generator learns the codec's utterance-level event budget without an explicit rate constraint.
\end{enumerate}

\section{Related Work}\label{sec:related-work}

\subsection{Neural speech codecs for language modeling}\label{sec:neural-speech-codecs-for-language-modeling}

SoundStream and EnCodec combine residual vector quantization with adversarial waveform reconstruction \cite{zeghidour2022soundstream,defossez2023encodec}; DAC improves the quantization and reconstruction components \cite{kumar2023dac}. Quantizer depth controls the information carried per temporal position. BigCodec instead studies high-capacity single-codebook representations at low bitrates \cite{xin2024bigcodec}.

Semantic supervision encourages codec tokens to retain linguistic content. SpeechTokenizer supervises the first RVQ layer with self-supervised features \cite{zhang2024speechtokenizer}, while Mimi combines a semantically distilled VQ with a parallel acoustic RVQ \cite{defossez2024moshi}. X-Codec-2 fuses an acoustic encoder with frozen w2v-BERT 2.0 features and uses semantic reconstruction supervision \cite{ye2025llasa}. RubatoCodec builds on this architecture, including its Vocos-style spectral decoder \cite{siuzdak2024vocos}, and adopts the synthesis upsampling approach of Inworld TTS \cite{atamanenko2025tts1}.

Low-frame-rate tokenizers reduce the temporal sequence length presented to a generative model. Mimi and the publicly released Qwen3-TTS-Tokenizer-12Hz operate at 12.5 Hz with semantic supervision \cite{defossez2024moshi,hu2026qwen3tts}, while MOSS-Audio-Tokenizer learns a low-rate representation through jointly optimized reconstruction, adversarial, and audio-to-text objectives without an external semantic distillation teacher \cite{gong2026mossaudiotokenizer}.

\subsection{Variable-frame-rate speech coding}\label{sec:variable-frame-rate-speech-coding}

Early variable-rate discrete representations combine slow autoencoders with run-length coding \cite{dieleman2021variablerate}. Recent codecs adapt temporal resolution through hierarchy selection, merging, or boundary selection. TFC uses waveform-derived temporal entropy and utterance-level granularity ratios \cite{zhang2025tfc}; VARSTok clusters contiguous latent frames and encodes duration in token identity \cite{zheng2026varstok}; FlexiCodec merges frames using frozen SenseVoice features and adjusts rate through its merge threshold \cite{li2026flexicodec}. RubatoCodec constructs its firing signal from fused-latent dissimilarity and waveform spectral flux.

CodecSlime constrains segment count through dynamic programming over feature distortion, adapting a pretrained fixed-rate codec through its Melt-and-Cool procedure \cite{wang2026codecslime}. RubatoCodec instead normalizes its firing signal to the event budget and places boundaries by cumulative firing, without searching candidate segmentations. Its trainable codec components learn from scratch with adaptive allocation active throughout training.

Elastic Time augments a frozen audio autoencoder with a learned latent predictor, using greedy or dynamic-programming boundary selection to retain a target number of anchor frames and reconstruct omitted frames \cite{bralios2026elastictime}. It adjusts the frame rate of continuous latents at inference; RubatoCodec instead produces discrete codes with explicit durations for causal decoding.

DTM-Codec's Path Length Equalization selects boundaries at successive thresholds of accumulated latent change \cite{sohn2026dtmcodec}. Its fixed threshold lets total change determine the event count; RubatoCodec normalizes bounded increments to a prescribed count. DyCAST uses character alignment or a learned boundary predictor and supports transmitted durations, length-conditioned duration prediction, or free duration prediction \cite{dellalibera2026dycast}. Its total-length constraint concerns decoded durations, not the number of encoded events.

Adaptive placement, inference-time rate adjustment, and exact event counts are distinct properties. FlexiCodec and DTM-Codec offer threshold-based adjustment; TFC, CodecSlime, and RubatoCodec enforce prescribed counts.

The dynamic tokenizer in The WER Trap normalizes frame weights to a prescribed token count before soft aggregation and FSQ \cite{zhang2026wertrap}. It follows Continuous Integrate-and-Fire (CIF), which accumulates learned weights into feature sums and uses target-length normalization during training \cite{dong2020cif}. Beyond waveform coding, Once-for-All compression extends CIF to inference-time rate adjustment \cite{chen2023onceforall}, and SpeechCLIP+ combines CIF with vector quantization and input-length-based targets \cite{wang2024speechclipplus}.

RubatoCodec shares cumulative budget allocation but separates boundary placement from event representation: deterministic statistics set event boundaries, and an aggregator combines the onset feature, the mean of the enclosed frame features, and the duration for each event. Table~\ref{tab:vfr-codec-comparison} compares VFR codecs in terms of temporal allocation, event-count control, timing representation, and training procedure.

\begin{table}[tbp]
\centering
\small
\caption{\textbf{Temporal allocation, event-count control, timing representations, and training procedures in VFR speech codecs.} Training procedures describe the recipes reported in the respective papers.}\label{tab:vfr-codec-comparison}
\begin{tabularx}{\linewidth}{P{2.1cm}LP{2.75cm}P{2.75cm}L}
\toprule
\textbf{System} & \textbf{Temporal allocation} & \textbf{Event-count control} & \textbf{Timing representation} & \textbf{Training procedure} \\
\midrule
TFC\newline \cite{zhang2025tfc} & Resolution selection on waveform entropy & Exact & Hierarchical mask & Single-stage \\
\rowrule
CodecSlime\newline \cite{wang2026codecslime} & DP segmentation on codec-feature distortion & Exact & Explicit lengths & Multi-stage (3 stages) \\
\rowrule
VARSTok\newline \cite{zheng2026varstok} & Contiguous clustering on codec latents & Content-dependent & In token ID & Adaptation (pretrained WavTokenizer) \\
\rowrule
FlexiCodec\newline \cite{li2026flexicodec} & Similarity merging on SenseVoice features & Content-dependent & Explicit lengths & Single-stage (frozen SenseVoice) \\
\rowrule
DTM-Codec\newline \cite{sohn2026dtmcodec} & Threshold crossing on accumulated codec-latent change & Content-dependent & Binary mask & Single-stage \\
\rowrule
DyCAST\newline \cite{dellalibera2026dycast} & MMS character alignment or boundary prediction on WavLM features & Content-dependent & Explicit or predicted lengths & Multi-stage (4 stages; frozen WavLM) \\
\rowrule
RubatoCodec & Budget-normalized firing on latent change and waveform flux & Exact & Explicit lengths & Single-stage (frozen w2v-BERT 2.0) \\
\bottomrule
\end{tabularx}
\end{table}

\subsection{Duration-aware and streaming speech generation}\label{sec:duration-aware-and-streaming-speech-generation}

Variable-duration speech generation must model both content and timing. Run-length Transformers model compressed event sequences \cite{dieleman2021variablerate}, while VARSTok includes duration in its vocabulary \cite{zheng2026varstok}. FlexiCodec predicts semantic tokens and frame lengths before flow-matching synthesis \cite{li2026flexicodec}; DyCAST also evaluates TTS with dynamic tokens and duration decoding \cite{dellalibera2026dycast}. QuarkAudio and FlexiSLM further explore adaptive representations for generation \cite{liu2025quarkaudio,li2026flexislm}.

Streaming synthesis additionally requires a decoding path that can render available codes incrementally. Mimi and Qwen3-TTS-Tokenizer-12Hz provide low-rate representations for streaming synthesis \cite{defossez2024moshi,hu2026qwen3tts}. MOSS-Audio-Tokenizer uses causal Transformer encoding and decoding \cite{gong2026mossaudiotokenizer}.

RubatoCodec supplies explicit event durations to a causal waveform decoder. A downstream model predicts code--duration pairs, which restore the decoding grid and can be synthesized as they become available.

\section{RubatoCodec Architecture}\label{sec:rubatocodec-architecture}

\begin{figure}[tbp]
\centering
\includegraphics[width=\linewidth,height=0.70\textheight,keepaspectratio]{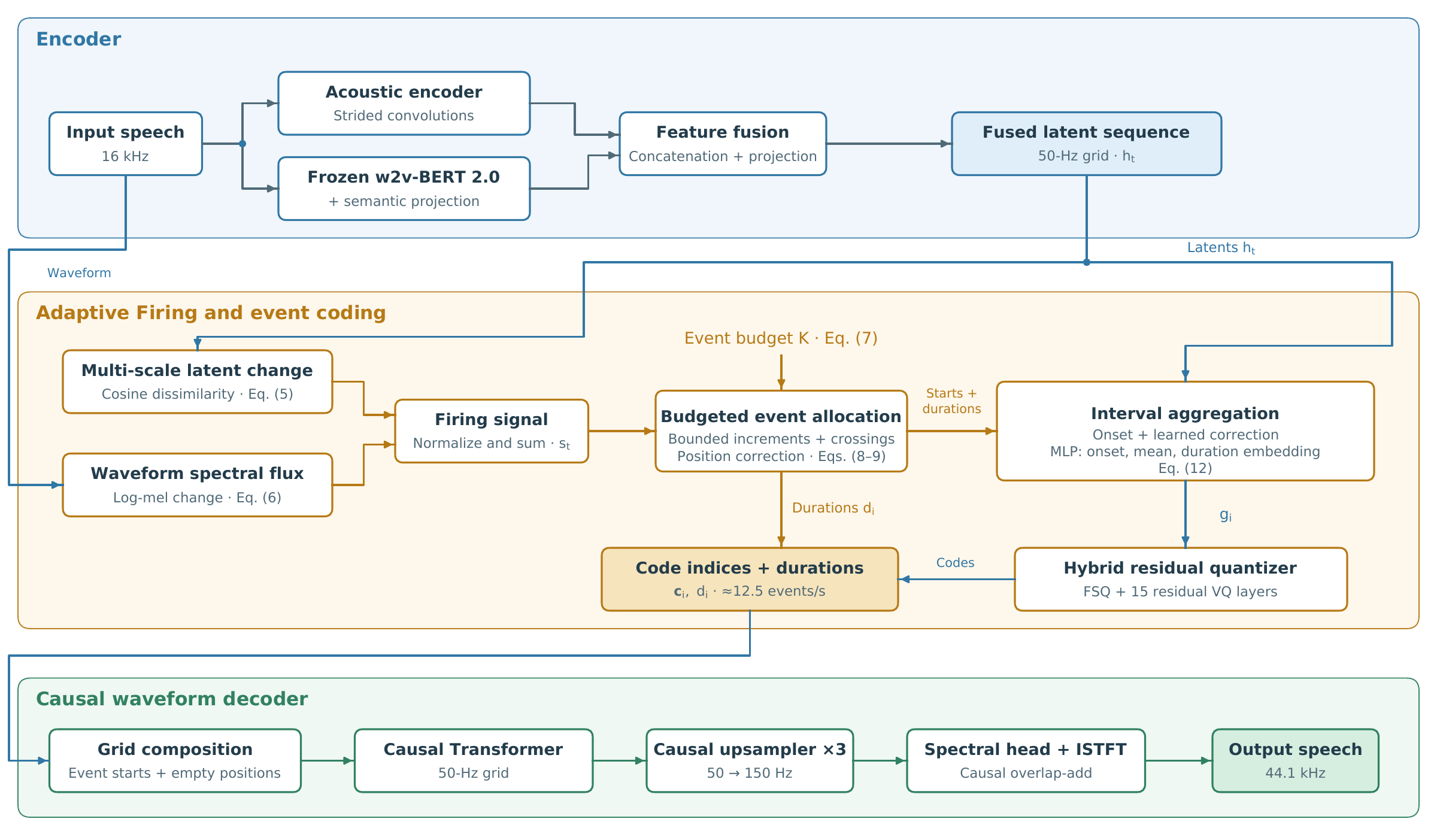}
\caption{\textbf{RubatoCodec architecture.} The codec builds on the X-Codec-2 encoder and decoder backbone \cite{ye2025llasa} and is shown at the default retention ratio of 0.25. Latent change and waveform spectral flux guide budgeted event allocation. The interval aggregator combines each event's onset feature with a learned correction based on that feature, the mean of its frame features, and its duration. Quantized event representations and durations restore the 50-Hz decoding grid; the Inworld-TTS-inspired upsampler increases the synthesis feature rate \cite{atamanenko2025tts1}.}\label{fig:codec-architecture}
\end{figure}

RubatoCodec retains X-Codec-2's acoustic/semantic encoding, feature fusion, semantic reconstruction pathway, and Transformer-based spectral synthesis backbone \cite{ye2025llasa}. We modify the temporal bottleneck through Adaptive Firing, interval aggregation, hybrid residual quantization, and duration-aware grid reconstruction (Figure~\ref{fig:codec-architecture}). Causal decoding enables incremental synthesis. RubatoCodec also uses a factor-three upsampler inspired by Inworld TTS that raises the synthesis feature rate \cite{atamanenko2025tts1}.

\subsection{Representation and signal path}\label{sec:representation-and-signal-path}

The encoder maps 16-kHz audio to \(T\) feature vectors on a nominal 50-Hz grid, with the aligned 44.1-kHz waveform serving as the reconstruction target. RubatoCodec partitions this grid into \(K\) nonempty intervals, each called an event. Event \(i\) is represented by \(Q\) quantizer indices \(\mathbf c_i=(c_{i,\allowbreak 0},\ldots,c_{i,Q-1})\), where \(Q\) is the number of quantizer layers (Section~\ref{sec:hybrid-residual-quantization-and-semantic-supervision}), and a duration \(d_i\in\mathbb N^+\) giving the number of 20-ms grid frames it spans. Let \(p_i\) be the index of the first grid frame of event \(i\), with \(p_0=0\):

\begin{equation}\label{eq:event-starts}
p_i=\sum_{j<i}d_j,\qquad \sum_{i=0}^{K-1}d_i=T. 
\end{equation}

Table~\ref{tab:codec-architecture} summarizes the architecture. The decoder emits \(882T\) output samples because \(44100/50=882\).

\begin{table}[tbp]
\centering
\small
\caption{\textbf{Main architecture in the default configuration.} Tensor shapes omit the batch dimension.}\label{tab:codec-architecture}
\begin{tabularx}{\linewidth}{P{0.28\linewidth}P{0.19\linewidth}L}
\toprule
\textbf{Stage} & \textbf{Temporal rate} & \textbf{Shape / specification} \\
\midrule
Waveform input & 16 kHz & Mono waveform \\
\rowrule
Acoustic convolutional encoder & 50 Hz & \(T\times1024\); strides \(2,\allowbreak 2,\allowbreak 4,\allowbreak 4,\allowbreak 5\) \\
\rowrule
Frozen w2v-BERT 2.0 \cite{seamlesscommunication2023seamless} + semantic projection & 50 Hz & Layer-16 features projected to \(T\times1024\) \\
\rowrule
Concatenation and prior projection & 50 Hz & \(T\times2048\) \\
\rowrule
Adaptive segmentation and aggregation & Approximately 12.5 events/s & \(K\times2048\) \\
\rowrule
Hybrid residual quantizer & Same event sequence & 16 indices/event; 4,096 values/layer \\
\rowrule
Sparse grid composition & 50 Hz & \(T\times2048\), followed by \(2048\to1024\) projection \\
\rowrule
Causal Transformer backbone & 50 Hz & 12 layers, width 1,024, 16 heads, MLP expansion 4 \\
\rowrule
Causal transposed-convolution upsampler & 150 Hz & Factor 3, kernel 9 \\
\rowrule
Spectral projection and ISTFT & 44.1 kHz & FFT 1,176; hop 294; \(882T\) samples \\
\bottomrule
\end{tabularx}
\end{table}

\subsection{Acoustic and semantic encoders}\label{sec:acoustic-and-semantic-encoders}

Following X-Codec-2 \cite{ye2025llasa}, the acoustic encoder uses strided convolutions with base width 48, residual dilations \(1,\allowbreak 3,\allowbreak 9\), and Snake-beta activations to produce acoustic features \(a_t\). Its total waveform hop size is 320 samples at 16 kHz, corresponding to a 50-Hz frame rate. A frozen w2v-BERT 2.0 model \cite{seamlesscommunication2023seamless} supplies layer-16 features \(b_t^\star\) at the same 50-Hz frame rate. A learned convolutional encoder transforms these into \(b_t\), which is fused with the acoustic features. The frozen features \(b_t^\star\) also serve as the semantic reconstruction targets in Section~\ref{sec:hybrid-residual-quantization-and-semantic-supervision}. Writing the aligned streams as \(a_t,b_t\in\mathbb R^{1024}\), their fusion is

\begin{equation}\label{eq:feature-fusion}
h_t=W_{\mathrm{prior}}[a_t;b_t]+b_{\mathrm{prior}}\in\mathbb R^{2048}. 
\end{equation}

\subsection{Adaptive Firing}\label{sec:adaptive-firing}

Adaptive Firing uses a firing signal derived from latent dissimilarity and waveform spectral flux to partition the 50-Hz latent sequence into \(K\) events. A separate aggregator combines the frame features within each event into one vector; unlike CIF \cite{dong2020cif}, the firing signal determines only the boundaries and does not weight the pooled features.

\subsubsection{Firing signal}\label{sec:firing-signal}

The firing signal sums two mean-normalized components, multi-scale latent dissimilarity and log-mel spectral flux:

\begin{equation}\label{eq:firing-signal}
s_t=s_t^{\mathrm{latent}}+\mathcal N(f)_t. 
\end{equation}

Here, \(\mathcal N\) denotes sequence-wise mean normalization, \(s_t^{\mathrm{latent}}\) is the normalized multi-scale latent-dissimilarity component, and \(f_t\) is the log-mel spectral flux. Each is defined below.

\textbf{Normalization.} Define sequence-wise mean normalization for a nonnegative scalar sequence \(v\) as

\begin{equation}\label{eq:mean-normalization}
\mathcal N(v)_t=\frac{v_t}{\max(\epsilon,T^{-1}\sum_u v_u)},\qquad \epsilon=10^{-8}. 
\end{equation}

\textbf{Latent dissimilarity.} The multi-scale latent-dissimilarity component uses offset pairs \(\mathcal O=\{(1,\allowbreak 0),(1,\allowbreak 1),(2,\allowbreak 1)\}\):

\begin{equation}\label{eq:latent-dissimilarity}
\begin{aligned}
\ell_t^{a,b}&=\max\{0,1-\cos(h_{t-a},h_{t+b})\},\\
s_t^{\mathrm{latent}}&=\mathcal N\!\left(\sum_{(a,b)\in\mathcal O}\mathcal N(\ell^{a,b})\right)_t.
\end{aligned} 
\end{equation}

Unavailable edge comparisons are set to zero. This latent-dissimilarity component follows the use of feature dissimilarity for dynamic chunking \cite{hwang2026hnet}, combining comparisons at multiple temporal offsets.

\textbf{Spectral flux.} The waveform component is log-mel spectral flux. Let \(M_{m,t}\) denote the mel-spectrogram magnitude at mel bin \(m\) and frame \(t\), computed with 80 mel bins, a 1,024-point STFT, and a 320-sample hop at 16 kHz. The flux is

\begin{equation}\label{eq:spectral-flux}
f_t=\frac1{80}\sum_{m=1}^{80}\left|\log(M_{m,t}+10^{-5})-\log(M_{m,t-1}+10^{-5})\right|. 
\end{equation}

Flux is computed using a centered STFT with reflection padding. Its first value and final two values are set to zero to suppress boundary artifacts, after which the sequence is truncated or zero-padded on the right to match the latent-grid length.

Normalization makes the combination depend on the relative temporal distributions of its components rather than their raw scales. The waveform component also provides an informative allocation signal early in training, before the acoustic encoder has learned meaningful features.

\subsubsection{Budget-constrained event allocation}\label{sec:budget-constrained-event-allocation}

For a target retention ratio \(0<R\leq1\), the integer event budget is

\begin{equation}\label{eq:event-budget}
K=\max\{1,\min(T,\operatorname{round}(RT))\}. 
\end{equation}

Here, \(\operatorname{round}(\cdot)\) rounds half to even. The first frame always fires and consumes one unit of this budget, leaving \(L=T-1\) frames to receive \(M=K-1\) further event starts. For \(M>0\), distribute the remaining budget over these \(L\) frames using bounded increments

\begin{equation}\label{eq:bounded-increments}
\delta_t=\min(\alpha s_t,1),\qquad
\sum_{t=1}^{T-1}\delta_t=M. 
\end{equation}

We find \(\alpha\) by bisection, after expanding the upper bracket until the constraint is met or a limit is reached.

For cumulative increments \(C_t=\sum_{u=1}^{t}\delta_u\), provisional event starts are the first crossings of thresholds \(j-\tfrac12\), for \(j=1,\ldots,M\). The last threshold lies half a unit below \(M\), providing a margin against endpoint rounding error. These cumulative threshold crossings follow the integrate-and-fire principle used in CIF \cite{dong2020cif}; here, bounded increments and the discrete position correction below enforce the prescribed number of distinct event starts.

Let \(q_j\), for \(j=0,\ldots,M-1\), be the zero-based position of the \((j+1)\)-th threshold crossing among these \(L\) frames. If a threshold is not reached, its position is set to \(L\). A monotone correction handles coincident positions and unreachable thresholds:

\begin{equation}\label{eq:boundary-correction}
u_j=j+\max_{0\le\ell\le j}(q_\ell-\ell),\qquad
v_j=\min(u_j,L-M+j),\qquad p_{j+1}=1+v_j. 
\end{equation}

\begin{figure}[tbp]
\centering
\includegraphics[width=\linewidth,height=0.70\textheight,keepaspectratio]{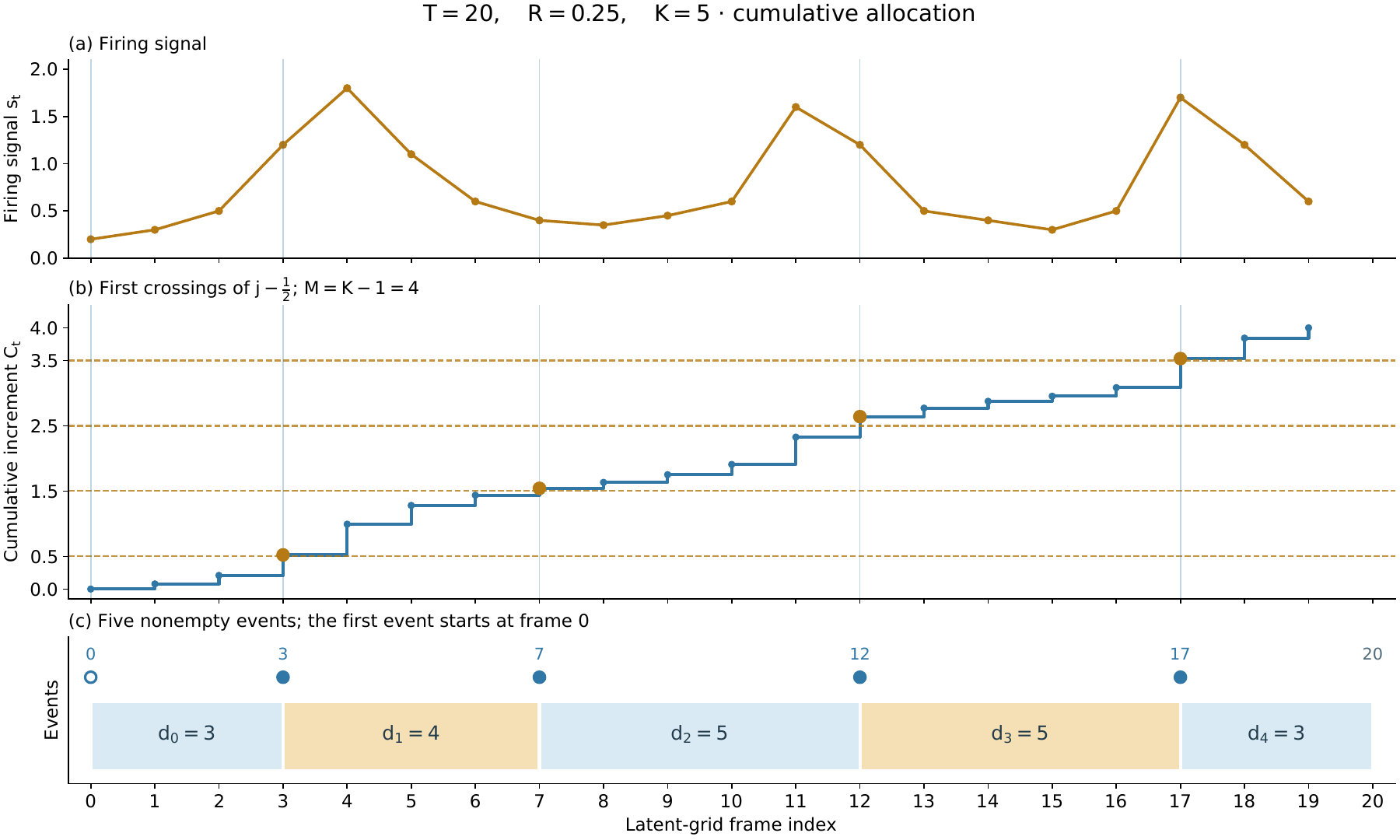}
\caption{\textbf{Cumulative allocation for a synthetic firing signal.} The example uses \(T=20\), \(R=0.25\), and \(K=5\). (a) Firing signal. (b) Cumulative bounded increments cross thresholds 0.5, 1.5, 2.5, and 3.5 at frames 3, 7, 12, and 17. (c) Together with the fixed start at frame 0, these crossings define five events spanning 3, 4, 5, 5, and 3 grid frames, respectively. All positions are computed from the displayed signal using equations (\ref{eq:bounded-increments}--\ref{eq:boundary-correction}); no position correction is needed in this example.}\label{fig:firing-timeline}
\end{figure}

The running maximum and capacity bound in \eqref{eq:boundary-correction} produce strictly increasing positions satisfying \(0\le v_j\le L-M+j<L\). Together with \(p_0=0\), these give exactly \(K\) distinct event starts. For \(K=1\), the single event covers the whole grid. Figure~\ref{fig:firing-timeline} illustrates the construction on a short synthetic firing signal.

Two cases need separate handling. When the firing signal over these \(L\) frames is negligible, we use uniform increments \(\delta_t=M/L\). When the equality in \eqref{eq:bounded-increments} is infeasible, we retain the increments returned by the bounded search and apply the same position correction in \eqref{eq:boundary-correction}, preserving exactly \(K\) nonempty events.

The budget is defined over the \(T\) encoder frames, corresponding to a represented duration of \(T/50\) seconds after any boundary padding. The utterance-average event rate on this grid is \(50K/T\). Away from the minimum-one-event clamp,

\begin{equation}\label{eq:event-rate-bound}
\left|\frac{50K}{T}-50R\right|\le\frac{25}{T}\ \text{events/s}. 
\end{equation}

When \(RT\) is integral, the grid-average event rate is exactly \(50R\). Allocation costs \(O(T)\) per bisection iteration, plus a linear-time threshold search and position correction.

Event allocation is deterministic and non-differentiable. Reconstruction and representation losses train the features and interval aggregator through the selected event representations. Boundaries therefore evolve during training as the features change, while the event count is enforced directly without a rate loss. The allocation constrains the total event count without imposing a maximum event duration. Sustained high-change regions can receive multiple events on different grid frames.

\subsubsection{Interval aggregation}\label{sec:interval-aggregation}

Define \(p_K=T\), \(d_i=p_{i+1}-p_i\), and the mean of the latent frame features within event \(i\):

\begin{equation}\label{eq:event-aggregation}
\bar h_i=\frac1{d_i}\sum_{t=p_i}^{p_{i+1}-1}h_t. 
\end{equation}

The quantizer input is the onset feature plus a learned correction:

\begin{equation}\label{eq:quantizer-input}
g_i=h_{p_i}+\operatorname{MLP}\!\left([h_{p_i};\bar h_i;E_d(\min(d_i,64))]\right). 
\end{equation}

The residual connection from the onset feature preserves information at the event start; the learned correction incorporates the mean and duration. The learned duration embedding \(E_d\) is indexed up to 64, without capping the actual event duration.

\subsection{Hybrid residual quantization and semantic supervision}\label{sec:hybrid-residual-quantization-and-semantic-supervision}

\subsubsection{Residual quantization}\label{sec:residual-quantization}

RubatoCodec uses \(Q=16\) quantizer layers. Layer zero is finite scalar quantization with levels \([8,\allowbreak 8,\allowbreak 8,\allowbreak 8]\), yielding \(8^4=4096\) indices \cite{mentzer2024fsq}. Layers 1--15 use DAC-style vector quantization with a factorized 8-dimensional codebook lookup, normalized lookup vectors, and 4,096 entries per layer \cite{kumar2023dac}. Each layer therefore requires 12 fixed-width index bits.

Each layer quantizes the residual left by the preceding layers. With \(r_{i,\allowbreak 0}=g_i\), the reconstructed event vector at depth \(D\) is

\begin{equation}\label{eq:residual-quantization}
q_{i,j}=Q_j(r_{i,j}),\quad r_{i,j+1}=r_{i,j}-q_{i,j},\quad
\hat g_i^{(D)}=\sum_{j=0}^{D-1}q_{i,j}. 
\end{equation}

Here, \(Q_j\) denotes the quantizer at layer \(j\). Using the first \(D\) layers adjusts the number of code indices per event while retaining the same event boundaries and durations.

\subsubsection{Semantic distillation}\label{sec:semantic-distillation}

Semantic supervision encourages the first layer to preserve linguistic information for downstream modeling. We adapt X-Codec-2's semantic reconstruction objective \cite{ye2025llasa}, sharing the motivation of Mimi \cite{defossez2024moshi} but using a serial residual chain rather than parallel semantic and acoustic quantizers.

A training-only semantic head predicts frozen w2v-BERT 2.0 features from the first-layer reconstruction. Its input uses the same sparse 50-Hz grid composition and offset embeddings as the waveform decoder (Section~\ref{sec:causal-grid-decoding}). Durations determine the grid placement; no separate duration embedding is added to the head. The semantic head consists of a linear projection from 2,048 to 1,024 dimensions followed by a residual convolutional network with four layers of kernel size three. Symmetric padding gives the head access to four grid frames on either side, including neighboring events within that range.

The head's output \(\hat b_t\) is trained to match the frozen teacher features \(b_t^\star\) at event starts using a dimension-normalized masked mean squared error:

\begin{equation}\label{eq:semantic-loss}
\mathcal L_{\mathrm{sem}}=
\frac{\sum_t m_t\|\hat b_t-b_t^\star\|_2^2}
     {D_s\sum_t m_t},\qquad m_t=\mathbf1[t\in\{p_i\}]. 
\end{equation}

Here, \(D_s\) is the feature dimension. The mask \(m_t\) assigns one teacher target at each event's onset, where its first-layer representation enters the grid, leaving the residual layers to encode additional waveform detail.

\subsection{Causal grid decoding}\label{sec:causal-grid-decoding}

The decoder places each dequantized event vector at its starting grid position and uses a learned empty vector between starts. Let \(E_r\) denote the learned offset embedding. With offset \(r=t-p_i\) inside event \(i\),

\begin{equation}\label{eq:decoding-grid}
z_t=\begin{cases}
\hat g_i^{(D)}+E_r(0),&r=0,\\
e_{\mathrm{empty}}+E_r(\min(r,64)),&r>0.
\end{cases} 
\end{equation}

Duration \(d_i\) sets the number of grid frames, and the offset embedding identifies their positions within the event. The causal decoder can attend to the onset feature to synthesize the audio for that event. Appendix~\ref{sec:sparse-versus-repeated-event-representations} compares this sparse input with repeating the event vector and supports the sparse design for signal fidelity.

The decoder retains the X-Codec-2 Transformer backbone \cite{ye2025llasa}, with causal attention and convolutional dependencies. A projection maps the 2,048-dimensional grid to this 1,024-dimensional backbone, which uses a causal embedding convolution, 12 Transformer layers, rotary positions, and 16 attention heads. A factor-three transposed-convolution upsampler, inspired by the super-resolution module in Inworld TTS \cite{atamanenko2025tts1}, produces 150-Hz synthesis features. A Vocos-style spectral head predicts magnitude and phase, followed by forward-window ISTFT with a hop size of 294 samples for 44.1-kHz output \cite{siuzdak2024vocos}. Convolutions are left-padded, transposed convolutions are right-trimmed, and attention excludes future frames, so synthesis depends only on the current and preceding events.

The ISTFT places each 1,176-sample synthesis window at the current frame's hop position and extends it forward in time. Each frame finalizes the next 294 samples, carrying waveform and window-normalization overlap buffers into subsequent frames. With three synthesis frames per grid step, an event of duration \(d_i\) accounts for \(882d_i\) output samples. At startup, normalization assumes preceding silent windows, yielding a 20-ms fade-in; at termination, the remaining window overhang is discarded rather than emitted as additional audio.

During inference, the decoder periodically discards its cached states and rebuilds them from a window of recent events beginning at an event onset, which limits long-utterance extrapolation. At each such history recomputation, a short crossfade blends the audio decoded before and after the recomputation. Streaming decoding processes incoming events incrementally, reusing decoder states between history recomputations and briefly buffering audio for the crossfade.

\subsection{Model size and bitrate}\label{sec:model-size-and-bitrate}

RubatoCodec contains approximately 246.1M codec parameters: 72.3M in the encoding path, 172.7M in the decoding path, and 1.1M in the quantizer. The additional frozen w2v-BERT 2.0 encoder has approximately 580M parameters, bringing the complete model to approximately 826M. Token-to-waveform synthesis uses the decoding path and quantizer without the encoding components.

Each event contains \(D\) code indices of 12 bits each. For \(K\) events over a represented duration of \(\tau=T/50\) seconds, the content bitrate is

\begin{equation}\label{eq:bitrate}
B_{\mathrm{content}}=\frac{12DK}{\tau}. 
\end{equation}

At 12.5 events per second, depths 1, 4, 8, and 16 correspond to content bitrates of 0.15, 0.60, 1.20, and 2.40 kbit/s, respectively. Timing bitrate is estimated from the empirical entropy \(\hat H(d)\) of event durations, pooled within each evaluation dataset (Section~\ref{sec:metrics}). The estimated total bitrate is \(B_{\mathrm{total}}=B_{\mathrm{content}}+(K/\tau)\hat H(d)\).

\section{Training Data and Optimization}\label{sec:training-data-and-optimization}

We train adaptive and fixed-rate architectures on a shared English subset at 6.25, 12.5, and 25 events/s, and adaptive and uniform-stride models at 12.5 events/s for the placement comparison. We then train a multilingual RubatoCodec on a larger speech mixture for comparison with external codecs.

\subsection{Objectives and phases}\label{sec:objectives-and-phases}

Both settings combine reconstruction, semantic supervision, quantization, adversarial, and perceptual objectives:

\begin{equation}\label{eq:training-objective}
\mathcal L_G=15\mathcal L_{\mathrm{mel}}+5\mathcal L_{\mathrm{sem}}+
\mathcal L_{\mathrm{VQ}}+
\mathbf1_{\mathrm{phase}\ge2}(\mathcal L_{\mathrm{adv}}+\mathcal L_{\mathrm{FM}})+
\mathbf1_{\mathrm{phase}=3}\mathcal L_{\mathrm{perc}}. 
\end{equation}

\textbf{Multi-resolution mel reconstruction.} Following DAC \cite{kumar2023dac}, \(\mathcal L_{\mathrm{mel}}\) sums mean absolute log-mel differences at 44.1 kHz over eight resolutions: window lengths \(32,\allowbreak 64,\allowbreak 128,\allowbreak 256,\allowbreak 512,\allowbreak 1024,\allowbreak 2048,\allowbreak 4096\), quarter-window hops, and corresponding mel-bin counts \(5,\allowbreak 10,\allowbreak 20,\allowbreak 40,\allowbreak 80,\allowbreak 160,\allowbreak 320,\allowbreak 640\). Magnitudes are clamped to \(10^{-5}\) before the base-10 logarithm.

\textbf{Semantic reconstruction.} Equation \eqref{eq:semantic-loss} supervises the first layer with frozen w2v-BERT 2.0 layer-16 features. RubatoCodec evaluates the loss at event starts. For fixed-rate controls, the 50-Hz teacher features are averaged over nonoverlapping groups of 8, 4, or 2 frames to match the 6.25-, 12.5-, or 25-Hz latent grid, respectively. These averaged features serve both as inputs to the semantic encoder for feature fusion and as targets for the semantic loss, which is evaluated at every latent position.

\textbf{Quantization.} \(\mathcal L_{\mathrm{VQ}}\) sums squared-error commitment and codebook losses over residual VQ layers \cite{kumar2023dac}, with weights 0.25 and 1.0 and stop-gradients on the code vector and encoder representation, respectively. FSQ has neither loss \cite{mentzer2024fsq}.

\textbf{Adversarial training and feature matching.} We use HiFi-GAN's multi-period discriminator with periods \(2,\allowbreak 3,\allowbreak 5,\allowbreak 7,\allowbreak 11,\allowbreak 17,\allowbreak 23,\allowbreak 37\) \cite{kong2020hifigan} and a UnivNet-style multi-resolution STFT-magnitude discriminator \cite{jang2021univnet} with FFT/window sizes \(216,\allowbreak 348,\allowbreak 568,\allowbreak 920,\allowbreak 1494,\allowbreak 2414,\allowbreak 3908,\allowbreak 6328\) and half-window hops. Least-squares GAN losses use target labels one for real audio and zero for reconstructed audio, while the generator targets one \cite{kong2020hifigan}. \(\mathcal L_{\mathrm{FM}}\) sums mean absolute differences of intermediate activations across both discriminator families, with detached target features. Discriminator parameters remain fixed during generator updates.

\textbf{Perceptual reconstruction.} Following Llasa's late-stage waveform supervision \cite{ye2025llasa}, \(\mathcal L_{\mathrm{perc}}\) compares frozen speech-model features of target and reconstructed audio resampled to 16 kHz. We use wav2vec2-XLS-R-300M \cite{babu2022xlsr}, averaging mean squared feature error over layers 10--21. Gradients propagate through the frozen extractor to the codec.

\textbf{Training phases.} The first 10\% of updates use mel, semantic, and quantization losses; the next 75\% add adversarial and feature-matching losses; the final 15\% also use perceptual loss. Trainable codec components start from scratch, while both pretrained feature extractors remain frozen.

\subsection{Training for architecture comparisons}\label{sec:training-for-architecture-comparisons}

Fixed-rate controls downsample before quantization to produce uniformly spaced codes. The adaptive architecture retains the 50-Hz grid and aggregates features within variable-duration events, restoring their representations to that grid for decoding.

The adaptive-versus-fixed-rate comparisons use the same approximately 5,000-hour English subset of HiFiTTS-2's 44-kHz partition \cite{langman2025hifitts2}. We verified that the full HiFiTTS-2 dataset has no speaker overlap with LibriSpeech test-clean. Models are trained separately at each target event rate for 100,000 steps using aligned six-second crops. Both models use a global batch of 64 and the objectives in Section~\ref{sec:objectives-and-phases}; detailed training settings are provided in Appendix~\ref{sec:codec-training}.

The adaptive models use the budget-constrained allocation in Section~\ref{sec:budget-constrained-event-allocation}, with retention ratios of 0.125, 0.25, and 0.5 for nominal event rates of 6.25, 12.5, and 25 Hz. Quantizer configuration, decoder depth and width, output sample rate, losses, and optimization are shared; encoder strides and synthesis upsampling differ to accommodate the temporal grids (Table~\ref{tab:architecture-training-configurations}).

\begin{table}[tbp]
\centering
\footnotesize
\caption{\textbf{Temporal configurations of the main architecture comparisons.} Encoder hop sizes, FFT sizes, and ISTFT hop sizes are given in samples. Each adaptive rate uses a separately trained model.}\label{tab:architecture-training-configurations}
\begin{tabularx}{\linewidth}{LP{2.5cm}P{2.3cm}R{1.8cm}P{1.6cm}P{1.65cm}P{1.9cm}}
\toprule
\textbf{Architecture} & \textbf{Target event rate} & \textbf{Retention ratio} & \textbf{Encoder hop at 16 kHz} & \textbf{Latent grid} & \textbf{Decoder upsampling factors} & \textbf{ISTFT FFT / hop} \\
\midrule
Adaptive & 6.25 / 12.5 / 25 Hz & 0.125 / 0.25 / 0.5 & 320 & 50 Hz & 3 & 1,176 / 294 \\
Fixed & 6.25 Hz & --- & 2,560 & 6.25 Hz & 4 \(\times\) 3 & 2,352 / 588 \\
Fixed & 12.5 Hz & --- & 1,280 & 12.5 Hz & 4 \(\times\) 3 & 1,176 / 294 \\
Fixed & 25 Hz & --- & 640 & 25 Hz & 3 \(\times\) 2 & 1,176 / 294 \\
\bottomrule
\end{tabularx}
\end{table}

To isolate the effect of adaptive event placement, we additionally train adaptive and uniform-stride aggregation models on the full English 44-kHz training partition of HiFiTTS-2 for one million steps, using a nominal event rate of 12.5 Hz. Both retain the same 50-Hz encoder and decoder grids, interval aggregator, and hybrid residual quantizer. The uniform control forms four-frame events. The global batch remains 64.

The firing-signal, decoder-causality, grid-composition, and firing-rule ablations are described in Appendix~\ref{sec:supporting-codec-ablations}.

\subsection{Multilingual model training}\label{sec:multilingual-model-training}

The multilingual model is trained on approximately 375,000 hours of public and internal speech data. Public sources include Emilia-YODAS (English, Japanese, Chinese, German, French, and Korean) \cite{he2025emilia}, the 44-kHz subset of HiFiTTS-2 \cite{langman2025hifitts2}, ReazonSpeech \cite{yin2023reazonspeech}, The People's Speech (excluding ShareAlike-licensed data) \cite{galvez2021peoplesspeech}, VoxPopuli \cite{wang2021voxpopuli}, Multilingual LibriSpeech (excluding English) \cite{pratap2020mls}, Omnilingual ASR Corpus \cite{omnilingualasrteam2025omnilingualasr}, and IndicVoices \cite{javed2024indicvoices}. Combined with internal data, the mixture covers read, spontaneous, conversational, broadcast, and expressive speech.

English, Japanese, and Chinese account for approximately 44\%, 27\%, and 13\% of the training data, respectively, with other languages making up the remaining 16\%.

Waveforms are resampled to 16 kHz for encoding and 44.1 kHz for reconstruction targets, with aligned six-second random crops; shorter examples are zero-padded.

The multilingual model uses retention ratio 0.25 and is trained for 1.5 million steps with a global batch of 160, following the loss phases in Section~\ref{sec:objectives-and-phases}. Optimization settings and loss schedules are detailed in Appendix~\ref{sec:codec-training}.

\section{Evaluation}\label{sec:evaluation}

\subsection{Evaluation setup}\label{sec:evaluation-setup}

\subsubsection{Datasets}\label{sec:datasets}

We evaluate speech reconstruction on five datasets covering English, Japanese, and Mandarin Chinese, with 10,563 utterances in total (Table~\ref{tab:evaluation-datasets}). The adaptive-versus-fixed comparisons use LibriSpeech test-clean, while the multilingual model and external codecs are evaluated across all five datasets.

\begin{table}[tbp]
\centering
\small
\caption{\textbf{Speech reconstruction evaluation datasets.}}\label{tab:evaluation-datasets}
\begin{tabularx}{\linewidth}{P{0.22\linewidth}P{0.15\linewidth}R{0.12\linewidth}L}
\toprule
\textbf{Dataset} & \textbf{Language} & \textbf{Utterances} & \textbf{Evaluation subset} \\
\midrule
LibriSpeech \cite{panayotov2015librispeech} & English & 2,620 & Full test-clean split \\
\rowrule
VCTK 0.92 \cite{yamagishi2019vctk} & English & 1,308 & Filtered mic1 recordings; 12 utterances per speaker \\
\rowrule
JVS \cite{takamichi2019jvs} & Japanese & 3,000 & nonpara30: 30 utterances from each of 100 speakers \\
\rowrule
JVNV \cite{xin2024jvnv} & Japanese & 1,615 & Full corpus: four speakers, six emotions, including nonverbal vocalizations \\
\rowrule
seed-tts-eval \cite{anastassiou2024seedtts} & Mandarin Chinese & 2,020 & test-zh recordings \\
\bottomrule
\end{tabularx}
\end{table}

For VCTK, we use the silence-trimmed mic1 recordings and retain utterances with nonempty transcripts and durations between 1 and 40 seconds. We then randomly select 12 eligible utterances per speaker, yielding 1,308 utterances from 109 speakers. The same subset is used for all codecs.

\subsubsection{Metrics}\label{sec:metrics}

Reconstruction-quality metrics use audio resampled to 16 kHz.

\textbf{Event rate and bitrate.} We compute event rate and bitrate per utterance using the duration represented by the codec's temporal grid, and report their arithmetic means within each dataset. Event rate is the number of encoded events per second; for fixed-rate codecs, each frame counts as one event. Bitrate includes code-index and timing costs (Section~\ref{sec:model-size-and-bitrate}). Explicit timing symbols are assigned their dataset-pooled empirical entropy per symbol; timing embedded in token indices is included in the code-index cost.

\textbf{Intelligibility.} We report dWER for English and dCER for Japanese and Mandarin Chinese, comparing the ASR transcript of reconstructed speech against that of the original waveform. English and Japanese speech are transcribed with Whisper-large-v3 \cite{radford2023whisper}, and Mandarin with Paraformer-zh \cite{gao2022paraformer}. The same normalization is applied to both transcripts: Whisper's English normalization for English, Whisper's basic normalization with spaces removed for Japanese, and the seed-tts-eval procedure for Mandarin \cite{anastassiou2024seedtts}. Lower dWER and dCER indicate better preservation of the original recording's ASR output.

\textbf{Speech quality.} We measure PESQ-WB \cite{itut2001pesq} and STOI \cite{taal2011stoi} against the original waveform, together with the reference-free predictors UTMOSv2 \cite{baba2024utmosv2} and DNSMOS \cite{reddy2022dnsmos}. The main tables use the overall-quality output of DNSMOS. Higher values are better for these metrics. For PESQ and STOI, the resampled waveforms are truncated to their common length from the start, without explicit delay correction or temporal warping.

\textbf{Speaker preservation.} Speaker similarity (SIM) is the cosine similarity between embeddings of the original and reconstructed utterances, extracted with a fine-tuned WavLM-Large speaker verification model \cite{chen2022wavlm}. Higher values indicate better preservation of speaker characteristics.

For codec reconstruction, dWER and dCER are micro-averaged within each dataset: total edit counts are divided by total reference words or characters. Other reconstruction-quality scores are averaged over utterances.

\subsubsection{Compared codecs}\label{sec:compared-codecs}

Table~\ref{tab:baseline-codec-specifications} summarizes the evaluated codecs, including training data, pretrained components, rates, streaming support, and approximate parameter counts. Encoding-time feature models are counted separately from codec parameters. Public code, checkpoint, dataset, and evaluation-tool links are collected in Appendix~\ref{sec:code-models-and-data}.

TFC \cite{zhang2025tfc} is excluded from the empirical comparisons because we could not locate publicly available code or pretrained weights. VARSTok's evaluated operating point falls below the bitrate range considered here and is omitted from the comparisons \cite{zheng2026varstok}.

\begin{table}[tbp]
\centering
\footnotesize
\setlength{\tabcolsep}{2pt}
\caption{\textbf{Specifications of the evaluated codecs.}}\label{tab:baseline-codec-specifications}
\begin{tabularx}{\linewidth}{P{2.3cm}LLLR{1.1cm}P{1.2cm}R{1.15cm}R{1.4cm}}
\toprule
\textbf{Codec} & \textbf{Training data} & \textbf{Semantic features or supervision} & \textbf{Event rate / allocation} & \textbf{Output (kHz)} & \textbf{Causal\newline enc. / dec.} & \textbf{Codec\newline params.\newline (M)} & \textbf{Extra enc.\newline params.\newline (M)\textsuperscript{a}} \\
\midrule
EnCodec \cite{defossez2023encodec} & Speech, music, and general audio & None & 75 Hz & 24 & Yes / Yes & 14.9 & --- \\
\rowrule
DAC (24 kHz) \cite{kumar2023dac} & Speech, music, and environmental audio, including VCTK & None & 75 Hz & 24 & No / No & 74.7 & --- \\
\rowrule
BigCodec \cite{xin2024bigcodec} & LibriSpeech, 960 h & None & 80 Hz & 16 & No / No & 159.4 & --- \\
\rowrule
WavTokenizer \cite{ji2025wavtokenizer} & Speech, music, and general audio & None & 75 Hz & 24 & No / No & 80.6 & --- \\
\rowrule
X-Codec-2 \cite{ye2025llasa} & Emilia and Multilingual LibriSpeech, approximately 150k h & Frozen w2v-BERT 2.0 features & 50 Hz & 16 & No / No & 242.2 & 580 \\
\rowrule
Mimi \cite{defossez2024moshi} & Moshi speech collection, 7M h\textsuperscript{b} & WavLM distillation & 12.5 Hz & 24 & Yes / Yes & 79.3 & --- \\
\rowrule
MiMo-Audio-Tokenizer \cite{xiaomillmcoreteam2025mimoaudio} & Multitask audio training, \textgreater11M h & Joint semantic and acoustic objectives & 25 Hz & 24 & No / No\textsuperscript{c} & 1,284.7 & --- \\
\rowrule
MOSS-Audio-Tokenizer \cite{gong2026mossaudiotokenizer} & Multilingual speech and general audio, 3M h & Joint audio--text language modeling & 12.5 Hz & 24 & Yes / Yes & 1,775 & --- \\
\rowrule
Qwen3-TTS-Tokenizer-12Hz \cite{hu2026qwen3tts} & Qwen3-TTS speech collection, 5M h\textsuperscript{b} & WavLM distillation & 12.5 Hz & 24 & Yes / Yes & 153.7 & --- \\
\rowrule
DTM-Codec \cite{sohn2026dtmcodec} & LibriSpeech, 960 h & None & 80-Hz target; dynamic token masking & 16 & No / No & 125.7 & --- \\
\rowrule
CodecSlime \cite{wang2026codecslime} & LibriSpeech, 960 h & None & 40 / 80 Hz targets; exact segment count & 16 & No / No & 159.4 & --- \\
\rowrule
FlexiCodec \cite{li2026flexicodec} & Libri-Light, 54k h & SenseVoice features & 12.5-Hz base; similarity-based merging & 16 & No / No & 215 & 234 \\
\rowrule
DyCAST \cite{dellalibera2026dycast} & LibriTTS, 585 h & WavLM features and CTC alignment & Character alignment (CA) or boundary prediction (BP) & 16 & No / No & 268.8 & 964.8 (CA); --- (BP)\textsuperscript{d} \\
\rowrule
RubatoCodec & Multilingual speech, 375k h & w2v-BERT 2.0 feature fusion and first-codebook supervision & 12.5 Hz nominal; exact utterance-level event budget & 44.1 & No / Yes & 246.1 & 580 \\
\bottomrule
\end{tabularx}

\medskip
\begin{minipage}{\linewidth}
\textsuperscript{a} Additional models required during encoding; training-only teachers are excluded.\\
\textsuperscript{b} System-level training data; tokenizer-specific hours are not separately reported.\\
\textsuperscript{c} Causal audio decoder followed by a vocoder with look-ahead.\\
\textsuperscript{d} MMS-1b-all CTC aligner, used only for character-aligned encoding (CA).
\end{minipage}
\end{table}

For DTM-Codec, we evaluate the released 80-Hz checkpoint using its stored threshold (approximately 0.1356), without recalibration on the evaluation datasets \cite{sohn2026dtmcodec}. This label indicates a nominal target rather than an enforced event rate. FlexiCodec is evaluated at merge thresholds of 0.91 and 1.0 \cite{li2026flexicodec}.

CodecSlime is evaluated with FSQ18K at a prescribed 40-Hz rate and at 80 Hz without merging \cite{wang2026codecslime}.

For DyCAST, we evaluate character-aligned encoding and boundary prediction with a minimum gap of one frame \cite{dellalibera2026dycast}. All DyCAST variants use the tokens-plus-utterance-length decoding mode: individual durations are predicted subject to the supplied total frame count, with 64 bits per event and a 16-bit length field per utterance. Character alignment uses MMS with the matching English, Japanese, or Mandarin language adapter and its audio-derived CTC predictions, without reference transcripts. Reconstruction uses the decoder's native output length and boundary handling, with retrieval disabled.

These encoding settings are held fixed across datasets; their realized event rates are reported in the result tables and compared in Section~\ref{sec:event-rate-control-across-datasets}.

\subsection{Controlled comparisons}\label{sec:controlled-comparisons}

\subsubsection{Adaptive versus fixed-rate architectures}\label{sec:adaptive-versus-fixed-rate-architectures}

We compare the adaptive architecture with fixed-rate downsampling controls on LibriSpeech test-clean under the matched training conditions in Section~\ref{sec:training-for-architecture-comparisons}. The architectures differ in temporal aggregation, latent-grid resolution, and decoder upsampling factors (Table~\ref{tab:architecture-training-configurations}).

\paragraph*{Comparison at 12.5 Hz}

At the default event rate of 12.5 Hz, the adaptive codec additionally transmits event durations, which add approximately 0.028 kbit/s to the reported bitrate (Table~\ref{tab:adaptive-fixed-rate-12hz}).

\begin{table}[tbp]
\centering
\footnotesize
\setlength{\tabcolsep}{3pt}
\caption{\textbf{Adaptive versus fixed-rate coding at 12.5 Hz on LibriSpeech test-clean.} Depth denotes the number of active quantizer layers. Bold marks the best score within each quantizer depth.}\label{tab:adaptive-fixed-rate-12hz}
\begin{tabularx}{\linewidth}{P{1.9cm}YYYYYYYY}
\toprule
\textbf{Codec} & \textbf{Depth} & \textbf{Bitrate (kbit/s)} & \textbf{UTMOSv2 \(\uparrow\)} & \textbf{DNSMOS \(\uparrow\)} & \textbf{PESQ-WB \(\uparrow\)} & \textbf{STOI \(\uparrow\)} & \textbf{dWER (\%) \(\downarrow\)} & \textbf{SIM \(\uparrow\)} \\
\midrule
Fixed-rate & 8 & 1.20 & 3.365 & 3.293 & 2.322 & 0.915 & 0.84 & \textbf{0.776} \\
Adaptive & 8 & 1.23 & \textbf{3.448} & \textbf{3.314} & \textbf{2.379} & \textbf{0.917} & \textbf{0.80} & 0.757 \\
\midrule
Fixed-rate & 16 & 2.40 & 3.363 & 3.288 & 2.442 & 0.923 & 0.72 & \textbf{0.810} \\
Adaptive & 16 & 2.43 & \textbf{3.444} & \textbf{3.307} & \textbf{2.556} & \textbf{0.929} & \textbf{0.62} & 0.807 \\
\bottomrule
\end{tabularx}
\end{table}

\begin{figure}[tbp]
\centering
\includegraphics[width=\linewidth,height=0.70\textheight,keepaspectratio]{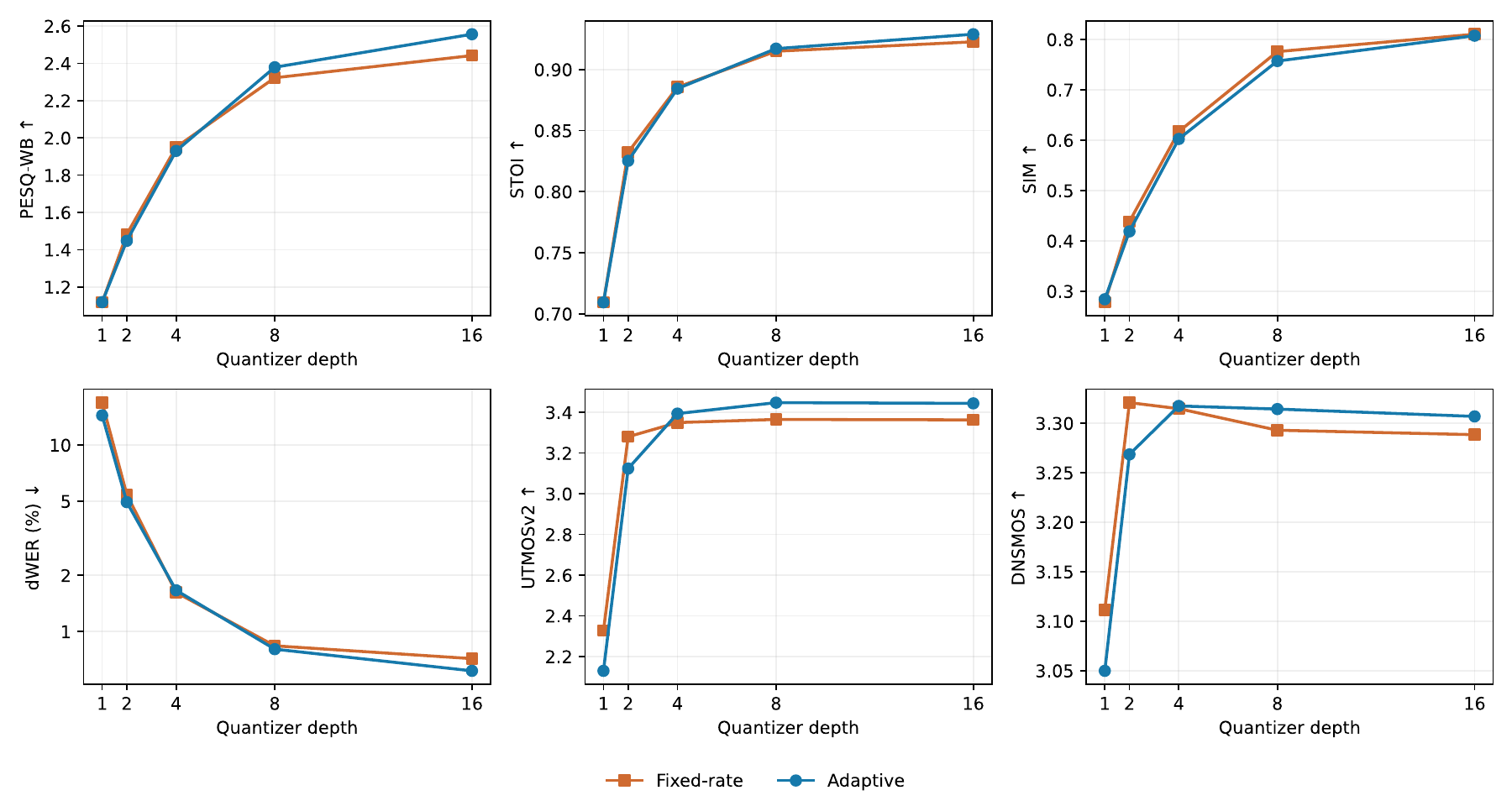}
\caption{\textbf{Adaptive and fixed-rate reconstruction at 12.5 Hz on LibriSpeech test-clean.} Each curve evaluates one trained model at quantizer depths 1, 2, 4, 8, and 16; the dWER axis is logarithmic.}\label{fig:reconstruction-12hz}
\end{figure}

At both depths, the adaptive codec improves every metric except SIM (Table~\ref{tab:adaptive-fixed-rate-12hz}). The gains grow with depth: PESQ-WB rises by 0.057 at depth 8 and by 0.114 at depth 16, where dWER falls from 0.72\% to 0.62\%. SIM is lower by 0.019 and 0.003, respectively. The intelligibility advantage extends to single-codebook reconstruction, where dWER falls from 16.86\% to 14.41\% (Figure~\ref{fig:reconstruction-12hz}). These gains reflect the combined effects of interval aggregation, adaptive placement, and decoding on a 50-Hz grid, compared with the control's 12.5-Hz grid.

\paragraph*{Comparisons at 6.25 and 25 Hz}

We also compare separately trained models at 6.25 and 25 Hz, using retention ratios \(R=0.125\) and \(R=0.5\) and one seed per configuration (Table~\ref{tab:adaptive-fixed-rate-multiple-rates}).

\begin{table}[tbp]
\centering
\footnotesize
\setlength{\tabcolsep}{3pt}
\caption{\textbf{Adaptive versus fixed-rate coding at 6.25 and 25 Hz on LibriSpeech test-clean.} Each codec is trained separately at its target event rate using one training seed. Bold marks the best score within each event rate and quantizer depth.}\label{tab:adaptive-fixed-rate-multiple-rates}
\begin{tabularx}{\linewidth}{R{1.25cm}LR{1cm}R{1cm}R{1.55cm}R{1.4cm}R{1.5cm}R{1cm}R{1cm}R{0.9cm}}
\toprule
\textbf{Event rate (Hz)} & \textbf{Codec} & \textbf{Depth} & \textbf{Bitrate (kbit/s)} & \textbf{UTMOSv2 \(\uparrow\)} & \textbf{DNSMOS \(\uparrow\)} & \textbf{PESQ-WB \(\uparrow\)} & \textbf{STOI \(\uparrow\)} & \textbf{dWER (\%) \(\downarrow\)} & \textbf{SIM \(\uparrow\)} \\
\midrule
6.25 & Fixed-rate & 8 & 0.60 & 3.290 & 3.292 & 1.715 & 0.870 & 1.78 & \textbf{0.582} \\
6.25 & Adaptive & 8 & 0.62 & \textbf{3.342} & \textbf{3.314} & \textbf{1.784} & \textbf{0.872} & \textbf{1.75} & 0.567 \\
\midrule
6.25 & Fixed-rate & 16 & 1.20 & 3.300 & 3.285 & 1.812 & 0.881 & 1.24 & \textbf{0.627} \\
6.25 & Adaptive & 16 & 1.22 & \textbf{3.381} & \textbf{3.314} & \textbf{1.932} & \textbf{0.889} & \textbf{1.08} & 0.621 \\
\midrule
25 & Fixed-rate & 8 & 2.40 & 3.380 & 3.302 & \textbf{2.828} & \textbf{0.945} & \textbf{0.52} & \textbf{0.885} \\
25 & Adaptive & 8 & 2.44 & \textbf{3.433} & \textbf{3.309} & 2.815 & 0.944 & 0.53 & 0.879 \\
\midrule
25 & Fixed-rate & 16 & 4.80 & 3.375 & 3.295 & 2.951 & 0.951 & \textbf{0.45} & \textbf{0.906} \\
25 & Adaptive & 16 & 4.84 & \textbf{3.430} & \textbf{3.302} & \textbf{2.980} & \textbf{0.952} & 0.47 & \textbf{0.906} \\
\bottomrule
\end{tabularx}
\end{table}

At 6.25 Hz, the pattern matches 12.5 Hz: the adaptive codec improves every metric except SIM at both depths (Table~\ref{tab:adaptive-fixed-rate-multiple-rates}), most clearly at depth 16 (PESQ-WB 1.812 \(\rightarrow\) 1.932, dWER 1.24\% \(\rightarrow\) 1.08\%).

\begin{figure}[tbp]
\centering
\includegraphics[width=\linewidth,height=0.70\textheight,keepaspectratio]{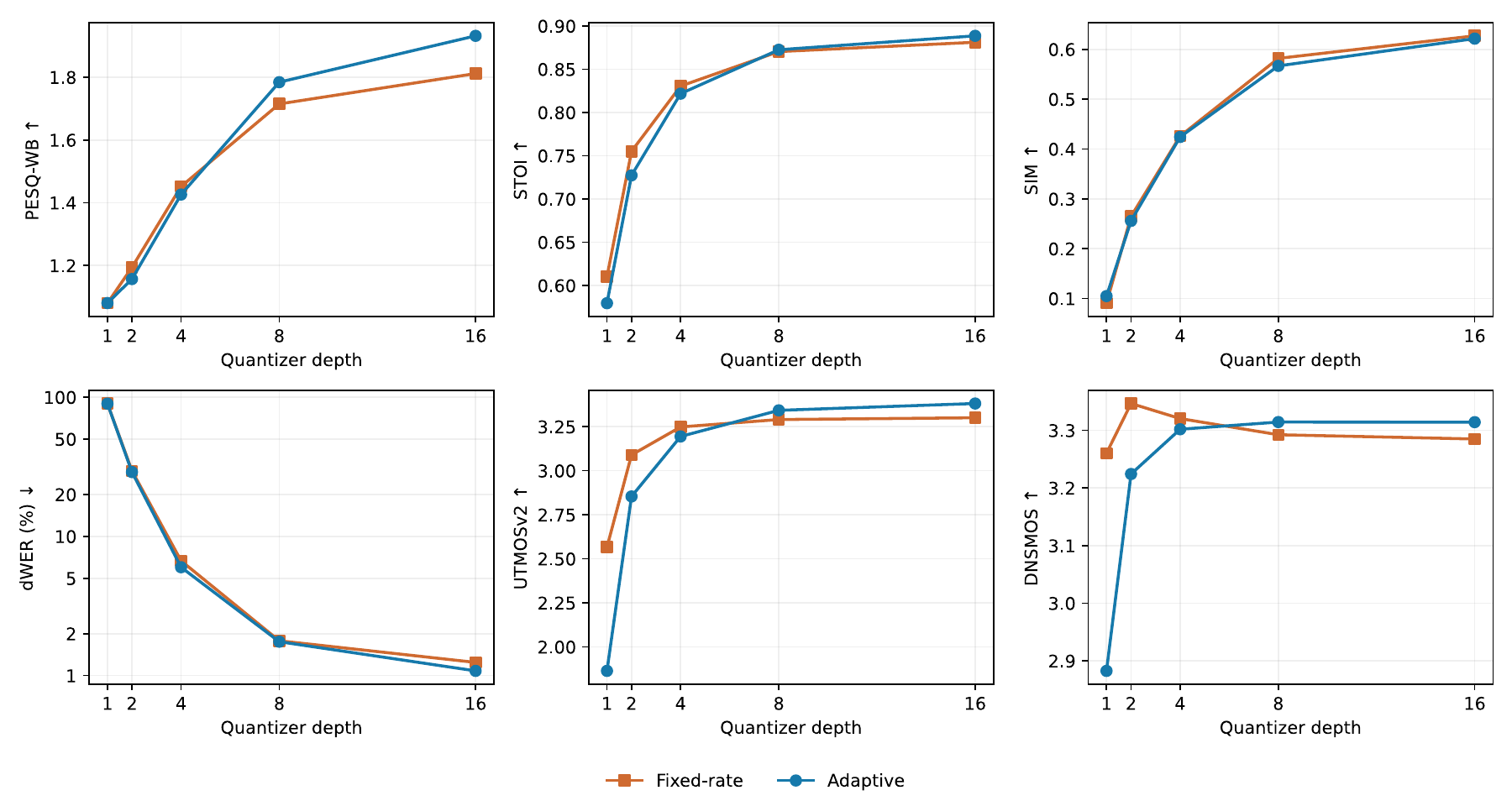}
\caption{\textbf{Adaptive and fixed-rate reconstruction at 6.25 Hz on LibriSpeech test-clean.} Each curve evaluates one trained model at quantizer depths 1, 2, 4, 8, and 16; the dWER axis is logarithmic.}\label{fig:reconstruction-6hz}
\end{figure}

\begin{figure}[tbp]
\centering
\includegraphics[width=\linewidth,height=0.70\textheight,keepaspectratio]{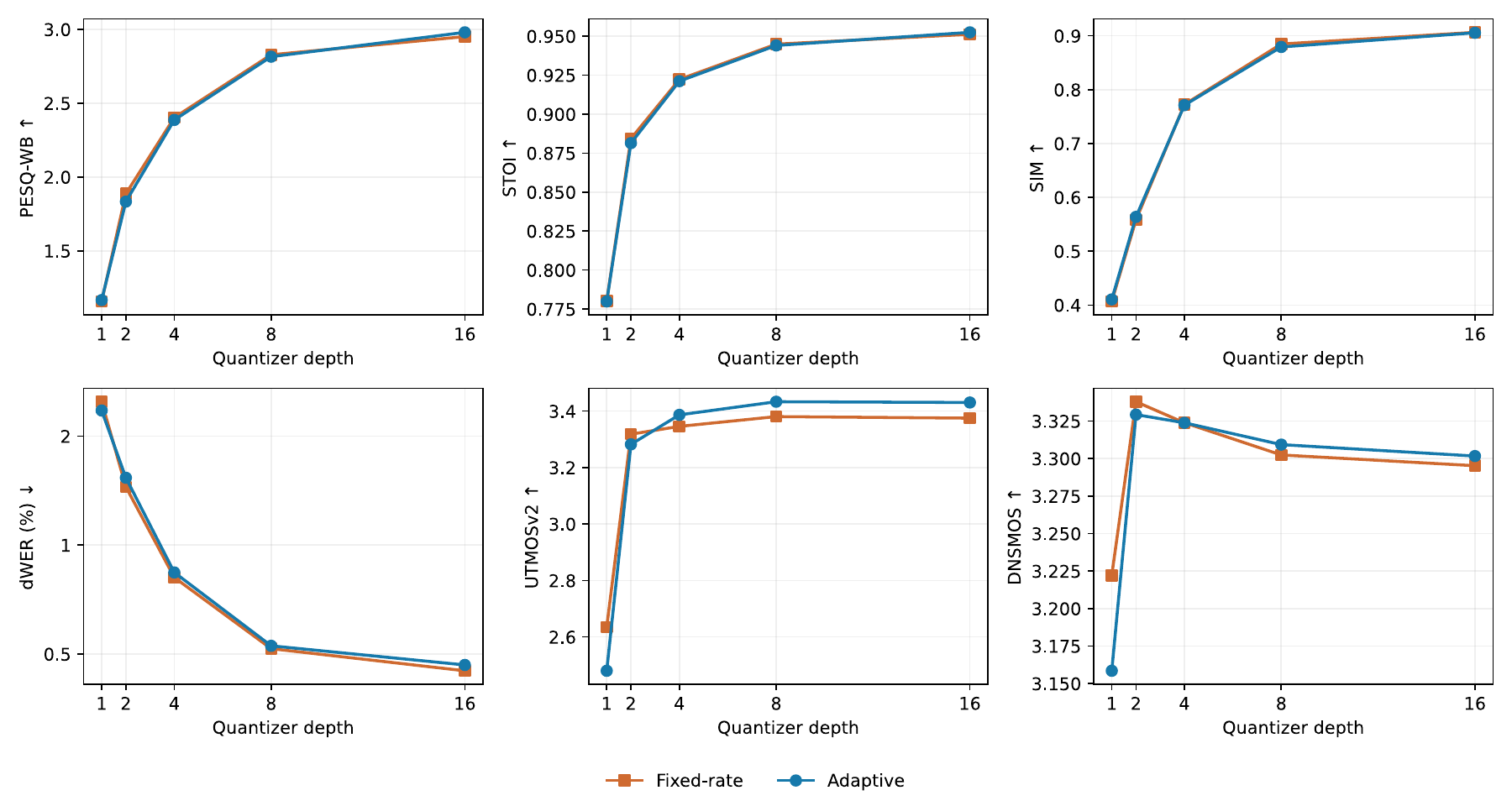}
\caption{\textbf{Adaptive and fixed-rate reconstruction at 25 Hz on LibriSpeech test-clean.} Each curve evaluates one trained model at quantizer depths 1, 2, 4, 8, and 16; the dWER axis is logarithmic.}\label{fig:reconstruction-25hz}
\end{figure}

At 25 Hz, the difference narrows. The adaptive codec keeps higher UTMOSv2 and DNSMOS at both depths and higher PESQ-WB and STOI at depth 16, but the fixed-rate control has slightly lower dWER at both depths (0.52\% versus 0.53\% and 0.45\% versus 0.47\%) and higher SIM at depth 8, with equal SIM at depth 16. Figures~\ref{fig:reconstruction-6hz}--\ref{fig:reconstruction-25hz} show the full depth sweeps.

\FloatBarrier
\subsubsection{Uniform-stride versus adaptive event placement}\label{sec:uniform-stride-versus-adaptive-event-placement}

To isolate event placement, we compare RubatoCodec's cumulative allocation (Section~\ref{sec:budget-constrained-event-allocation}) with a uniform-stride control that starts an event every four frames. Both models share the 50-Hz latent grid, onset-anchored interval aggregator, hybrid residual quantizer, and causal decoder, and are trained for one million updates on the full English 44-kHz partition of HiFiTTS-2 (Section~\ref{sec:training-for-architecture-comparisons}).

\begin{table}[htbp]
\centering
\footnotesize
\setlength{\tabcolsep}{3pt}
\caption{\textbf{Reconstruction with uniform-stride and adaptive event placement on LibriSpeech test-clean.} Bold marks the best score within each quantizer depth. Uniform-stride bitrates slightly exceed 1.2 and 2.4 kbit/s because a final interval shorter than four frames still requires a complete set of code indices.}\label{tab:event-placement}
\begin{tabularx}{\linewidth}{P{2.4cm}YYYYYYYY}
\toprule
\textbf{Event placement} & \textbf{Depth} & \textbf{Bitrate (kbit/s)} & \textbf{UTMOSv2 \(\uparrow\)} & \textbf{DNSMOS \(\uparrow\)} & \textbf{PESQ-WB \(\uparrow\)} & \textbf{STOI \(\uparrow\)} & \textbf{dWER (\%) \(\downarrow\)} & \textbf{SIM \(\uparrow\)} \\
\midrule
Uniform stride & 8 & 1.21 & 3.442 & 3.321 & 2.655 & 0.932 & \textbf{0.70} & 0.849 \\
Adaptive & 8 & 1.23 & \textbf{3.451} & \textbf{3.331} & \textbf{2.681} & \textbf{0.933} & 0.76 & \textbf{0.859} \\
\midrule
Uniform stride & 16 & 2.42 & \textbf{3.461} & 3.309 & 2.957 & \textbf{0.949} & 0.52 & 0.913 \\
Adaptive & 16 & 2.43 & 3.457 & \textbf{3.317} & \textbf{2.989} & \textbf{0.949} & \textbf{0.44} & \textbf{0.914} \\
\bottomrule
\end{tabularx}
\end{table}

\begin{figure}[tbp]
\centering
\includegraphics[width=\linewidth,height=0.70\textheight,keepaspectratio]{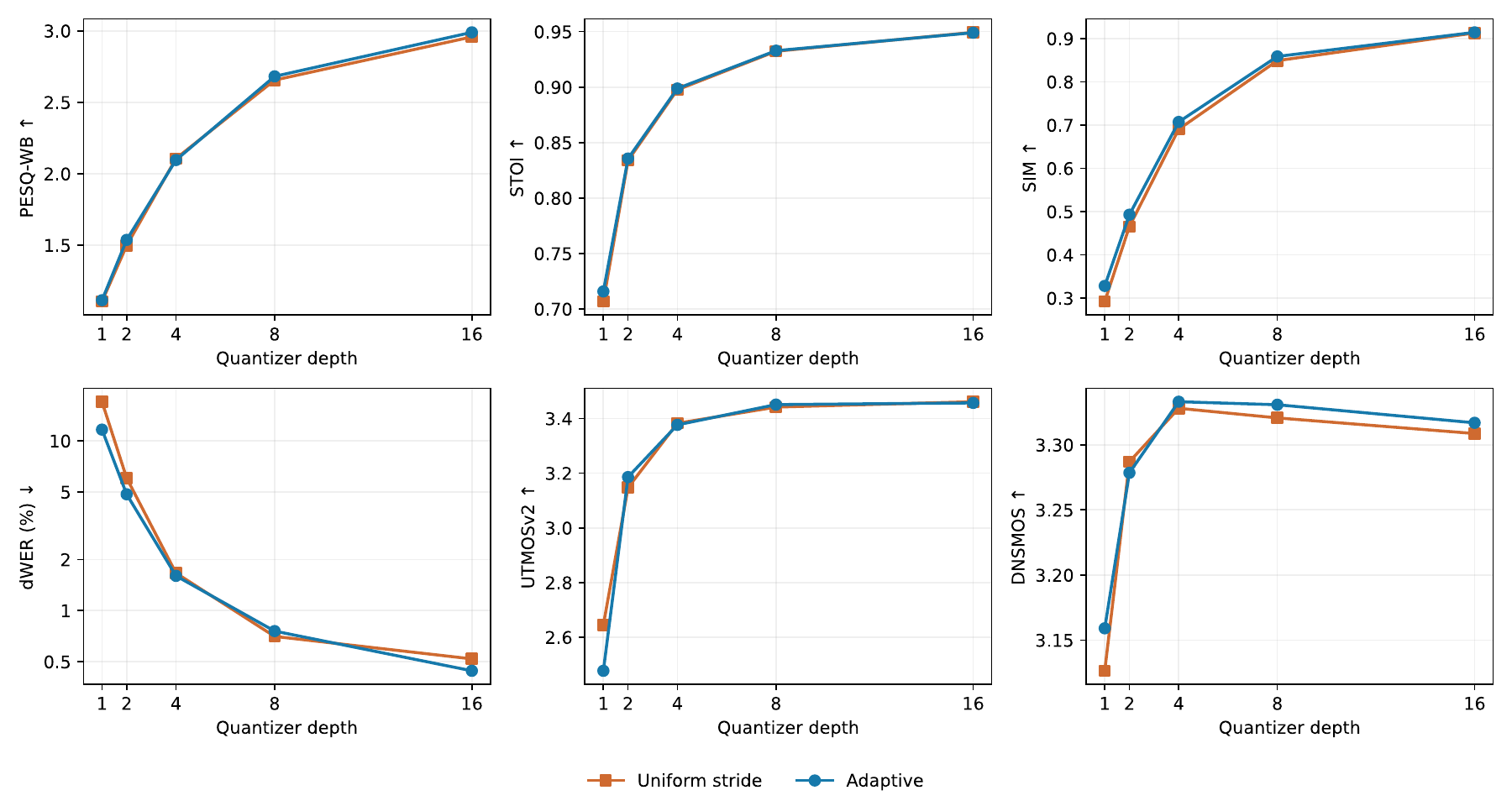}
\caption{\textbf{Uniform-stride and adaptive event placement across quantizer depths on LibriSpeech test-clean.} Both models use a nominal event rate of 12.5 Hz. Each curve evaluates one trained model at quantizer depths 1, 2, 4, 8, and 16; the dWER axis is logarithmic.}\label{fig:event-placement}
\end{figure}

Adaptive placement gives higher SIM at every depth and higher PESQ-WB at every depth except 4 (Table~\ref{tab:event-placement}; Figure~\ref{fig:event-placement}). It also lowers dWER at every depth except 8, most clearly with one codebook (17.00\% \(\rightarrow\) 11.64\%); at depths 8 and 16, dWER is close (0.76\% versus 0.70\% and 0.44\% versus 0.52\%). STOI and MOS predictors show no consistent ordering. With aggregation and decoding held fixed, adaptive placement thus yields modest reconstruction gains, most consistently in speaker similarity, while uniform placement remains a strong control.

\subsubsection{Event rate and quantizer depth at similar bitrates}\label{sec:event-rate-and-quantizer-depth-at-similar-bitrates}

Table~\ref{tab:matched-bitrate} compares adaptive operating points with code-index rates of 1.2 or 2.4 kbit/s, trading event rate against quantizer depth. Total bitrates additionally include duration information.

\begin{table}[tbp]
\centering
\footnotesize
\setlength{\tabcolsep}{3pt}
\caption{\textbf{Adaptive coding at similar bitrates on LibriSpeech test-clean.} Bold marks the best score within each bitrate group.}\label{tab:matched-bitrate}
\begin{tabularx}{\linewidth}{YYYYYYYYY}
\toprule
\textbf{Event rate (Hz)} & \textbf{Depth} & \textbf{Bitrate (kbit/s)} & \textbf{UTMOSv2 \(\uparrow\)} & \textbf{DNSMOS \(\uparrow\)} & \textbf{PESQ-WB \(\uparrow\)} & \textbf{STOI \(\uparrow\)} & \textbf{dWER (\%) \(\downarrow\)} & \textbf{SIM \(\uparrow\)} \\
\midrule
6.25 & 16 & 1.22 & 3.381 & 3.314 & 1.932 & 0.889 & 1.08 & 0.621 \\
12.5 & 8 & 1.23 & \textbf{3.448} & 3.314 & 2.379 & 0.917 & \textbf{0.80} & 0.757 \\
25 & 4 & 1.24 & 3.387 & \textbf{3.324} & \textbf{2.387} & \textbf{0.921} & 0.84 & \textbf{0.771} \\
\midrule
12.5 & 16 & 2.43 & \textbf{3.444} & 3.307 & 2.556 & 0.929 & 0.62 & 0.807 \\
25 & 8 & 2.44 & 3.433 & \textbf{3.309} & \textbf{2.815} & \textbf{0.944} & \textbf{0.53} & \textbf{0.879} \\
\bottomrule
\end{tabularx}
\end{table}

Near 1.2 kbit/s, 12.5 Hz at depth 8 nearly matches 25 Hz at depth 4 in PESQ-WB (2.379 versus 2.387), with slightly lower dWER (0.80\% versus 0.84\%) and higher UTMOSv2, but lower SIM (0.757 versus 0.771). At 6.25 Hz, even depth 16 falls well behind in fidelity and SIM. Near 2.4 kbit/s, 25 Hz at depth 8 is clearly ahead of 12.5 Hz at depth 16 (PESQ-WB 2.815 versus 2.556), with only UTMOSv2 favoring 12.5 Hz.

The 12.5-Hz configuration thus halves the temporal sequence length relative to 25 Hz at nearly the same signal fidelity and intelligibility near 1.2 kbit/s. Hierarchical generators can trade temporal-backbone steps for codebook predictions in a lighter within-event module \cite{defossez2024moshi,hu2026qwen3tts,gong2026mosstts}; Section~\ref{sec:temporal-resolution-and-quantizer-depth-for-speech-generation} discusses this trade-off.

We also vary the retention ratio at inference for the multilingual model trained at 12.5 Hz, without retraining. Higher event rates improve reconstruction fidelity and speaker similarity, whereas substantial rate reductions sharply degrade quality even at greater quantizer depths (Appendix~\ref{sec:inference-time-event-rate-variation}).

\FloatBarrier
\subsection{Comparison with existing codecs}\label{sec:comparison-with-existing-codecs}

We evaluate the multilingual RubatoCodec model trained at scale using the settings in Section~\ref{sec:multilingual-model-training}, comparing its reconstruction quality with existing codecs across English, Japanese, and Mandarin Chinese.

\subsubsection{English}\label{sec:english}

\begin{figure}[tbp]
\centering
\includegraphics[width=\linewidth,height=0.70\textheight,keepaspectratio]{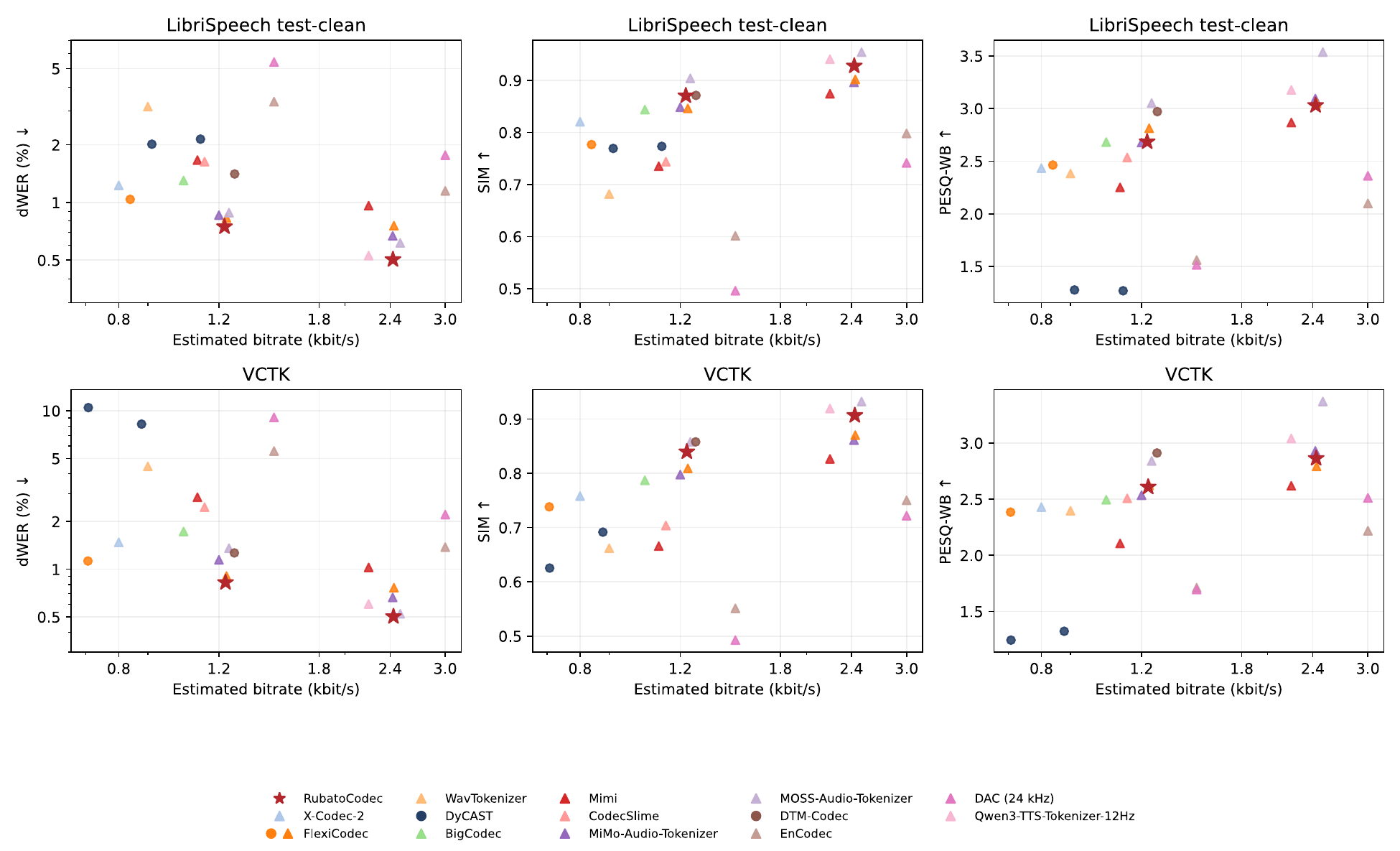}
\caption{\textbf{English reconstruction quality versus estimated bitrate.} Stars denote RubatoCodec at depths 8 and 16; circles denote adaptive allocation and triangles fixed allocation for the other codecs. Colors identify codec families. The bitrate and dWER axes are logarithmic. Operating points match Tables~\ref{tab:librispeech-reconstruction}--\ref{tab:vctk-reconstruction}.}\label{fig:english-reconstruction}
\end{figure}

Tables~\ref{tab:librispeech-reconstruction} and \ref{tab:vctk-reconstruction} compare the multilingual model with external codecs on LibriSpeech and VCTK. Operating points were selected within 0.8--1.5 and 2.2--3.0 kbit/s on LibriSpeech, and the same configurations are used for VCTK. Figure~\ref{fig:english-reconstruction} plots reconstruction quality against bitrate for these operating points.

The allocation column describes the evaluated encoding setting: \textbf{Adaptive} denotes content-dependent event placement or durations, and \textbf{Fixed} denotes a fixed temporal grid. FlexiCodec at \(\tau\)=1 and CodecSlime at 80 Hz use no temporal merging and are therefore labeled Fixed, although both codec families support adaptive allocation at other settings. RubatoCodec maintains adaptive placement at both depths.

DyCAST's predicted durations can change local timing even when the total frame count is preserved. Its PESQ and STOI scores therefore also reflect temporal mismatch with the reference waveform.

\paragraph*{LibriSpeech test-clean}

\begin{table}[tbp]
\centering
\footnotesize
\setlength{\tabcolsep}{2pt}
\renewcommand{\arraystretch}{1.1}
\caption{\textbf{Reconstruction on LibriSpeech test-clean.} Event rate is measured in coded events per second. Thick rules separate the bitrate bands described above. Bold and underline indicate the best and second-best scores in each of the two comparison bands, excluding GT; ties at displayed precision receive the same formatting. \(D\) denotes the number of active quantizer layers; DTM-Codec labels are nominal targets, and CA/BP denote DyCAST's character-aligned/boundary-prediction settings.}\label{tab:librispeech-reconstruction}
\begin{tabularx}{\linewidth}{LP{1.45cm}R{1.25cm}R{1cm}R{1.55cm}R{1.4cm}R{1.5cm}R{1cm}R{1cm}R{0.9cm}}
\toprule
\textbf{Codec / configuration} & \textbf{Allocation} & \textbf{Event rate (Hz)} & \textbf{Bitrate (kbit/s)} & \textbf{UTMOSv2 \(\uparrow\)} & \textbf{DNSMOS \(\uparrow\)} & \textbf{PESQ-WB \(\uparrow\)} & \textbf{STOI \(\uparrow\)} & \textbf{dWER (\%) \(\downarrow\)} & \textbf{SIM \(\uparrow\)} \\
\midrule
GT & --- & --- & --- & 3.233 & 3.272 & 4.644 & 1.000 & 0.00 & 1.000 \\
\midrule[\heavyrulewidth]
X-Codec-2 & Fixed & 50.0 & 0.80 & 3.233 & 3.299 & 2.434 & 0.919 & 1.23 & 0.821 \\
FlexiCodec, \(\tau\)=0.91, \(D\)=8 & Adaptive & 8.3 & 0.84 & 3.174 & 3.307 & 2.463 & 0.919 & 1.04 & 0.777 \\
WavTokenizer, 75 Hz & Fixed & 75.0 & 0.90 & 2.772 & \textbf{3.342} & 2.382 & 0.912 & 3.16 & 0.682 \\
DyCAST, CA & Adaptive & 14.2 & 0.91 & 2.868 & 3.316 & 1.277 & 0.213 & 2.01 & 0.769 \\
BigCodec & Fixed & 80.0 & 1.04 & 3.362 & 3.333 & 2.681 & 0.935 & 1.30 & 0.844 \\
Mimi, \(D\)=8 & Fixed & 12.5 & 1.10 & 2.810 & 3.159 & 2.251 & 0.906 & 1.67 & 0.735 \\
DyCAST, BP minimum gap 1 & Adaptive & 17.4 & 1.11 & 2.894 & 3.318 & 1.269 & 0.212 & 2.14 & 0.773 \\
CodecSlime, FSQ18K, 80 Hz & Fixed & 80.0 & 1.13 & 3.341 & 3.327 & 2.535 & 0.926 & 1.63 & 0.744 \\
MiMo-Audio-Tokenizer, \(D\)=6 & Fixed & 25.0 & 1.20 & 3.183 & 3.291 & 2.677 & 0.927 & 0.86 & 0.848 \\
FlexiCodec, \(\tau\)=1, \(D\)=8 & Fixed & 12.5 & 1.24 & 3.194 & 3.284 & 2.813 & 0.940 & \underline{0.83} & 0.846 \\
MOSS-Audio-Tokenizer, \(D\)=10 & Fixed & 12.5 & 1.25 & 3.230 & 3.217 & \textbf{3.052} & \textbf{0.949} & 0.89 & \textbf{0.904} \\
DTM-Codec, 80-Hz target & Adaptive & 81.5 & 1.28 & \underline{3.421} & \underline{3.334} & \underline{2.971} & \underline{0.947} & 1.41 & \underline{0.871} \\
EnCodec, 1.5 kbit/s & Fixed & 75.0 & 1.50 & 1.398 & 2.425 & 1.560 & 0.845 & 3.36 & 0.601 \\
DAC (24 kHz), \(D\)=2 & Fixed & 75.0 & 1.50 & 1.413 & 2.717 & 1.515 & 0.836 & 5.40 & 0.496 \\
\textbf{\boldmath RubatoCodec, \(D\)=8} & Adaptive & 12.5 & 1.23 & \textbf{3.478} & 3.327 & 2.684 & 0.931 & \textbf{0.75} & \underline{0.871} \\
\midrule[\heavyrulewidth]
Mimi, \(D\)=16 & Fixed & 12.5 & 2.20 & 2.943 & 3.175 & 2.867 & 0.939 & 0.96 & 0.874 \\
Qwen3-TTS-Tokenizer-12Hz & Fixed & 12.5 & 2.20 & 3.230 & 3.273 & \underline{3.177} & \underline{0.958} & \underline{0.53} & \underline{0.941} \\
MiMo-Audio-Tokenizer, \(D\)=13 & Fixed & 25.0 & 2.42 & 3.210 & \underline{3.288} & 3.094 & 0.947 & 0.67 & 0.896 \\
FlexiCodec, \(\tau\)=1, \(D\)=16 & Fixed & 12.5 & 2.44 & 3.235 & 3.277 & 3.052 & 0.953 & 0.76 & 0.902 \\
MOSS-Audio-Tokenizer, \(D\)=20 & Fixed & 12.5 & 2.50 & \underline{3.239} & 3.203 & \textbf{3.536} & \textbf{0.968} & 0.62 & \textbf{0.954} \\
EnCodec, 3 kbit/s & Fixed & 75.0 & 3.00 & 1.786 & 2.765 & 2.098 & 0.900 & 1.15 & 0.798 \\
DAC (24 kHz), \(D\)=4 & Fixed & 75.0 & 3.00 & 2.005 & 3.116 & 2.361 & 0.905 & 1.77 & 0.742 \\
\textbf{\boldmath RubatoCodec, \(D\)=16} & Adaptive & 12.5 & 2.43 & \textbf{3.486} & \textbf{3.314} & 3.029 & 0.950 & \textbf{0.50} & 0.928 \\
\bottomrule
\end{tabularx}
\end{table}

RubatoCodec's strengths are intelligibility preservation and predicted quality: on LibriSpeech, it has the lowest dWER and highest UTMOSv2 in both bitrate bands, while the strongest comparators retain higher signal fidelity and speaker similarity (Table~\ref{tab:librispeech-reconstruction}). At depth 8, its dWER of 0.75\% is below that of FlexiCodec at \(\tau\)=1 (0.83\%), MiMo-Audio-Tokenizer at depth 6 (0.86\%), and MOSS-Audio-Tokenizer at depth 10 (0.89\%). MOSS, DTM-Codec, and FlexiCodec have higher PESQ-WB and STOI, MOSS has higher SIM, and DTM-Codec has the same SIM at displayed precision (0.871).

At depth 16, dWER falls to 0.50\%, slightly below Qwen3-TTS-Tokenizer-12Hz (0.53\%), and DNSMOS is also highest. MOSS and Qwen retain higher PESQ-WB and SIM, whereas MiMo has higher PESQ-WB but lower SIM.

\paragraph*{VCTK}

\begin{table}[tbp]
\centering
\footnotesize
\setlength{\tabcolsep}{2pt}
\renewcommand{\arraystretch}{1.1}
\caption{\textbf{Reconstruction on VCTK.} Allocation labels, bitrate bands, and rank formatting follow Table~\ref{tab:librispeech-reconstruction}. Configurations are grouped using their bitrate on this dataset.}\label{tab:vctk-reconstruction}
\begin{tabularx}{\linewidth}{LP{1.45cm}R{1.25cm}R{1cm}R{1.55cm}R{1.4cm}R{1.5cm}R{1cm}R{1cm}R{0.9cm}}
\toprule
\textbf{Codec / configuration} & \textbf{Allocation} & \textbf{Event rate (Hz)} & \textbf{Bitrate (kbit/s)} & \textbf{UTMOSv2 \(\uparrow\)} & \textbf{DNSMOS \(\uparrow\)} & \textbf{PESQ-WB \(\uparrow\)} & \textbf{STOI \(\uparrow\)} & \textbf{dWER (\%) \(\downarrow\)} & \textbf{SIM \(\uparrow\)} \\
\midrule
GT & --- & --- & --- & 3.566 & 3.202 & 4.644 & 1.000 & 0.00 & 1.000 \\
\midrule[\heavyrulewidth]
FlexiCodec, \(\tau\)=0.91, \(D\)=8 & Adaptive & 7.0 & 0.71 & 3.354 & 3.191 & 2.384 & 0.846 & 1.13 & 0.738 \\
DyCAST, CA & Adaptive & 11.0 & 0.71 & 2.594 & 3.179 & 1.245 & 0.202 & 10.47 & 0.625 \\
X-Codec-2 & Fixed & 50.0 & 0.80 & 3.443 & 3.190 & 2.430 & 0.858 & 1.48 & 0.758 \\
DyCAST, BP minimum gap 1 & Adaptive & 13.6 & 0.88 & 2.805 & 3.226 & 1.324 & 0.242 & 8.24 & 0.692 \\
WavTokenizer, 75 Hz & Fixed & 75.0 & 0.90 & 3.198 & 3.208 & 2.397 & 0.853 & 4.45 & 0.662 \\
BigCodec & Fixed & 80.0 & 1.04 & 3.368 & \textbf{3.243} & 2.495 & 0.873 & 1.73 & 0.787 \\
Mimi, \(D\)=8 & Fixed & 12.5 & 1.10 & 3.036 & 2.908 & 2.106 & 0.808 & 2.85 & 0.666 \\
CodecSlime, FSQ18K, 80 Hz & Fixed & 80.0 & 1.13 & 3.429 & 3.223 & 2.507 & 0.867 & 2.46 & 0.704 \\
MiMo-Audio-Tokenizer, \(D\)=6 & Fixed & 25.0 & 1.20 & \underline{3.495} & 3.087 & 2.535 & 0.875 & 1.15 & 0.798 \\
FlexiCodec, \(\tau\)=1, \(D\)=8 & Fixed & 12.5 & 1.24 & 3.357 & 3.166 & 2.622 & 0.879 & \underline{0.91} & 0.809 \\
MOSS-Audio-Tokenizer, \(D\)=10 & Fixed & 12.5 & 1.25 & 3.351 & 3.029 & \underline{2.840} & \underline{0.885} & 1.36 & \textbf{0.858} \\
DTM-Codec, 80-Hz target & Adaptive & 80.2 & 1.28 & 3.492 & 3.230 & \textbf{2.911} & \textbf{0.886} & 1.27 & \textbf{0.858} \\
EnCodec, 1.5 kbit/s & Fixed & 75.0 & 1.50 & 1.523 & 2.498 & 1.711 & 0.768 & 5.57 & 0.551 \\
DAC (24 kHz), \(D\)=2 & Fixed & 75.0 & 1.50 & 1.620 & 2.930 & 1.696 & 0.808 & 9.08 & 0.493 \\
\textbf{\boldmath RubatoCodec, \(D\)=8} & Adaptive & 12.5 & 1.23 & \textbf{3.533} & \underline{3.235} & 2.608 & 0.868 & \textbf{0.82} & \underline{0.840} \\
\midrule[\heavyrulewidth]
Mimi, \(D\)=16 & Fixed & 12.5 & 2.20 & 3.252 & 2.926 & 2.619 & 0.840 & 1.03 & 0.827 \\
Qwen3-TTS-Tokenizer-12Hz & Fixed & 12.5 & 2.20 & 3.494 & 3.107 & \underline{3.040} & 0.900 & 0.60 & \underline{0.919} \\
MiMo-Audio-Tokenizer, \(D\)=13 & Fixed & 25.0 & 2.42 & \underline{3.501} & 3.104 & 2.930 & \underline{0.901} & 0.66 & 0.861 \\
FlexiCodec, \(\tau\)=1, \(D\)=16 & Fixed & 12.5 & 2.44 & 3.367 & \underline{3.168} & 2.791 & 0.891 & 0.76 & 0.871 \\
MOSS-Audio-Tokenizer, \(D\)=20 & Fixed & 12.5 & 2.50 & 3.392 & 3.009 & \textbf{3.370} & \textbf{0.917} & \underline{0.52} & \textbf{0.932} \\
EnCodec, 3 kbit/s & Fixed & 75.0 & 3.00 & 1.981 & 2.790 & 2.219 & 0.813 & 1.38 & 0.750 \\
DAC (24 kHz), \(D\)=4 & Fixed & 75.0 & 3.00 & 2.348 & 3.112 & 2.511 & 0.874 & 2.21 & 0.722 \\
\textbf{\boldmath RubatoCodec, \(D\)=16} & Adaptive & 12.5 & 2.43 & \textbf{3.524} & \textbf{3.222} & 2.863 & 0.886 & \textbf{0.50} & 0.907 \\
\bottomrule
\end{tabularx}
\end{table}

VCTK follows the same pattern, with the highest UTMOSv2 and lowest dWER in both bands (Table~\ref{tab:vctk-reconstruction}). At depth 8, dWER is 0.82\% against 0.91\% for FlexiCodec at \(\tau\)=1, and SIM (0.840) is below MOSS-Audio-Tokenizer and DTM-Codec (both 0.858) but above MiMo-Audio-Tokenizer at depth 6 (0.798). At depth 16, dWER (0.50\%) is slightly below MOSS-Audio-Tokenizer (0.52\%) and DNSMOS is highest, while MOSS and Qwen3-TTS-Tokenizer-12Hz retain higher fidelity and SIM, and MiMo-Audio-Tokenizer retains higher fidelity.

\subsubsection{Japanese and Mandarin Chinese}\label{sec:japanese-and-mandarin-chinese}

\begin{figure}[htbp]
\centering
\includegraphics[width=\linewidth,height=0.70\textheight,keepaspectratio]{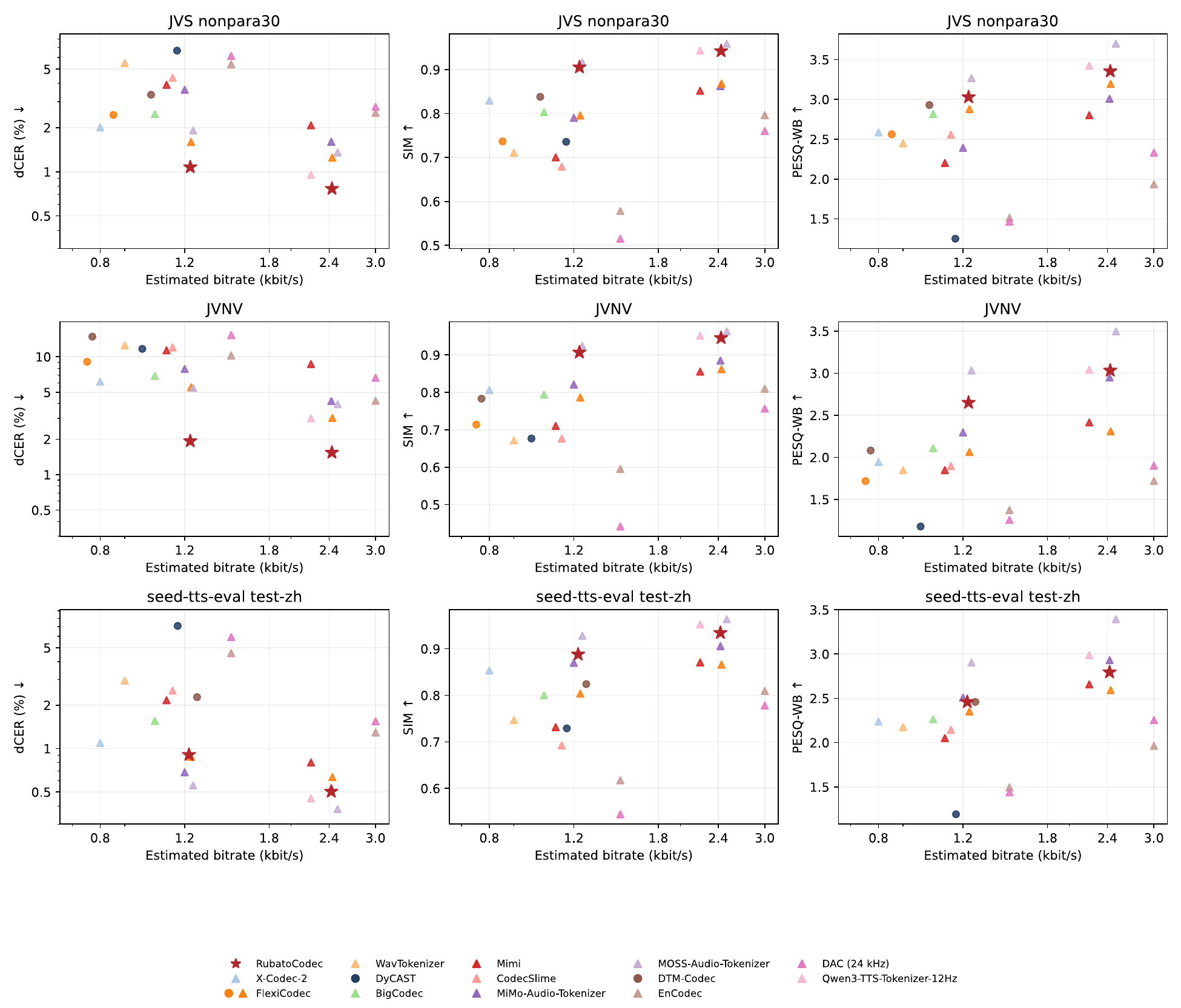}
\caption{\textbf{Japanese and Mandarin reconstruction quality versus estimated bitrate.} Marker and color conventions follow Figure~\ref{fig:english-reconstruction}. The bitrate and dCER axes are logarithmic. Operating points match Tables~\ref{tab:jvs-reconstruction}--\ref{tab:jvnv-reconstruction} and \ref{tab:mandarin-reconstruction}.}\label{fig:multilingual-reconstruction}
\end{figure}

Tables~\ref{tab:jvs-reconstruction}--\ref{tab:jvnv-reconstruction} and \ref{tab:mandarin-reconstruction} evaluate the same model on Japanese and Mandarin, using dCER and the English tables' allocation labels and comparison groups. The same codec configurations are retained across datasets, even when content-dependent rates move outside the nominal bitrate bands. Figure~\ref{fig:multilingual-reconstruction} plots the corresponding quality--bitrate comparisons.

\begin{table}[tbp]
\centering
\footnotesize
\setlength{\tabcolsep}{2pt}
\renewcommand{\arraystretch}{1.1}
\caption{\textbf{Reconstruction on JVS nonpara30.} Event rates, bitrate grouping, and score highlighting follow Table~\ref{tab:librispeech-reconstruction}.}\label{tab:jvs-reconstruction}
\begin{tabularx}{\linewidth}{LP{1.45cm}R{1.25cm}R{1cm}R{1.55cm}R{1.4cm}R{1.5cm}R{1cm}R{1cm}R{0.9cm}}
\toprule
\textbf{Codec / configuration} & \textbf{Allocation} & \textbf{Event rate (Hz)} & \textbf{Bitrate (kbit/s)} & \textbf{UTMOSv2 \(\uparrow\)} & \textbf{DNSMOS \(\uparrow\)} & \textbf{PESQ-WB \(\uparrow\)} & \textbf{STOI \(\uparrow\)} & \textbf{dCER (\%) \(\downarrow\)} & \textbf{SIM \(\uparrow\)} \\
\midrule
GT & --- & --- & --- & 3.621 & 3.334 & 4.644 & 0.999 & 0.00 & 1.000 \\
\midrule[\heavyrulewidth]
X-Codec-2 & Fixed & 50.0 & 0.80 & 3.234 & 3.312 & 2.585 & 0.924 & 2.01 & 0.829 \\
FlexiCodec, \(\tau\)=0.91, \(D\)=8 & Adaptive & 8.5 & 0.85 & 2.772 & 3.292 & 2.562 & 0.925 & 2.44 & 0.736 \\
WavTokenizer, 75 Hz & Fixed & 75.0 & 0.90 & 2.656 & 3.336 & 2.448 & 0.913 & 5.51 & 0.711 \\
DTM-Codec, 80-Hz target & Adaptive & 63.3 & 1.02 & \underline{3.275} & \underline{3.354} & 2.929 & 0.944 & 3.35 & 0.838 \\
BigCodec & Fixed & 80.0 & 1.04 & 3.244 & 3.348 & 2.816 & 0.938 & 2.47 & 0.803 \\
Mimi, \(D\)=8 & Fixed & 12.5 & 1.10 & 2.686 & 3.096 & 2.202 & 0.904 & 3.91 & 0.700 \\
CodecSlime, FSQ18K, 80 Hz & Fixed & 80.0 & 1.13 & 3.185 & 3.345 & 2.557 & 0.923 & 4.37 & 0.679 \\
DyCAST, BP minimum gap 1 & Adaptive & 18.0 & 1.16 & 2.573 & 3.321 & 1.251 & 0.131 & 6.69 & 0.736 \\
MiMo-Audio-Tokenizer, \(D\)=6 & Fixed & 25.0 & 1.20 & 3.171 & 3.304 & 2.393 & 0.906 & 3.62 & 0.790 \\
FlexiCodec, \(\tau\)=1, \(D\)=8 & Fixed & 12.5 & 1.24 & 2.802 & 3.276 & 2.878 & 0.939 & \underline{1.60} & 0.795 \\
MOSS-Audio-Tokenizer, \(D\)=10 & Fixed & 12.5 & 1.25 & 3.248 & 3.240 & \textbf{3.267} & \textbf{0.959} & 1.91 & \textbf{0.916} \\
EnCodec, 1.5 kbit/s & Fixed & 75.0 & 1.50 & 1.230 & 2.426 & 1.514 & 0.843 & 5.40 & 0.578 \\
DAC (24 kHz), \(D\)=2 & Fixed & 75.0 & 1.50 & 1.204 & 2.821 & 1.465 & 0.834 & 6.16 & 0.515 \\
\textbf{\boldmath RubatoCodec, \(D\)=8} & Adaptive & 12.5 & 1.23 & \textbf{3.408} & \textbf{3.358} & \underline{3.028} & \underline{0.949} & \textbf{1.07} & \underline{0.905} \\
\midrule[\heavyrulewidth]
Mimi, \(D\)=16 & Fixed & 12.5 & 2.20 & 2.869 & 3.124 & 2.802 & 0.943 & 2.07 & 0.852 \\
Qwen3-TTS-Tokenizer-12Hz & Fixed & 12.5 & 2.20 & \textbf{3.434} & 3.275 & \underline{3.422} & \underline{0.965} & \underline{0.95} & \underline{0.943} \\
MiMo-Audio-Tokenizer, \(D\)=13 & Fixed & 25.0 & 2.42 & 3.285 & \underline{3.317} & 3.009 & 0.939 & 1.60 & 0.862 \\
FlexiCodec, \(\tau\)=1, \(D\)=16 & Fixed & 12.5 & 2.44 & 2.884 & 3.274 & 3.194 & 0.957 & 1.25 & 0.868 \\
MOSS-Audio-Tokenizer, \(D\)=20 & Fixed & 12.5 & 2.50 & 3.249 & 3.224 & \textbf{3.700} & \textbf{0.976} & 1.35 & \textbf{0.959} \\
EnCodec, 3 kbit/s & Fixed & 75.0 & 3.00 & 1.575 & 2.763 & 1.934 & 0.898 & 2.52 & 0.796 \\
DAC (24 kHz), \(D\)=4 & Fixed & 75.0 & 3.00 & 1.919 & 3.146 & 2.332 & 0.906 & 2.76 & 0.760 \\
\textbf{\boldmath RubatoCodec, \(D\)=16} & Adaptive & 12.5 & 2.43 & \underline{3.407} & \textbf{3.349} & 3.353 & \underline{0.965} & \textbf{0.77} & 0.942 \\
\bottomrule
\end{tabularx}
\end{table}

\paragraph*{JVS nonpara30 and JVNV}

\begin{table}[tbp]
\centering
\footnotesize
\setlength{\tabcolsep}{2pt}
\renewcommand{\arraystretch}{1.1}
\caption{\textbf{Reconstruction on JVNV.} Event rates, bitrate grouping, and score highlighting follow Table~\ref{tab:librispeech-reconstruction}.}\label{tab:jvnv-reconstruction}
\begin{tabularx}{\linewidth}{LP{1.45cm}R{1.25cm}R{1cm}R{1.55cm}R{1.4cm}R{1.5cm}R{1cm}R{1cm}R{0.9cm}}
\toprule
\textbf{Codec / configuration} & \textbf{Allocation} & \textbf{Event rate (Hz)} & \textbf{Bitrate (kbit/s)} & \textbf{UTMOSv2 \(\uparrow\)} & \textbf{DNSMOS \(\uparrow\)} & \textbf{PESQ-WB \(\uparrow\)} & \textbf{STOI \(\uparrow\)} & \textbf{dCER (\%) \(\downarrow\)} & \textbf{SIM \(\uparrow\)} \\
\midrule
GT & --- & --- & --- & 2.526 & 3.098 & 4.644 & 1.000 & 0.00 & 1.000 \\
\midrule[\heavyrulewidth]
FlexiCodec, \(\tau\)=0.91, \(D\)=8 & Adaptive & 7.5 & 0.75 & 1.742 & 3.133 & 1.718 & 0.817 & 9.06 & 0.714 \\
DTM-Codec, 80-Hz target & Adaptive & 46.1 & 0.77 & 2.523 & 3.154 & 2.081 & 0.862 & 14.82 & 0.783 \\
X-Codec-2 & Fixed & 50.0 & 0.80 & 2.369 & 3.148 & 1.944 & 0.858 & 6.15 & 0.807 \\
WavTokenizer, 75 Hz & Fixed & 75.0 & 0.90 & 1.769 & 3.150 & 1.846 & 0.842 & 12.48 & 0.672 \\
DyCAST, BP minimum gap 1 & Adaptive & 15.3 & 0.98 & 1.797 & 3.071 & 1.178 & 0.156 & 11.67 & 0.677 \\
BigCodec & Fixed & 80.0 & 1.04 & 2.439 & 3.154 & 2.109 & 0.871 & 6.87 & 0.794 \\
Mimi, \(D\)=8 & Fixed & 12.5 & 1.10 & 1.913 & 2.907 & 1.848 & 0.826 & 11.34 & 0.710 \\
CodecSlime, FSQ18K, 80 Hz & Fixed & 80.0 & 1.13 & 2.225 & 3.140 & 1.896 & 0.851 & 12.01 & 0.676 \\
MiMo-Audio-Tokenizer, \(D\)=6 & Fixed & 25.0 & 1.20 & \textbf{2.681} & \textbf{3.173} & 2.295 & 0.872 & 7.88 & 0.821 \\
FlexiCodec, \(\tau\)=1, \(D\)=8 & Fixed & 12.5 & 1.24 & 1.811 & 3.128 & 2.062 & 0.862 & 5.52 & 0.786 \\
MOSS-Audio-Tokenizer, \(D\)=10 & Fixed & 12.5 & 1.25 & 2.396 & 3.089 & \textbf{3.032} & \textbf{0.917} & \underline{5.44} & \textbf{0.924} \\
EnCodec, 1.5 kbit/s & Fixed & 75.0 & 1.50 & 1.070 & 2.196 & 1.373 & 0.779 & 10.27 & 0.595 \\
DAC (24 kHz), \(D\)=2 & Fixed & 75.0 & 1.50 & 0.860 & 2.520 & 1.256 & 0.759 & 15.26 & 0.441 \\
\textbf{\boldmath RubatoCodec, \(D\)=8} & Adaptive & 12.5 & 1.23 & \underline{2.628} & \underline{3.164} & \underline{2.650} & \underline{0.907} & \textbf{1.92} & \underline{0.907} \\
\midrule[\heavyrulewidth]
Mimi, \(D\)=16 & Fixed & 12.5 & 2.20 & 1.974 & 2.924 & 2.414 & 0.877 & 8.66 & 0.856 \\
Qwen3-TTS-Tokenizer-12Hz & Fixed & 12.5 & 2.20 & 2.638 & 3.097 & \underline{3.041} & \underline{0.931} & \underline{3.01} & \underline{0.951} \\
MiMo-Audio-Tokenizer, \(D\)=13 & Fixed & 25.0 & 2.42 & \textbf{2.708} & \textbf{3.176} & 2.949 & 0.911 & 4.21 & 0.885 \\
FlexiCodec, \(\tau\)=1, \(D\)=16 & Fixed & 12.5 & 2.44 & 1.930 & 3.119 & 2.309 & 0.885 & 3.02 & 0.862 \\
MOSS-Audio-Tokenizer, \(D\)=20 & Fixed & 12.5 & 2.50 & 2.383 & 3.057 & \textbf{3.496} & \textbf{0.941} & 3.96 & \textbf{0.963} \\
EnCodec, 3 kbit/s & Fixed & 75.0 & 3.00 & 1.266 & 2.480 & 1.719 & 0.841 & 4.23 & 0.810 \\
DAC (24 kHz), \(D\)=4 & Fixed & 75.0 & 3.00 & 1.280 & 2.868 & 1.901 & 0.861 & 6.64 & 0.756 \\
\textbf{\boldmath RubatoCodec, \(D\)=16} & Adaptive & 12.5 & 2.43 & \underline{2.651} & \underline{3.138} & 3.032 & 0.929 & \textbf{1.54} & 0.946 \\
\bottomrule
\end{tabularx}
\end{table}

RubatoCodec achieves the lowest dCER in both comparison groups on both Japanese datasets (Tables~\ref{tab:jvs-reconstruction}--\ref{tab:jvnv-reconstruction}). Its advantage is particularly pronounced on JVNV: dCER is 1.92\% versus 5.44\% for the next-best codec at depth 8, and 1.54\% versus 3.01\% at depth 16, extending its intelligibility preservation advantage to emotional speech containing nonverbal vocalizations. Predicted-quality rankings vary by dataset: RubatoCodec leads DNSMOS in both groups and UTMOSv2 at depth 8 on JVS, whereas MiMo-Audio-Tokenizer leads both predictors on JVNV. MOSS-Audio-Tokenizer retains an advantage in signal fidelity and speaker similarity.

\paragraph*{seed-tts-eval test-zh}

\begin{table}[tbp]
\centering
\footnotesize
\setlength{\tabcolsep}{2pt}
\renewcommand{\arraystretch}{1.1}
\caption{\textbf{Reconstruction on seed-tts-eval test-zh.} Event rates, bitrate grouping, and score highlighting follow Table~\ref{tab:librispeech-reconstruction}.}\label{tab:mandarin-reconstruction}
\begin{tabularx}{\linewidth}{LP{1.45cm}R{1.25cm}R{1cm}R{1.55cm}R{1.4cm}R{1.5cm}R{1cm}R{1cm}R{0.9cm}}
\toprule
\textbf{Codec / configuration} & \textbf{Allocation} & \textbf{Event rate (Hz)} & \textbf{Bitrate (kbit/s)} & \textbf{UTMOSv2 \(\uparrow\)} & \textbf{DNSMOS \(\uparrow\)} & \textbf{PESQ-WB \(\uparrow\)} & \textbf{STOI \(\uparrow\)} & \textbf{dCER (\%) \(\downarrow\)} & \textbf{SIM \(\uparrow\)} \\
\midrule
GT & --- & --- & --- & 3.003 & 3.203 & 4.644 & 1.000 & 0.00 & 1.000 \\
\midrule[\heavyrulewidth]
X-Codec-2 & Fixed & 50.0 & 0.80 & 2.783 & 3.217 & 2.237 & 0.902 & 1.09 & 0.853 \\
WavTokenizer, 75 Hz & Fixed & 75.0 & 0.90 & 2.406 & \textbf{3.258} & 2.175 & 0.892 & 2.96 & 0.747 \\
BigCodec & Fixed & 80.0 & 1.04 & \underline{2.850} & 3.237 & 2.264 & 0.909 & 1.55 & 0.800 \\
Mimi, \(D\)=8 & Fixed & 12.5 & 1.10 & 2.397 & 3.035 & 2.052 & 0.882 & 2.16 & 0.732 \\
CodecSlime, FSQ18K, 80 Hz & Fixed & 80.0 & 1.13 & 2.840 & 3.236 & 2.144 & 0.897 & 2.52 & 0.692 \\
DyCAST, BP minimum gap 1 & Adaptive & 18.1 & 1.16 & 2.337 & \underline{3.253} & 1.191 & 0.334 & 7.08 & 0.729 \\
MiMo-Audio-Tokenizer, \(D\)=6 & Fixed & 25.0 & 1.20 & \textbf{2.862} & 3.218 & \underline{2.508} & 0.915 & \underline{0.69} & 0.870 \\
FlexiCodec, \(\tau\)=1, \(D\)=8 & Fixed & 12.5 & 1.24 & 2.528 & 3.175 & 2.353 & 0.911 & 0.88 & 0.804 \\
MOSS-Audio-Tokenizer, \(D\)=10 & Fixed & 12.5 & 1.25 & 2.842 & 3.108 & \textbf{2.905} & \textbf{0.942} & \textbf{0.56} & \textbf{0.928} \\
DTM-Codec, 80-Hz target & Adaptive & 80.3 & 1.27 & 2.783 & 3.213 & 2.459 & \underline{0.917} & 2.27 & 0.824 \\
EnCodec, 1.5 kbit/s & Fixed & 75.0 & 1.50 & 1.225 & 2.357 & 1.494 & 0.830 & 4.58 & 0.617 \\
DAC (24 kHz), \(D\)=2 & Fixed & 75.0 & 1.50 & 1.209 & 2.662 & 1.439 & 0.821 & 5.93 & 0.544 \\
\textbf{\boldmath RubatoCodec, \(D\)=8} & Adaptive & 12.5 & 1.22 & 2.811 & 3.240 & 2.459 & 0.916 & 0.91 & \underline{0.888} \\
\midrule[\heavyrulewidth]
Mimi, \(D\)=16 & Fixed & 12.5 & 2.20 & 2.562 & 3.068 & 2.658 & 0.924 & 0.80 & 0.871 \\
Qwen3-TTS-Tokenizer-12Hz & Fixed & 12.5 & 2.20 & \textbf{2.930} & \underline{3.209} & \underline{2.987} & \underline{0.951} & \underline{0.45} & \underline{0.952} \\
MiMo-Audio-Tokenizer, \(D\)=13 & Fixed & 25.0 & 2.42 & \underline{2.893} & 3.205 & 2.931 & 0.938 & 0.53 & 0.906 \\
FlexiCodec, \(\tau\)=1, \(D\)=16 & Fixed & 12.5 & 2.44 & 2.619 & 3.171 & 2.592 & 0.930 & 0.63 & 0.866 \\
MOSS-Audio-Tokenizer, \(D\)=20 & Fixed & 12.5 & 2.50 & 2.860 & 3.085 & \textbf{3.393} & \textbf{0.963} & \textbf{0.38} & \textbf{0.964} \\
EnCodec, 3 kbit/s & Fixed & 75.0 & 3.00 & 1.485 & 2.655 & 1.963 & 0.889 & 1.29 & 0.810 \\
DAC (24 kHz), \(D\)=4 & Fixed & 75.0 & 3.00 & 1.763 & 3.043 & 2.255 & 0.895 & 1.55 & 0.778 \\
\textbf{\boldmath RubatoCodec, \(D\)=16} & Adaptive & 12.5 & 2.42 & 2.858 & \textbf{3.231} & 2.795 & 0.939 & 0.50 & 0.934 \\
\bottomrule
\end{tabularx}
\end{table}

On seed-tts-eval test-zh, RubatoCodec shows competitive intelligibility preservation and speaker similarity, although the strongest comparators lead both bands (Table~\ref{tab:mandarin-reconstruction}). At depth 8, its dCER of 0.91\% is close to FlexiCodec's 0.88\% with substantially higher SIM (0.888 versus 0.804); only MOSS-Audio-Tokenizer achieves both lower dCER and higher SIM. At depth 16, dCER falls to 0.50\%, behind MOSS-Audio-Tokenizer (0.38\%) and Qwen3-TTS-Tokenizer-12Hz (0.45\%) but ahead of MiMo-Audio-Tokenizer at depth 13 (0.53\%). In this band, RubatoCodec also exceeds FlexiCodec on every metric and has the highest DNSMOS.

\FloatBarrier

\subsection{Event allocation analysis}\label{sec:event-allocation-analysis}

\subsubsection{Event-rate control across datasets}\label{sec:event-rate-control-across-datasets}

Table~\ref{tab:event-rates} compares event rates at the fixed settings in Section~\ref{sec:compared-codecs}. RubatoCodec and CodecSlime prescribe event counts from input length \cite{wang2026codecslime}; DTM-Codec's accumulated-change thresholds \cite{sohn2026dtmcodec}, FlexiCodec's similarity thresholds \cite{li2026flexicodec}, and DyCAST's alignment or boundary predictions \cite{dellalibera2026dycast} allow counts to vary with content.

\begin{table}[tbp]
\centering
\small
\caption{\textbf{Mean event rates (Hz) across datasets at fixed encoding settings.} Rates follow Section~\ref{sec:metrics}. DTM-Codec's 80-Hz label denotes a nominal target.}\label{tab:event-rates}
\begin{tabularx}{\linewidth}{LR{1.9cm}R{1.1cm}R{1.7cm}R{1.1cm}R{1.9cm}}
\toprule
\textbf{System / setting} & \textbf{LibriSpeech test-clean} & \textbf{VCTK} & \textbf{JVS nonpara30} & \textbf{JVNV} & \textbf{seed-tts-eval test-zh} \\
\midrule
RubatoCodec & 12.5 & 12.5 & 12.5 & 12.5 & 12.5 \\
CodecSlime, FSQ18K, 40 Hz & 40.0 & 40.0 & 40.0 & 40.0 & 40.0 \\
DTM-Codec, 80-Hz target & 81.5 & 80.2 & 63.3 & 46.1 & 80.3 \\
FlexiCodec, \(\tau\)=0.91 & 8.3 & 7.0 & 8.5 & 7.5 & 7.2 \\
DyCAST, character-aligned & 14.2 & 11.0 & 4.4 & 3.9 & 3.7 \\
DyCAST, boundary predictor, min-gap 1 & 17.4 & 13.6 & 18.0 & 15.3 & 18.1 \\
\bottomrule
\end{tabularx}
\end{table}

RubatoCodec and CodecSlime maintain their prescribed rates up to integer rounding. DTM-Codec's 80-Hz setting ranges from 46.1 to 81.5 events/s, and FlexiCodec at a fixed threshold ranges from 7.0 to 8.5 events/s.

These measurements illustrate the calibration problem noted in Section~\ref{sec:introduction}; an explicit budget instead determines counts directly from utterance length and target ratio.

DyCAST's character-aligned rate is 11.0--14.2 Hz on LibriSpeech test-clean and VCTK, 4.4 Hz on JVS nonpara30, 3.9 Hz on JVNV, and 3.7 Hz on seed-tts-eval test-zh; its boundary predictor does not show this reduction. These alignment rates depend on audio-derived CTC predictions and language-specific character units.

\FloatBarrier
\subsubsection{Adaptive Firing on real speech}\label{sec:adaptive-firing-on-real-speech}

\begin{figure}[tbp]
\centering
\includegraphics[width=\linewidth,height=0.70\textheight,keepaspectratio]{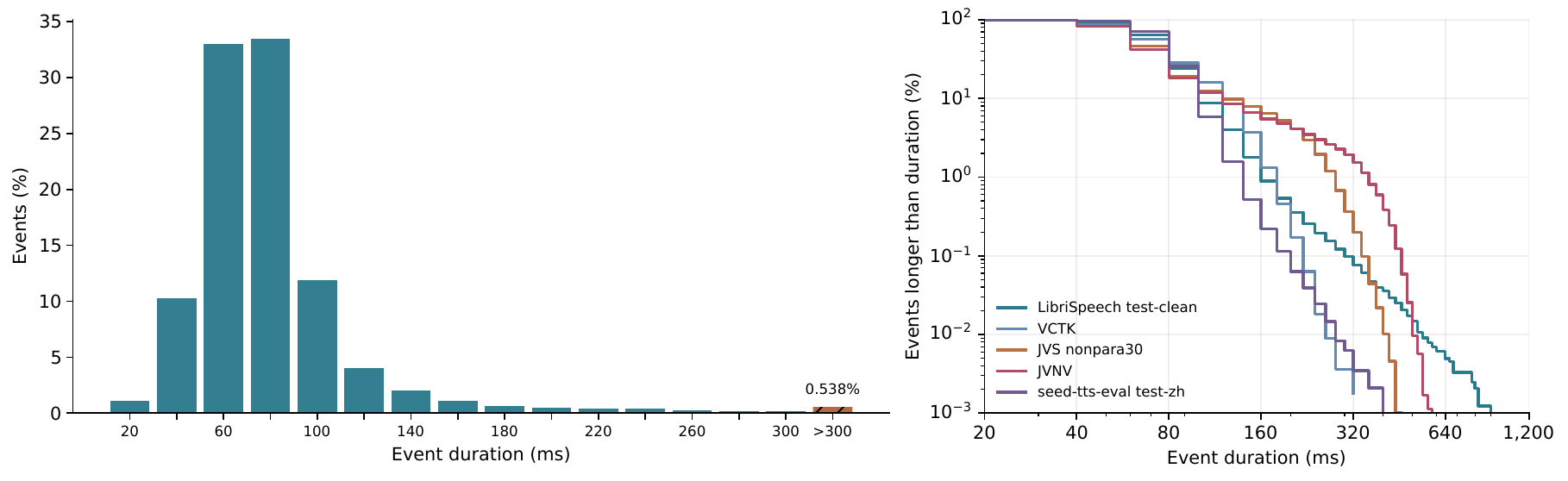}
\caption{\textbf{Event durations in reconstructed-speech evaluation.} Left: event-weighted distribution across all five datasets, with durations above 300 ms grouped in the final bin. Right: the percentage of events longer than each duration, shown separately for each dataset; both axes are logarithmic.}\label{fig:codec-duration-distribution}
\end{figure}

We analyze the multilingual model's event durations across all five evaluation datasets: 10,563 utterances comprising 817,359 events. Each event contributes one duration, including the final event of each utterance. Figure~\ref{fig:codec-duration-distribution} shows the pooled duration distribution and the fraction of events exceeding each duration within each dataset; Table~\ref{tab:event-durations} summarizes duration statistics by dataset.

\begin{table}[tbp]
\centering
\small
\caption{\textbf{Event-duration statistics by dataset.} Percentages are computed over encoded events.}\label{tab:event-durations}
\begin{tabularx}{\linewidth}{P{0.26\linewidth}YYYY}
\toprule
\textbf{Dataset} & \textbf{Median (ms)} & \textbf{95th percentile (ms)} & \textbf{99th percentile (ms)} & \textbf{Events \textgreater300 ms (\%)} \\
\midrule
LibriSpeech test-clean & 80 & 120 & 160 & 0.098 \\
VCTK & 80 & 140 & 180 & 0.004 \\
JVS nonpara30 & 60 & 200 & 280 & 0.363 \\
JVNV & 60 & 180 & 360 & 1.932 \\
seed-tts-eval test-zh & 80 & 120 & 140 & 0.006 \\
\bottomrule
\end{tabularx}
\end{table}

The pooled median duration is 80 ms, with 95\% of events lasting at most 140 ms and 99\% at most 260 ms. Longer events are uncommon: 4,395 events (0.538\%) exceed 300 ms, accounting for 2.476\% of the represented audio duration. The longest event lasts 1.14 s. The approximately 80-ms mean follows from the prescribed 12.5-Hz event budget; the distribution around that mean characterizes the nonuniform allocation.

JVS and JVNV have shorter median durations (60 ms) and higher 95th percentiles (200 and 180 ms, respectively) than the English and Mandarin datasets (80-ms medians and 120--140-ms 95th percentiles), despite nearly identical average event rates. JVNV has the largest proportion of events exceeding 300 ms (1.932\%, compared with 0.363\% on JVS) and the highest 99th percentile (360 ms).

\begin{figure}[tbp]
\centering
\includegraphics[width=0.85\linewidth,height=0.70\textheight,keepaspectratio]{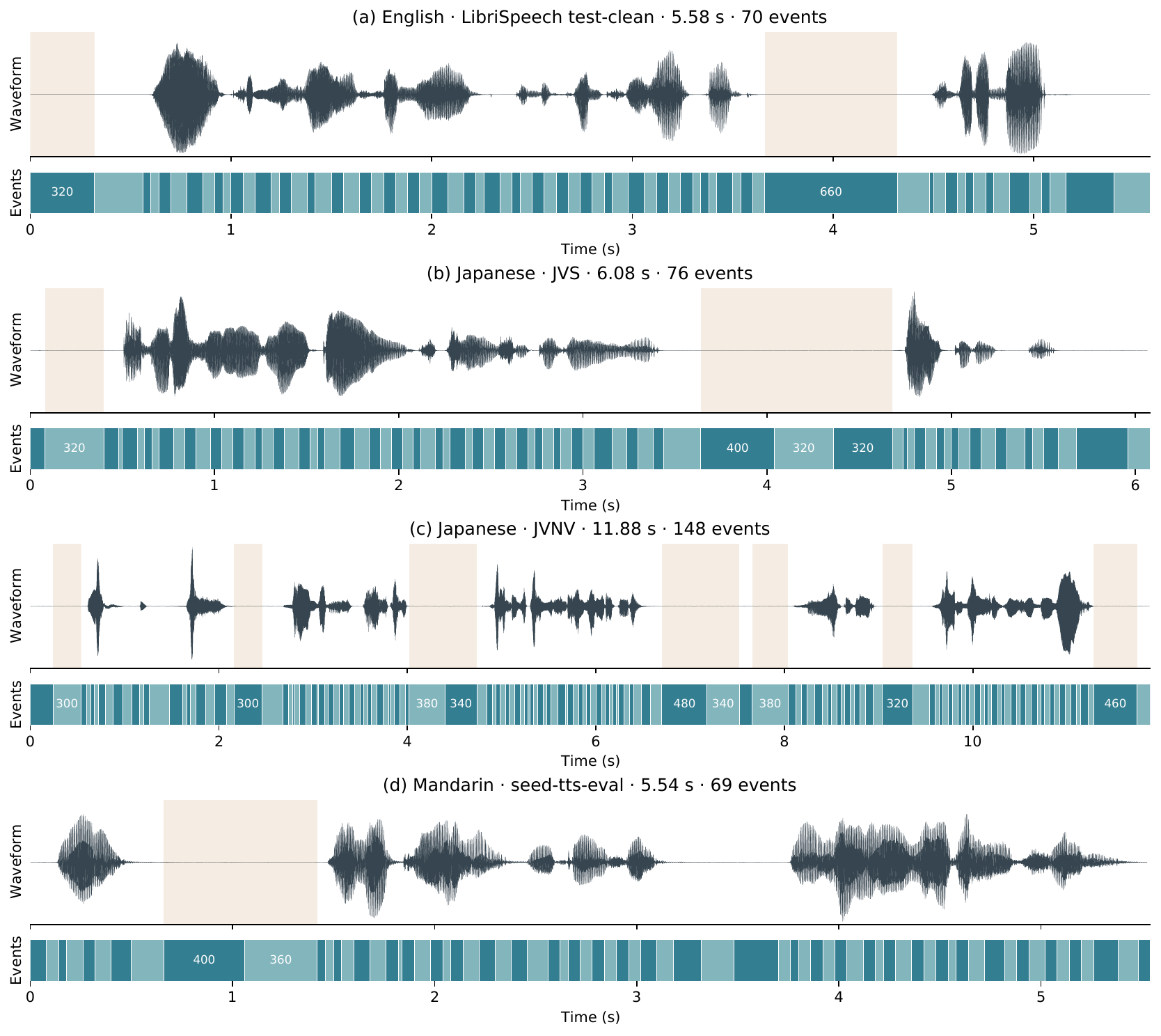}
\caption{\textbf{Adaptive event placement on English, Japanese, and Mandarin speech.} The upper trace in each example is the input waveform, normalized by its peak amplitude; the lower strip partitions the same timeline into encoded events. Each rectangle begins at a firing position and has width proportional to its duration. Alternating colors distinguish neighboring events. Events lasting at least 300 ms are labeled in milliseconds and shaded on the waveform.}\label{fig:firing-examples}
\end{figure}

Figure~\ref{fig:firing-examples} shows longer events spanning low-amplitude portions of the input, with shorter events distributed through the surrounding speech. To quantify this relationship across all five datasets, we compute input RMS amplitude in nonoverlapping 20-ms windows aligned to the codec grid. We classify an event as predominantly low-energy when at least 80\% of its windows fall below \(-50\) dBFS. This amplitude-based measure is a proxy for near-silence rather than an annotated speech-activity or phonetic category.

Of the 4,395 events longer than 300 ms, 4,394 (99.98\%) are predominantly low-energy. Varying the threshold from \(-40\) to \(-60\) dBFS yields proportions of 94.0--100.0\% across the five datasets. At \(-50\) dBFS, 99.97\% of frames within these long events are low-energy, compared with 25.2\% of frames in the remaining events. These observations are consistent with allocating longer-duration events to low-energy regions while retaining the prescribed utterance-level event budget.

\FloatBarrier
\subsection{Computational efficiency}\label{sec:computational-efficiency}

We measure encoding and decoding throughput on a single NVIDIA H200 with an utterance batch size of one, using the same 100 variable-length LibriSpeech test-clean utterances for each model. RTFx is the total input audio duration divided by the total processing time; higher values indicate faster processing. Timings exclude two warmup utterances and input resampling, and synchronize the GPU around each measured call. Encoding includes required feature extraction and temporal allocation. We use the released inference implementations, including the Transformers implementation for Mimi. Throughput depends on numerical precision, software, CPU processing, and output rate.

\begin{table}[tbp]
\centering
\small
\caption{\textbf{Full-utterance encoding and decoding throughput for selected codecs.}}\label{tab:codec-throughput}
\begin{tabularx}{\linewidth}{LR{0.22\linewidth}R{0.22\linewidth}}
\toprule
\textbf{Codec / configuration} & \textbf{Encoding RTFx \(\uparrow\)} & \textbf{Decoding RTFx \(\uparrow\)} \\
\midrule
X-Codec-2 \cite{ye2025llasa} & 72.7 & 271.9 \\
Mimi, \(D\)=16 \cite{defossez2024moshi} & 219.7 & 335.4 \\
MOSS-Audio-Tokenizer, \(D\)=20 \cite{gong2026mossaudiotokenizer} & 161.9 & 199.2 \\
Qwen3-TTS-Tokenizer-12Hz \cite{hu2026qwen3tts} & 416.3 & 279.8 \\
DTM-Codec, 80-Hz target \cite{sohn2026dtmcodec} & 471.7 & 518.6 \\
CodecSlime, FSQ18K, 40 Hz \cite{wang2026codecslime} & 25.7 & 190.4 \\
FlexiCodec, \(\tau\)=0.91, \(D\)=8 \cite{li2026flexicodec} & 95.8 & 196.7 \\
DyCAST, BP minimum gap 1 \cite{dellalibera2026dycast} & 165.8 & 1019.6 \\
RubatoCodec, \(D\)=8 & 70.9 & 446.8 \\
RubatoCodec, \(D\)=16 & 71.2 & 427.4 \\
\bottomrule
\end{tabularx}
\end{table}

RubatoCodec decodes 44.1-kHz audio at 446.8\(\times\) and 427.4\(\times\) real time at depths 8 and 16 (Table~\ref{tab:codec-throughput}), so quantizer depth has little effect on decoding throughput.

Encoding reaches approximately 71\(\times\) real time, including the frozen semantic feature extractor. Compared with X-Codec-2, which uses the same encoder backbone, RubatoCodec at depth 8 takes approximately 2.5\% longer to encode the same audio. This indicates a small overall encoding overhead, although the comparison does not separate Adaptive Firing from quantization differences. CodecSlime's 40-Hz encoding time includes its released CPU implementation of dynamic-programming segmentation.

In a separate 250-event streaming test using codes obtained from synthetic audio, event-by-event RubatoCodec decoding takes an average of 11.07 ms per event, including history recomputation and crossfading. At the nominal rate of 12.5 events/s, each event represents approximately 80 ms of audio on average, indicating sufficient average decoding throughput for real-time streaming synthesis on the tested hardware.

\section{Duration-Aware Text-to-Speech Modeling}\label{sec:duration-aware-text-to-speech-modeling}

\begin{figure}[tbp]
\centering
\includegraphics[width=\linewidth,height=0.70\textheight,keepaspectratio]{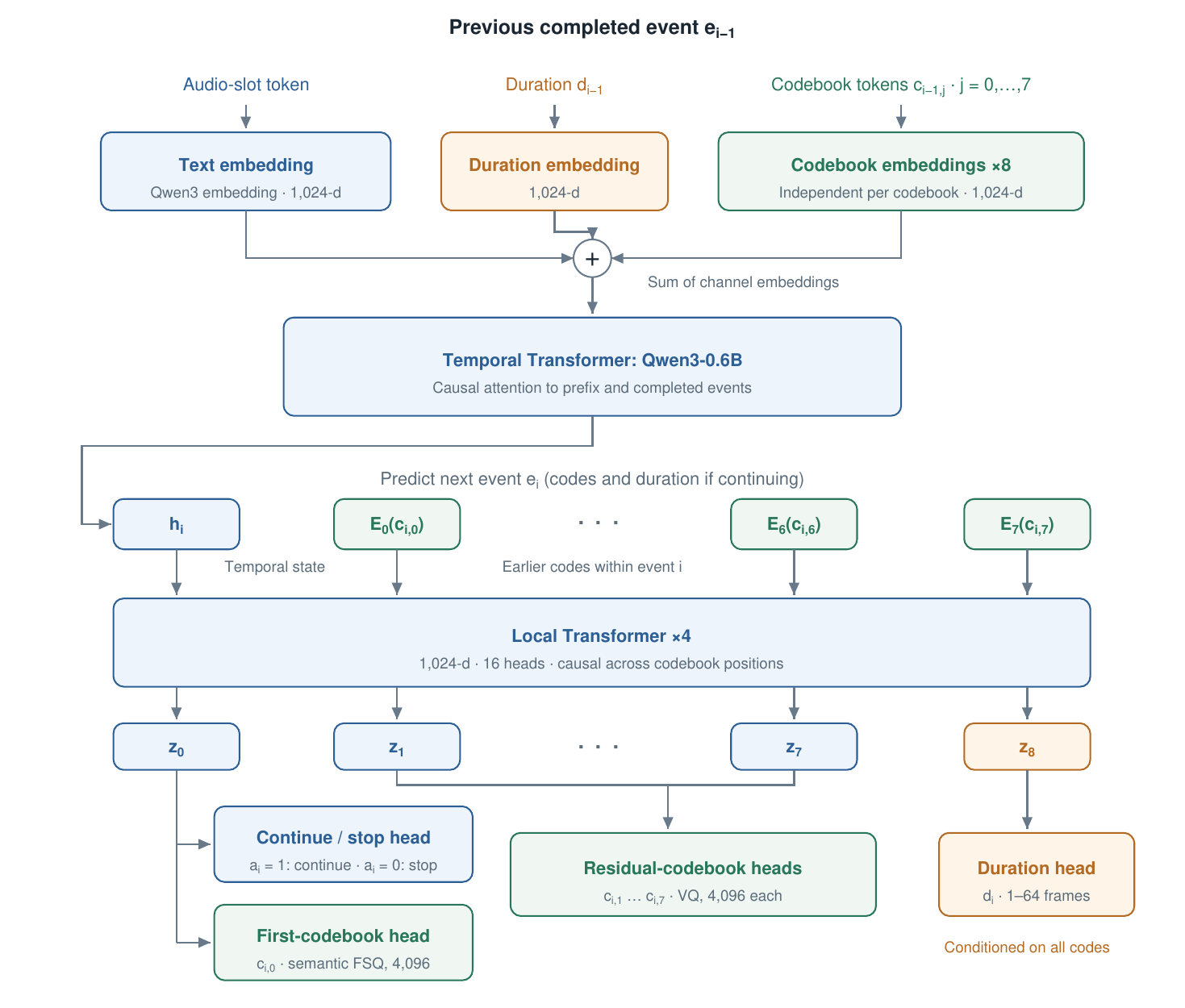}
\caption{\textbf{MOSS-TTS adapted to RubatoCodec, shown with eight codebooks.} The temporal Transformer predicts each event from the conditioning prefix and the completed preceding events; the local Transformer then predicts continuation or stop, the codebook tokens, and finally the duration. Completed code--duration pairs can be passed to the causal RubatoCodec decoder for streaming synthesis.}\label{fig:tts-architecture}
\end{figure}

\subsection{Generative interface}\label{sec:generative-interface}

We adapt the MOSS-TTS-Local-Transformer v1.5 implementation \cite{gong2026mosstts} to RubatoCodec (Figure~\ref{fig:tts-architecture}), retaining its temporal--local autoregressive structure, reference-and-text input format, and continuation/stop prediction. We add duration embeddings to the temporal input and predict each event's duration at the final local position, conditioned on all codebook tokens for that event. We use Qwen3-0.6B \cite{yang2025qwen3} in place of the Qwen3-4B temporal backbone and a four-layer local Transformer.

Reference speech and target text form the conditioning prefix, with the reference preceding the text. At text positions, only the text-token embedding contributes; at speech-event positions, the audio-slot token embedding is summed with the codebook and duration embeddings. Each completed event is fed back to the temporal Transformer to predict the next event. The first event is predicted from the state at the audio-start token following the conditioning prefix.

Each event is represented by \(e_i=(\mathbf c_i,d_i)\). Given text \(y\), optional reference \(r\), and preceding events \(e_{<i}\), the first local state predicts the continuation probability \(p(a_i=1\mid y,r,e_{<i})\), where \(a_i=1\) denotes continuation and \(a_i=0\) denotes stop. Conditional on continuation, the next event follows

\begin{equation}\label{eq:tts-factorization}
\begin{aligned}
p(\mathbf c_i,d_i\mid y,r,e_{<i},a_i=1)
&=\left[\prod_{j=0}^{D-1}p(c_{i,j}\mid y,r,e_{<i},a_i=1,c_{i,<j})\right]\\
&\quad\times p(d_i\mid y,r,e_{<i},a_i=1,\mathbf c_i).
\end{aligned}
\end{equation}

During training, speech events are labeled as continuation and the following end token as stop; codebook and duration losses apply only to speech events. During inference, generation terminates when stop is sampled.

We also compared alternative positions for duration prediction in a small ablation study; details are provided in Appendix~\ref{sec:duration-prediction-order}.

Once the local Transformer has generated an event's codebook tokens and duration, the causal codec decoder (Section~\ref{sec:causal-grid-decoding}) incrementally synthesizes the corresponding \(d_i/50\) seconds of audio. Decoder state is reused across events, with brief buffering at history recomputations (Section~\ref{sec:causal-grid-decoding}).

\subsection{TTS evaluation}\label{sec:tts-evaluation}

We compare TTS using the 100,000-update adaptive and fixed-rate codecs from Section~\ref{sec:training-for-architecture-comparisons}, both with eight quantizer layers. Each encodes approximately 31,300 hours from HiFiTTS-2's English 44-kHz partition after reserving held-out speakers \cite{langman2025hifitts2}. Both TTS models are trained for three epochs with a Qwen3-0.6B temporal backbone and a four-layer local Transformer, using matched initialization and training settings (Appendix~\ref{sec:tts-training-and-sampling}). For adaptive coding, utterances containing durations above 64 frames (1.28 s) are excluded, affecting approximately 0.07\% of training utterances.

We evaluate zero-shot voice cloning on the 1,088 English utterances in seed-tts-eval \cite{anastassiou2024seedtts}, using the same target texts and reference recordings for both systems. Following the benchmark protocol, WER is computed from Whisper-large-v3 transcriptions \cite{radford2023whisper}, and SIM is the cosine similarity of embeddings from the same fine-tuned WavLM-Large speaker verification model \cite{chen2022wavlm} between generated speech and the original reference recording. We additionally assess predicted speech quality with UTMOS \cite{saeki2022utmos} and UTMOSv2 \cite{baba2024utmosv2}. Sampling settings are listed in Appendix~\ref{sec:tts-training-and-sampling}. TTS WER is macro-averaged over utterances, following the seed-tts-eval protocol; SIM, UTMOS, and UTMOSv2 are also averaged over utterances.

\begin{table}[tbp]
\centering
\small
\caption{\textbf{TTS results on the English seed-tts-eval benchmark.} Both systems use eight codebooks at a nominal event rate of 12.5 Hz. Bold marks the better score; ties share the formatting.}\label{tab:tts-results}
\begin{tabularx}{\linewidth}{LYYYY}
\toprule
\textbf{Codec} & \textbf{WER (\%) \(\downarrow\)} & \textbf{SIM \(\uparrow\)} & \textbf{UTMOS \(\uparrow\)} & \textbf{UTMOSv2 \(\uparrow\)} \\
\midrule
Fixed-rate codec & 3.20 & \textbf{0.392} & 4.080 & 3.239 \\
Adaptive codec & \textbf{3.10} & \textbf{0.392} & \textbf{4.162} & \textbf{3.295} \\
\bottomrule
\end{tabularx}
\end{table}

The adaptive codec achieves lower WER and comparable speaker similarity, with higher UTMOS and UTMOSv2 scores (Table~\ref{tab:tts-results}).

\subsection{Generated event timing}\label{sec:generated-event-timing}

\begin{figure}[tbp]
\centering
\includegraphics[width=\linewidth,height=0.70\textheight,keepaspectratio]{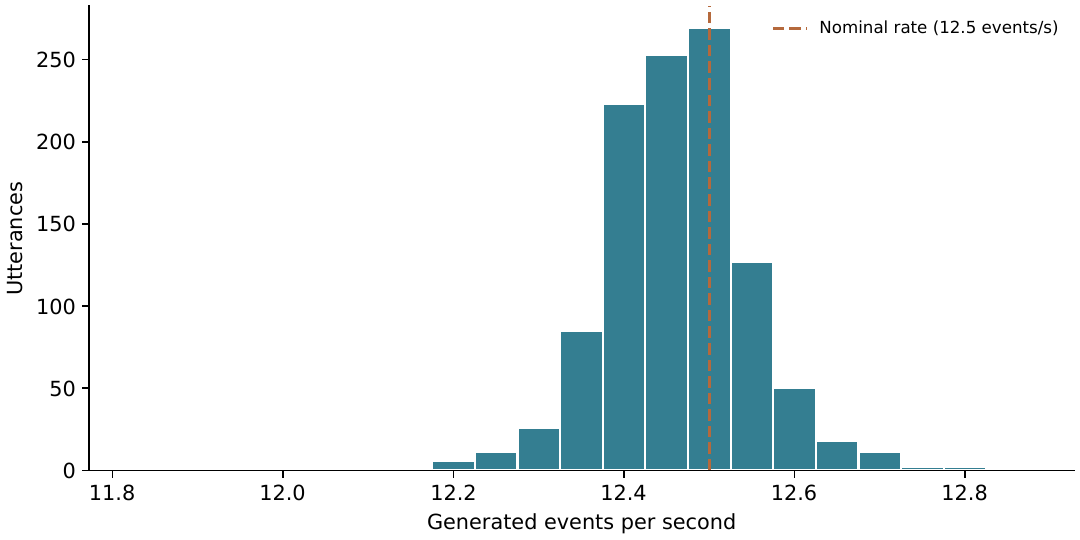}
\caption{\textbf{Generated event rates on English seed-tts-eval.} Each observation is one utterance. The histogram uses 0.05-events/s bins centered on multiples of 0.05, and the dashed line marks the codec's nominal rate of 12.5 events/s. All 1,088 utterances are included.}\label{fig:generated-event-rates}
\end{figure}

We examine generated event timing on the 1,088 English utterances in seed-tts-eval \cite{anastassiou2024seedtts}, drawing a separate set of samples from the adaptive TTS model with the sampling settings in Appendix~\ref{sec:tts-training-and-sampling}. For an utterance with \(N\) generated events and durations \(d_i\) on the 50-Hz grid, the event rate is \(50N/\sum_i d_i\) events/s. Figure~\ref{fig:generated-event-rates} shows that these rates concentrate near the codec's nominal rate of 12.5 events/s: the mean is 12.461, the median is 12.454, and the 5th--95th percentile interval is 12.329--12.609 events/s. Individual utterance rates range from 11.897 to 12.849 events/s. Generated sequences also follow the codec's budget rule: for 88.3\% of utterances, the generated event count equals \(\operatorname{round}(0.25\,T)\) for the generated grid length \(T=\sum_i d_i\), and 99.9\% are within one event, without an explicit generation-time rate constraint.

\begin{figure}[tbp]
\centering
\includegraphics[width=\linewidth,height=0.70\textheight,keepaspectratio]{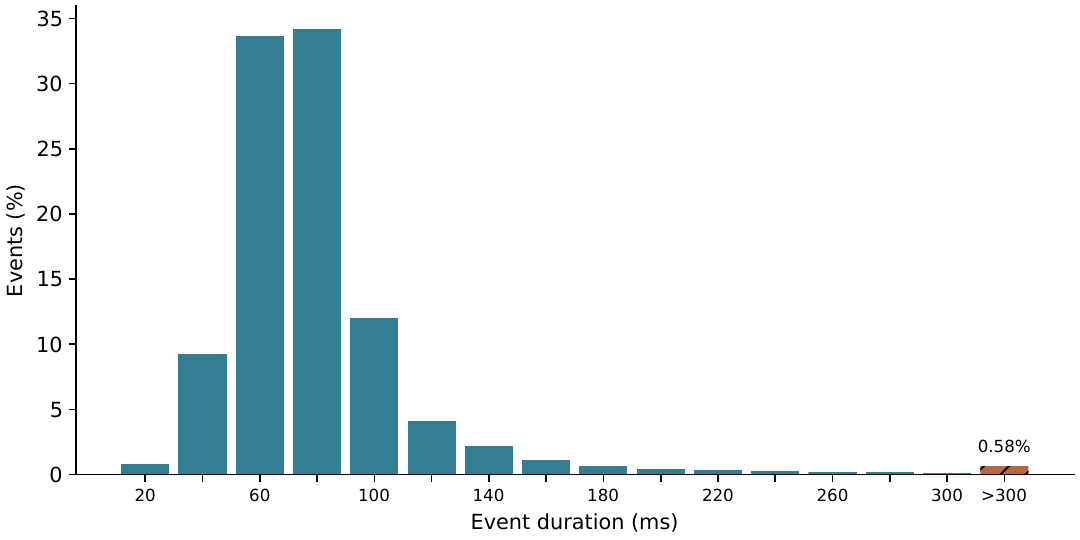}
\caption{\textbf{Generated event durations on English seed-tts-eval.} Bars show the percentage of events at each 20-ms duration increment through 300 ms. The final, hatched bin groups all durations above 300 ms. All events, including the long tail, contribute to the reported mean, median, and range.}\label{fig:generated-event-durations}
\end{figure}

Across 61,965 generated events, mean duration is 80.24 ms, the median is 80 ms, and the 5th--95th percentile interval is 40--140 ms (Figure~\ref{fig:generated-event-durations}). A sparse tail contains 362 events (0.58\%) above 300 ms, including two above one second; the maximum is 1.22 s.

To characterize the long-duration tail, we use the same predominantly low-energy criterion as in Section~\ref{sec:adaptive-firing-on-real-speech}: at least 80\% of an event's aligned 20-ms windows must have RMS amplitude below \(-50\) dBFS. Of the 362 events longer than 300 ms, 361 (99.7\%) meet this criterion; varying the threshold from \(-40\) to \(-60\) dBFS yields proportions of 99.4--100\%. All 98 first events and 32 last events in this subset are predominantly low-energy, as are 231 of the 232 interior events. Long events are disproportionately common at utterance boundaries: they account for 9.0\% of first events and 2.9\% of last events, compared with 0.39\% of interior events.

The long tail predominantly represents near-silence, concentrated at utterance boundaries. Informal listening also revealed an unnatural within-phrase pause associated with a long event.

\section{Discussion}\label{sec:discussion}

\subsection{Adaptive allocation under an explicit budget}\label{sec:adaptive-allocation-under-an-explicit-budget}

An exact event budget leaves useful freedom to distribute temporal resolution within an utterance. This freedom matters most at lower event rates: the gains over fixed-rate downsampling appear at 6.25 and 12.5 Hz and narrow at 25 Hz (Section~\ref{sec:adaptive-versus-fixed-rate-architectures}). Those comparisons change placement, interval aggregation, and decoding grid together, and the adaptive codec has lower speaker similarity there. With aggregation and grid held fixed, however, adaptive placement raises speaker similarity relative to uniform stride (Section~\ref{sec:uniform-stride-versus-adaptive-event-placement}). Although the two comparisons use different training data and lengths, this suggests that the SIM gap in Section~\ref{sec:adaptive-versus-fixed-rate-architectures} stems from interval aggregation and grid design rather than from placement itself.

The uniform-stride comparison (Section~\ref{sec:uniform-stride-versus-adaptive-event-placement}) supports content-dependent event placement with aggregation and decoding grid held fixed. The firing-rule comparison (Appendix~\ref{sec:firing-decision-rules}) separately favors cumulative allocation over the tested level-thresholding rule. Cumulative allocation produces approximately uniform spacing under a constant firing signal and adjusts that spacing with the cumulative bounded increments. Broad temporal coverage combined with content-dependent placement is one possible explanation for its performance, but these experiments do not isolate that mechanism.

The firing analysis shows how the budget is used: events longer than 300 ms predominantly cover low-energy regions across all five datasets, leaving more events for the remaining speech (Section~\ref{sec:adaptive-firing-on-real-speech}). Budgeted and threshold-based allocation also make different trade-offs. A prescribed count gives predictable sequence lengths without dataset-specific calibration, but equally long utterances receive the same budget regardless of acoustic complexity. Threshold-based methods can instead spend more events on difficult utterances.

\subsection{Temporal resolution and quantizer depth for speech generation}\label{sec:temporal-resolution-and-quantizer-depth-for-speech-generation}

Temporal resolution and quantizer depth are not interchangeable ways of spending a bitrate budget. In the evaluated configurations, additional residual layers do not fully compensate for coarser segmentation: 12.5 Hz approaches 25 Hz near 1.2 kbit/s, but 6.25 Hz falls clearly behind, and 25 Hz regains a fidelity advantage near 2.4 kbit/s (Section~\ref{sec:event-rate-and-quantizer-depth-at-similar-bitrates}).

For generation, these choices also distribute computation differently. Hierarchical models such as Moshi, Qwen3-TTS, and MOSS-TTS predict codebook tokens with a smaller within-event module \cite{defossez2024moshi,hu2026qwen3tts,gong2026mosstts}. Increasing depth adds predictions there, whereas increasing event rate adds steps to the larger temporal backbone. The reconstruction results therefore motivate 12.5 Hz as a practical compromise between quality and sequence length, without establishing it as optimal. Because RubatoCodec's waveform decoder still operates on a 50-Hz grid, the shorter sequence reduces downstream modeling cost but not waveform-decoder computation.

The inference-time ratio experiment (Appendix~\ref{sec:inference-time-event-rate-variation}) shows asymmetric tolerance to event rates outside the training setting: raising the rate improves reconstruction at additional bitrate, whereas aggressive reduction degrades it sharply even at greater quantizer depth. The model thus tolerates higher, but not lower, inference rates.

\subsection{From encoded durations to generated timing}\label{sec:from-encoded-durations-to-generated-timing}

The matched TTS comparison shows that codes and durations can be generated jointly while improving intelligibility and predicted quality over the fixed-rate codec (Section~\ref{sec:tts-evaluation}). Each completed event can be passed directly to the causal waveform decoder, so the same representation serves both temporal compression and incremental synthesis.

The generator also learns the codec's utterance-level event budget. Generated event counts match the codec's budget rule for 88.3\% of utterances and fall within one event for 99.9\% (Section~\ref{sec:generated-event-timing}), so the generator offsets long events with short ones within each utterance. Generated sequence length therefore remains predictable, although no rate constraint is imposed during generation. Long events also predominantly correspond to low-energy regions in both encoded speech and generated audio under the same measurement criterion (Sections~\ref{sec:adaptive-firing-on-real-speech} and \ref{sec:generated-event-timing}), indicating that the generator reproduces the codec's use of long events for near-silence. The concentration of long events at utterance boundaries is consistent with learning leading and trailing silence from HiFiTTS-2, although that training-data correspondence has not been measured directly.

Pause placement, however, is not always natural: informal listening revealed an unnatural within-phrase pause associated with a long event (Section~\ref{sec:generated-event-timing}). Efficiently representing silence and placing pauses naturally are therefore distinct modeling concerns.

\subsection{Reconstruction quality and multilingual performance}\label{sec:reconstruction-quality-and-multilingual-performance}

On the English benchmarks, RubatoCodec combines strong intelligibility preservation with high predicted-quality scores, while the strongest comparators retain higher signal fidelity and speaker similarity (Section~\ref{sec:comparison-with-existing-codecs}). On the Japanese benchmarks, its intelligibility advantage extends to both read and emotional speech, whereas predicted-quality rankings depend on the dataset. In Mandarin, intelligibility preservation remains competitive, although the strongest alternatives achieve lower dCER. These patterns distinguish preservation of linguistic content, speaker characteristics, and waveform detail; improvements in one need not imply improvements in the others. RubatoCodec's UTMOSv2 scores exceed those of the original recordings on LibriSpeech, indicating that higher predicted quality should not be interpreted as greater fidelity to the source. The 44.1-kHz output rate is a synthesis capability, not evidence of full-band source recovery. Encoding uses 16-kHz input, so higher frequencies must be inferred.

The English, Japanese, and Mandarin results establish reconstruction across languages with a shared model and allocation rule. Combined with the cross-dataset rate analysis (Section~\ref{sec:event-rate-control-across-datasets}), they demonstrate both predictable encoded sequence lengths and content preservation across the evaluated languages.

\subsection{Scope and limitations}\label{sec:scope-and-limitations}

The reported architecture comparisons use one trained model per configuration, and supporting architectural ablations have narrower coverage. In particular, firing-signal selection uses an early single-FSQ, low-bitrate design and a smaller training budget; its ranking may change with the current residual quantizer and training scale (Appendix~\ref{sec:firing-signal-selection}). The placement comparison uses one seed and English read speech, so its robustness across training runs, languages, and recording conditions remains untested. Firing statistics characterize duration and energy associations, not phonetic alignment or optimal allocation, and do not separate the quality contribution of allocation within speech from that of allocation to low-energy regions.

External codecs differ in training resources and operating points. Reporting realized rates and timing costs makes these differences visible but does not control them. All evaluated languages occur in RubatoCodec's training mixture, so the multilingual results do not establish unseen-language transfer or language robustness attributable solely to Adaptive Firing. Dataset differences also confound language, speakers, and recording conditions.

TTS evidence covers one English audiobook corpus and codecs from the controlled comparison (Section~\ref{sec:training-for-architecture-comparisons}). The comparison evaluates the complete representation and does not isolate the TTS contribution of adaptive placement. It also does not establish the quality attainable with the multilingual codec, broader data, or a larger generator. Long events are more frequent in TTS than in the reconstruction evaluation, but the corpora differ. Human evaluation remains necessary to assess reconstruction and TTS preferences and pause naturalness.

\section{Future Work}\label{sec:future-work}

For the codec, the priority is signal fidelity at low event rates: how far improved acoustic representations and reconstruction objectives can close PESQ/STOI gaps without longer sequences, and when finer temporal resolution becomes necessary. Training with multiple retention ratios could broaden the inference-time tolerance observed in Appendix~\ref{sec:inference-time-event-rate-variation}, following adjustable-rate coding and sequence compression \cite{li2026flexicodec,chen2023onceforall}. Online encoding would further require a causal acoustic--semantic path and a windowed or running budget in place of the utterance-level guarantee, trading buffering latency against placement quality and rate control.

For generation, larger-scale TTS with the multilingual codec, broader data, and larger backbones would test whether reconstruction improvements carry over, and explicit durations could support utterance-length and pause control. Finally, applying Adaptive Firing to other codec backbones and to music or environmental sound would test its generality.

\section{Conclusion}\label{sec:conclusion}

We introduced RubatoCodec, a speech codec that combines content-adaptive event placement, an exact utterance-level event budget, and streaming waveform decoding. By representing speech as discrete codes paired with durations, it allows temporal resolution to follow speech content while keeping sequence length predictable.

Controlled experiments support the value of adaptive allocation for reconstruction at low event rates. Multilingual evaluation demonstrates competitive reconstruction quality, with the best intelligibility preservation among the compared codecs on the evaluated English and Japanese benchmarks at comparable bitrates. Signal fidelity remains an area for improvement relative to the strongest alternatives. In matched text-to-speech experiments, the adaptive representation improves intelligibility and predicted speech quality with comparable speaker similarity. The generator also learns the codec's utterance-level event budget without an explicit rate constraint.

These results support budgeted adaptive speech tokenization as a practical representation for both reconstruction and speech generation. Combining content-dependent timing with predictable sequence length provides a basis for compact speech modeling and streaming synthesis.

\section*{Generative AI Use Disclosure}

Generative AI tools were used to help express the authors' technical descriptions and arguments in English and improve the manuscript's clarity and readability. Their use was limited to writing and editing assistance; the original ideas, methods, and conclusions were developed by the authors. The authors reviewed and revised the text and take responsibility for its content.

\clearpage
\bibliographystyle{IEEEtran}
\bibliography{bibliography-control,references}

\clearpage
\appendix
\counterwithin*{table}{section}
\counterwithin*{figure}{section}
\renewcommand{\thetable}{\thesection\arabic{table}}
\renewcommand{\thefigure}{\thesection\arabic{figure}}
\renewcommand{\theHtable}{\thesection.\arabic{table}}
\renewcommand{\theHfigure}{\thesection.\arabic{figure}}
\section{Supporting Codec Ablations}\label{sec:supporting-codec-ablations}

\subsection{Firing-signal selection}\label{sec:firing-signal-selection}

We conducted this preliminary study before introducing residual vector quantization, using a single FSQ codebook to compare candidate firing signals (Table~\ref{tab:firing-signal-ablation}). Each model was trained for 100,000 steps and evaluated on 200 utterances from a HiFiTTS-2 validation split constructed by holding out speakers from the training subset. The scores serve primarily to compare signal choices within this early configuration; their absolute values do not characterize the final codec and are not directly comparable with the LibriSpeech test-clean results in Section~\ref{sec:adaptive-versus-fixed-rate-architectures}.

\begin{table}[tbp]
\centering
\small
\caption{\textbf{Firing-signal ablations with a single FSQ codebook.} Bold marks the best score in each column.}\label{tab:firing-signal-ablation}
\begin{tabularx}{\linewidth}{LR{1.7cm}R{1.2cm}R{1.2cm}R{1cm}}
\toprule
\textbf{Firing signal} & \textbf{PESQ-WB \(\uparrow\)} & \textbf{STOI \(\uparrow\)} & \textbf{dWER (\%) \(\downarrow\)} & \textbf{SIM \(\uparrow\)} \\
\midrule
Adjacent latent dissimilarity & 1.262 & 0.787 & 3.30 & 0.397 \\
Multi-scale latent dissimilarity & 1.307 & 0.804 & 2.81 & \textbf{0.458} \\
Waveform flux & 1.327 & 0.810 & 2.20 & 0.447 \\
Adjacent latent dissimilarity + flux & 1.324 & 0.808 & 2.53 & 0.426 \\
Multi-scale latent dissimilarity + flux & \textbf{1.345} & 0.810 & 2.31 & 0.454 \\
Learned difficulty & 1.335 & 0.811 & 2.31 & 0.412 \\
Learned difficulty + flux & 1.336 & \textbf{0.813} & \textbf{1.36} & 0.418 \\
Learned difficulty + multi-scale latent dissimilarity & 1.330 & 0.810 & 1.86 & 0.412 \\
Learned difficulty + multi-scale latent dissimilarity + flux & 1.319 & 0.811 & 2.04 & 0.404 \\
\bottomrule
\end{tabularx}
\end{table}

\textbf{Latent dissimilarity and flux.} Multi-scale latent dissimilarity and waveform flux are the two components described in Section~\ref{sec:firing-signal}: the latent-dissimilarity component in Equation~\eqref{eq:latent-dissimilarity} and log-mel spectral flux in Equation~\eqref{eq:spectral-flux}. Adjacent latent dissimilarity is the single-offset variant of Equation~\eqref{eq:latent-dissimilarity}, using only \((a,b)=(1,\allowbreak 0)\), so it compares each latent frame with its immediate predecessor. The multi-scale-plus-flux row uses the firing signal defined in Equation~\eqref{eq:firing-signal}.

\textbf{Learned difficulty.} A separate convolutional network takes the current and preceding latent frames and predicts a nonnegative scalar reconstruction difficulty. During training, its target is the decoder's framewise absolute log-mel reconstruction error, averaged over mel bins and normalized by its utterance mean. The network learns this target with a mean-squared regression loss; during encoding, its prediction supplies the firing signal without requiring waveform reconstruction. In mixed-signal variants, each component is normalized to unit utterance mean before summation.

We prioritize firing signals that require no additional trainable predictor. Among these candidates, multi-scale latent dissimilarity combined with spectral flux achieves the highest PESQ-WB, with STOI and speaker similarity comparable to flux alone, although dWER is slightly higher. Learned difficulty combined with flux yields lower dWER but requires an additional predictor and has lower speaker similarity. We therefore adopt the combined firing signal as a balance between architectural simplicity and reconstruction quality in this exploratory comparison.

\subsection{Effect of causal decoding}\label{sec:effect-of-causal-decoding}

We compared the fully causal decoder architecture described in Section~\ref{sec:causal-grid-decoding} with a fully non-causal variant, keeping the rest of the codec architecture unchanged, at 12.5 events per second. The non-causal variant uses bidirectional attention, symmetric padding in its convolutional and transposed-convolutional layers, and symmetrically cropped overlap-add synthesis in the ISTFT. Each model was trained for 100,000 steps on the approximately 5,000-hour HiFiTTS-2 subset and evaluated on the 2,620 LibriSpeech test-clean utterances at quantizer depths 8 and 16. The causal model is the adaptive codec from Section~\ref{sec:adaptive-versus-fixed-rate-architectures}.

\begin{table}[tbp]
\centering
\footnotesize
\setlength{\tabcolsep}{3pt}
\caption{\textbf{Reconstruction quality with fully non-causal and fully causal decoders.} Bold marks the best score within each quantizer depth.}\label{tab:decoder-causality}
\begin{tabularx}{\linewidth}{P{0.28\linewidth}YYYYYY}
\toprule
\textbf{Decoder} & \textbf{Depth} & \textbf{UTMOSv2 \(\uparrow\)} & \textbf{PESQ-WB \(\uparrow\)} & \textbf{STOI \(\uparrow\)} & \textbf{dWER (\%) \(\downarrow\)} & \textbf{SIM \(\uparrow\)} \\
\midrule
Non-causal & 8 & \textbf{3.465} & \textbf{2.478} & \textbf{0.923} & \textbf{0.76} & \textbf{0.792} \\
Causal & 8 & 3.448 & 2.379 & 0.917 & 0.80 & 0.757 \\
\midrule
Non-causal & 16 & \textbf{3.471} & \textbf{2.651} & \textbf{0.933} & \textbf{0.60} & \textbf{0.836} \\
Causal & 16 & 3.444 & 2.556 & 0.929 & 0.62 & 0.807 \\
\bottomrule
\end{tabularx}
\end{table}

In Table~\ref{tab:decoder-causality}, causal decoding reduces PESQ-WB relative to fully non-causal decoding by 0.099 at depth 8 and 0.095 at depth 16, with corresponding STOI decreases of 0.006 and 0.004 and SIM decreases of 0.035 and 0.029. UTMOSv2 is slightly lower. dWER is also slightly higher with causal decoding at both depths (0.80\% versus 0.76\% at depth 8 and 0.62\% versus 0.60\% at depth 16). These results support fully causal waveform synthesis with a modest loss in reference-based reconstruction quality.

\subsection{Sparse versus repeated event representations}\label{sec:sparse-versus-repeated-event-representations}

This ablation tests whether the decoder benefits from receiving the event representation at every frame of its interval, rather than only at its start. We compare the sparse grid composition in Equation~\eqref{eq:decoding-grid} with a variant that replaces the learned empty vector at interior positions with the corresponding dequantized event representation, giving \(z_t=\hat g_i^{(D)}+E_r(\min(r,\allowbreak 64))\) throughout the interval. Both variants retain the same offset embeddings and fully causal decoder architecture. Each model was trained for 100,000 steps on the approximately 5,000-hour HiFiTTS-2 subset at 12.5 events per second and evaluated on the 2,620 LibriSpeech test-clean utterances at quantizer depths 8 and 16. The sparse model is the same adaptive codec used in Appendix~\ref{sec:effect-of-causal-decoding}.

\begin{table}[tbp]
\centering
\footnotesize
\setlength{\tabcolsep}{3pt}
\caption{\textbf{Sparse grid composition versus repeating the event representation across its interval.} Bold marks the best score within each quantizer depth.}\label{tab:grid-composition}
\begin{tabularx}{\linewidth}{P{0.28\linewidth}YYYYYY}
\toprule
\textbf{Grid composition} & \textbf{Depth} & \textbf{UTMOSv2 \(\uparrow\)} & \textbf{PESQ-WB \(\uparrow\)} & \textbf{STOI \(\uparrow\)} & \textbf{dWER (\%) \(\downarrow\)} & \textbf{SIM \(\uparrow\)} \\
\midrule
Repeated event vector & 8 & \textbf{3.463} & 2.324 & 0.915 & 0.84 & \textbf{0.759} \\
Sparse grid & 8 & 3.448 & \textbf{2.379} & \textbf{0.917} & \textbf{0.80} & 0.757 \\
\midrule
Repeated event vector & 16 & \textbf{3.475} & 2.507 & 0.926 & 0.65 & \textbf{0.809} \\
Sparse grid & 16 & 3.444 & \textbf{2.556} & \textbf{0.929} & \textbf{0.62} & 0.807 \\
\bottomrule
\end{tabularx}
\end{table}

Sparse composition improves PESQ-WB and STOI at both depths (Table~\ref{tab:grid-composition}), while repeated representations yield slightly higher SIM and UTMOSv2. Sparse composition also yields lower dWER at both depths. These results support sparse placement for signal fidelity.

\subsection{Firing decision rules}\label{sec:firing-decision-rules}

This ablation compares cumulative allocation with level thresholding of the firing signal. Both models use the architecture in Section~\ref{sec:rubatocodec-architecture}, including the firing signal defined in Equation~\eqref{eq:firing-signal}, interval aggregation, hybrid residual quantizer, sparse decoding grid, and fully causal waveform decoder. Each model is trained from scratch for 100,000 steps on the approximately 5,000-hour HiFiTTS-2 subset using the settings in Section~\ref{sec:training-for-architecture-comparisons}, with retention ratio \(R=0.25\). Evaluation uses LibriSpeech test-clean.

\textbf{Level thresholding} selects frames using an utterance-specific threshold at the signal's \((1-R)\)-quantile, with the initial frame always retained. Training uses a sigmoid gate with a straight-through estimator, while inference uses hard decisions. \textbf{Cumulative allocation} uses the bounded cumulative allocation in Section~\ref{sec:budget-constrained-event-allocation} and enforces the prescribed event count without differentiating through boundary selection. The two rules differ in both event-count enforcement and the training gradient through boundary selection, so this comparison evaluates the complete firing rules rather than isolating placement alone.

\begin{table}[tbp]
\centering
\footnotesize
\setlength{\tabcolsep}{3pt}
\caption{\textbf{Firing-rule comparison with the residual quantizer and causal decoder.} Bold marks the best score within each quantizer depth. dWER follows Section~\ref{sec:metrics}.}\label{tab:firing-rules}
\begin{tabularx}{\linewidth}{P{0.25\linewidth}YYYYYYY}
\toprule
\textbf{Firing rule} & \textbf{Depth} & \textbf{UTMOSv2 \(\uparrow\)} & \textbf{DNSMOS \(\uparrow\)} & \textbf{PESQ-WB \(\uparrow\)} & \textbf{STOI \(\uparrow\)} & \textbf{dWER (\%) \(\downarrow\)} & \textbf{SIM \(\uparrow\)} \\
\midrule
Level thresholding & 8 & 3.299 & 3.311 & 1.657 & 0.859 & 5.62 & 0.613 \\
Cumulative allocation & 8 & \textbf{3.448} & \textbf{3.314} & \textbf{2.379} & \textbf{0.917} & \textbf{0.80} & \textbf{0.757} \\
\midrule
Level thresholding & 16 & 3.319 & 3.306 & 1.751 & 0.871 & 3.33 & 0.649 \\
Cumulative allocation & 16 & \textbf{3.444} & \textbf{3.307} & \textbf{2.556} & \textbf{0.929} & \textbf{0.62} & \textbf{0.807} \\
\bottomrule
\end{tabularx}
\end{table}

Cumulative allocation achieves higher PESQ-WB, STOI, speaker similarity, UTMOSv2, and DNSMOS and lower dWER than level thresholding at both depths (Table~\ref{tab:firing-rules}). These results favor cumulative allocation over the tested level-thresholding rule in this setting; they do not identify which difference between the rules causes the improvement.

\subsection{Inference-time event-rate variation}\label{sec:inference-time-event-rate-variation}

We test whether the multilingual model trained at retention ratio \(R=0.25\) can reconstruct speech at other event rates without retraining. We re-encode the same 2,620 LibriSpeech test-clean utterances with \(R\in\{0.10,\allowbreak 0.15,\allowbreak 0.20,\allowbreak 0.25,\allowbreak 0.30,\allowbreak 0.35,\allowbreak 0.40\}\), corresponding to nominal rates of 5--20 events/s on the 50-Hz grid. Only the retention ratio changes; model weights, the firing-signal definition, and decoding settings remain fixed. Each representation is decoded at depths 4, 8, and 16. Metrics and bitrate accounting follow Section~\ref{sec:metrics}.

\begin{figure}[tbp]
\centering
\includegraphics[width=\linewidth,height=0.70\textheight,keepaspectratio]{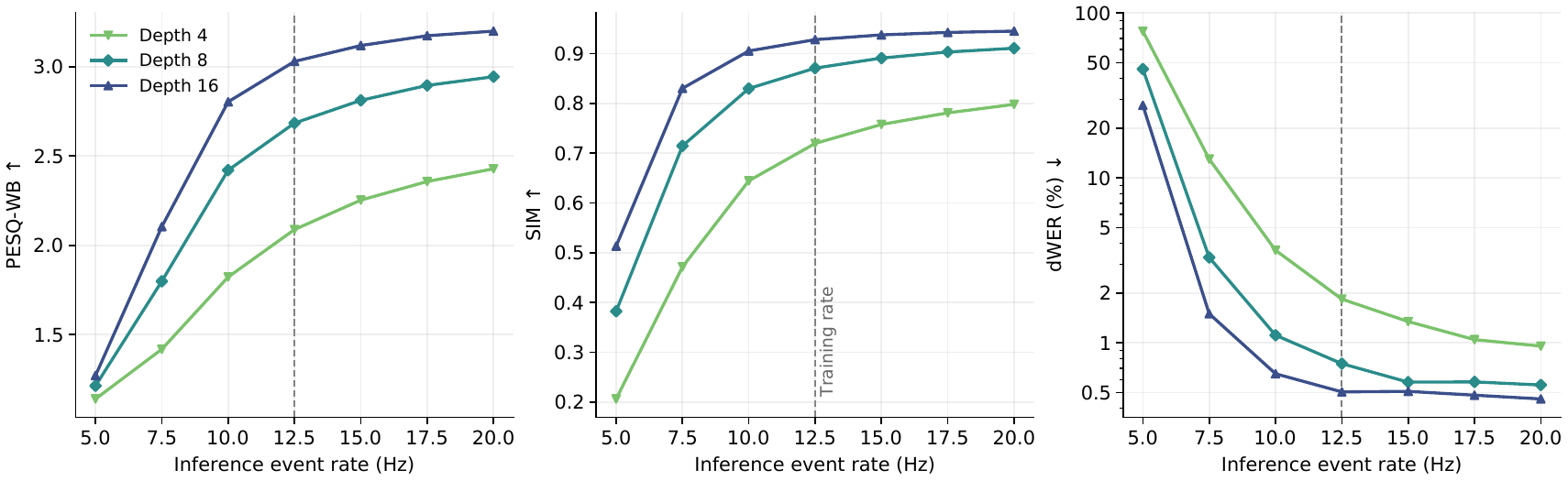}
\caption{\textbf{Inference-time event-rate variation without retraining.} All curves use the same model trained at 12.5 Hz, marked by the dashed vertical line. The horizontal axis shows the nominal inference event rate; colors and markers identify quantizer depth. The dWER axis is logarithmic.}\label{fig:inference-rate-variation}
\end{figure}

\begin{table}[tbp]
\centering
\footnotesize
\setlength{\tabcolsep}{3pt}
\caption{\textbf{Reconstruction at inference ratios different from the training ratio.} The training ratio is 0.25. Event rates and bitrates are measured utterance averages; bitrate includes duration information. Depth-4 results are shown in Figure~\ref{fig:inference-rate-variation}.}\label{tab:inference-rate-variation}
\begin{tabularx}{\linewidth}{YYYYYYYY}
\toprule
\textbf{Inference ratio} & \textbf{Depth} & \textbf{Event rate (Hz)} & \textbf{Bitrate (kbit/s)} & \textbf{PESQ-WB \(\uparrow\)} & \textbf{STOI \(\uparrow\)} & \textbf{dWER (\%) \(\downarrow\)} & \textbf{SIM \(\uparrow\)} \\
\midrule
0.10 & 8 & 5.0 & 0.50 & 1.214 & 0.708 & 45.66 & 0.383 \\
0.15 & 8 & 7.5 & 0.74 & 1.799 & 0.872 & 3.29 & 0.714 \\
0.20 & 8 & 10.0 & 0.98 & 2.421 & 0.918 & 1.11 & 0.830 \\
0.25 (trained) & 8 & 12.5 & 1.23 & 2.684 & 0.931 & 0.75 & 0.871 \\
0.30 & 8 & 15.0 & 1.47 & 2.811 & 0.938 & 0.58 & 0.891 \\
0.35 & 8 & 17.5 & 1.71 & 2.895 & 0.942 & 0.58 & 0.903 \\
0.40 & 8 & 20.0 & 1.96 & 2.944 & 0.945 & 0.56 & 0.911 \\
\midrule
0.10 & 16 & 5.0 & 0.98 & 1.271 & 0.739 & 27.55 & 0.513 \\
0.15 & 16 & 7.5 & 1.46 & 2.105 & 0.901 & 1.50 & 0.830 \\
0.20 & 16 & 10.0 & 1.94 & 2.803 & 0.939 & 0.65 & 0.905 \\
0.25 (trained) & 16 & 12.5 & 2.43 & 3.029 & 0.950 & 0.50 & 0.928 \\
0.30 & 16 & 15.0 & 2.91 & 3.118 & 0.954 & 0.51 & 0.937 \\
0.35 & 16 & 17.5 & 3.39 & 3.173 & 0.957 & 0.48 & 0.942 \\
0.40 & 16 & 20.0 & 3.88 & 3.199 & 0.958 & 0.46 & 0.945 \\
\bottomrule
\end{tabularx}
\end{table}

Table~\ref{tab:inference-rate-variation} shows an asymmetric response around the training rate. At depth 8, reducing the rate from 12.5 to 10 Hz lowers PESQ-WB from 2.684 to 2.421 and raises dWER from 0.75\% to 1.11\%. At 7.5 Hz, dWER reaches 3.29\%; at 5 Hz, it rises to 45.66\%. Greater quantizer depth mitigates the loss at intermediate rates, but depth 16 still has dWER of 27.55\% at 5 Hz. Depth 4 is more sensitive to rate reduction (Figure~\ref{fig:inference-rate-variation}).

Increasing the rate above 12.5 Hz improves PESQ-WB, STOI, and SIM at all three depths, with diminishing gains. At depth 8, dWER decreases to 0.56\% at 20 Hz. At depth 16, dWER remains near the training-rate value of 0.50\%, ranging from 0.46\% to 0.51\% across the higher rates. These quality gains also increase bitrate: depth-8 coding rises from 1.23 to 1.96 kbit/s between 12.5 and 20 Hz.

The model therefore exhibits partial tolerance to unseen inference ratios, particularly when increasing the event rate, while aggressive rate reduction degrades reconstruction sharply.

\section{Duration Prediction Order}\label{sec:duration-prediction-order}

This ablation compares four ways to predict event duration alongside codebook tokens. The independent-head variant predicts continuation, duration, and the first codebook token through three separate heads at the first local Transformer position; the remaining codes are predicted autoregressively without conditioning on the current duration. The other variants insert duration into the local autoregressive sequence before all codes, after the first (semantically supervised) code, or after all codes (Table~\ref{tab:duration-prediction-order}). Duration is conditioned on preceding codes, and subsequent codes are conditioned on duration. The duration-last variant corresponds to Equation~\eqref{eq:tts-factorization}.

All four models share a Qwen3-0.6B temporal backbone \cite{yang2025qwen3}, a single-layer local Transformer, eight codebooks, and the same frozen codec. Each is trained for 20,000 steps on approximately 4,000 hours from the HiFiTTS-2 44-kHz subset \cite{langman2025hifitts2}, with a global batch of 128 and training seed 42. AdamW \cite{loshchilov2019adamw} uses learning rates of \(10^{-4}\) for the pretrained backbone and \(3\times10^{-4}\) for newly initialized modules, with 1,000 warmup steps followed by cosine decay. Reference speech is supplied with probability 0.9 and limited to 375 events.

We evaluate duration cross-entropy on a HiFiTTS-2 validation split with speakers held out from TTS training, and TTS WER on the 1,088 English utterances in seed-tts-eval \cite{anastassiou2024seedtts}.

\begin{table}[tbp]
\centering
\small
\caption{\textbf{Effect of duration prediction structure and order.} The codebook indices follow Equation~\eqref{eq:tts-factorization}. Lower values are better for both metrics; bold marks the best score in each column.}\label{tab:duration-prediction-order}
\begin{tabularx}{\linewidth}{P{3.5cm}LR{2cm}R{1.2cm}}
\toprule
\textbf{Duration prediction} & \textbf{Local prediction order} & \textbf{Duration CE (nats) \(\downarrow\)} & \textbf{WER (\%) \(\downarrow\)} \\
\midrule
Independent head & Parallel duration prediction; autoregressive codes & 0.9698 & 4.71 \\
Before all codes & \(d_i \rightarrow c_{i,\allowbreak 0} \rightarrow \cdots \rightarrow c_{i,\allowbreak 7}\) & 0.9737 & 4.43 \\
After the semantic code & \(c_{i,\allowbreak 0} \rightarrow d_i \rightarrow c_{i,\allowbreak 1} \rightarrow \cdots \rightarrow c_{i,\allowbreak 7}\) & 0.8538 & 4.40 \\
After all codes & \(c_{i,\allowbreak 0} \rightarrow \cdots \rightarrow c_{i,\allowbreak 7} \rightarrow d_i\) & \textbf{0.4265} & \textbf{4.00} \\
\bottomrule
\end{tabularx}
\end{table}

Duration cross-entropy is not directly comparable across orders: duration-last conditions on all current-event codes, which may retain duration information from interval aggregation (Equation~\eqref{eq:quantizer-input}). Its lower cross-entropy therefore reflects easier prediction under this conditioning rather than better modeling of temporal structure.

We adopt duration-last prediction based on its lower duration cross-entropy and TTS WER in this small-scale comparison (Table~\ref{tab:duration-prediction-order}).

\section{Training Settings}\label{sec:training-settings}

\subsection{Codec training}\label{sec:codec-training}

Table~\ref{tab:codec-training-settings} provides the detailed settings for the architecture and placement comparisons in Section~\ref{sec:training-for-architecture-comparisons} and the multilingual model in Section~\ref{sec:multilingual-model-training}. Each optimizer advances its own learning-rate schedule; discriminator updates begin in the adversarial phase.

\begin{table}[tbp]
\centering
\small
\caption{\textbf{Training settings for the architecture and placement comparisons and multilingual model.}}\label{tab:codec-training-settings}
\begin{tabularx}{\linewidth}{P{0.2\linewidth}LLL}
\toprule
\textbf{Setting} & \textbf{Adaptive-versus-fixed comparisons} & \textbf{Uniform-stride-versus-adaptive placement} & \textbf{Multilingual model} \\
\midrule
Training data & HiFiTTS-2, 44-kHz subset; approximately 5,000 h & HiFiTTS-2, full English 44-kHz training partition & Public and internal multilingual speech; approximately 375,000 h \\
\rowrule
Input / target sample rate & 16 / 44.1 kHz & 16 / 44.1 kHz & 16 / 44.1 kHz \\
\rowrule
Crop duration & 6 s & 6 s & 6 s \\
\rowrule
Training length & 100,000 steps & 1,000,000 steps & 1,500,000 steps \\
\rowrule
Distributed training & 2 GPUs; 32 examples/GPU; global batch 64; no gradient accumulation & 8 GPUs; 8 examples/GPU; global batch 64; no gradient accumulation & 8 GPUs; 20 examples/GPU; global batch 160; no gradient accumulation \\
\rowrule
Precision & bfloat16 mixed precision; residual quantization in float32 & bfloat16 mixed precision; residual quantization in float32 & bfloat16 mixed precision; residual quantization in float32 \\
\rowrule
Optimizers & Separate AdamW \cite{loshchilov2019adamw} for generator and discriminators; \(\beta=(0.8,\allowbreak 0.9)\); weight decay 0.01 & Separate AdamW \cite{loshchilov2019adamw} for generator and discriminators; \(\beta=(0.8,\allowbreak 0.9)\); weight decay 0.01 & Separate AdamW \cite{loshchilov2019adamw} for generator and discriminators; \(\beta=(0.8,\allowbreak 0.9)\); weight decay 0.01 \\
\rowrule
Gradient clipping & 1.0 for each optimizer & 1.0 for each optimizer & 1.0 for each optimizer \\
\rowrule
Learning-rate schedule & Quadratic warmup for 3,000 updates, then linear decay over 97,000 updates, for each optimizer & Quadratic warmup for 3,000 updates, then linear decay over 997,000 updates, for each optimizer & Quadratic warmup for 15,000 updates; linear decay over 1,485,000 generator / 1,335,000 discriminator updates \\
\rowrule
Learning rates & Initial \(10^{-5}\); peak \(10^{-4}\); decay endpoint \(2\times10^{-5}\) & Initial \(10^{-5}\); peak \(10^{-4}\); decay endpoint \(2\times10^{-5}\) & Initial \(10^{-5}\); peak \(10^{-4}\); decay endpoint \(2\times10^{-5}\) \\
\rowrule
Loss phases & 0--10k: mel, semantic, and VQ; 10k--85k: add adversarial and feature matching; 85k--100k: add perceptual loss & 0--100k: mel, semantic, and VQ; 100k--850k: add adversarial and feature matching; 850k--1,000k: add perceptual loss & 0--150k: mel, semantic, and VQ; 150k--1,275k: add adversarial and feature matching; 1,275k--1,500k: add perceptual loss \\
\rowrule
Quantization & First-layer FSQ with 4,096 values; 15 residual VQ layers with 4,096 entries each & First-layer FSQ with 4,096 values; 15 residual VQ layers with 4,096 entries each & First-layer FSQ with 4,096 values; 15 residual VQ layers with 4,096 entries each \\
\rowrule
Quantizer dropout & Retained prefix depth sampled uniformly from 1--16 for each example & Retained prefix depth sampled uniformly from 1--16 for each example & Retained prefix depth sampled uniformly from 1--16 for each example \\
\rowrule
Decoder backbone & 12 causal Transformer layers; width 1,024; 16 attention heads; MLP expansion 4 & 12 causal Transformer layers; width 1,024; 16 attention heads; MLP expansion 4 & 12 causal Transformer layers; width 1,024; 16 attention heads; MLP expansion 4 \\
\rowrule
Decoder regularization & Dropout 0.1; attention dropout 0 & Dropout 0.1; attention dropout 0 & Dropout 0.1; attention dropout 0 \\
\bottomrule
\end{tabularx}
\end{table}

\subsection{TTS training and sampling}\label{sec:tts-training-and-sampling}

Table~\ref{tab:tts-training-settings} provides the shared settings for the TTS comparison in Section~\ref{sec:tts-evaluation}. The temporal backbone and text embeddings are initialized from pretrained Qwen3-0.6B weights; the local Transformer and codec-specific embeddings and prediction heads are newly initialized. Losses are computed only on the target sequence, excluding the conditioning prefix and padding. Cross-entropy is averaged over valid positions within each batch for each head, and the resulting head losses are averaged with equal weights.

\begin{table}[tbp]
\centering
\small
\caption{\textbf{TTS training settings shared by the adaptive and fixed-rate comparison.}}\label{tab:tts-training-settings}
\begin{tabularx}{\linewidth}{P{0.29\linewidth}L}
\toprule
\textbf{Setting} & \textbf{Configuration} \\
\midrule
Training data & HiFiTTS-2, 44-kHz English partition; approximately 31,300 h \cite{langman2025hifitts2} \\
\rowrule
Codec & 12.5-Hz adaptive and fixed-rate codecs (Section~\ref{sec:training-for-architecture-comparisons}) \\
\rowrule
Temporal backbone & Qwen3-0.6B \cite{yang2025qwen3}; 28 layers; hidden size 1,024; 16 query / 8 key-value heads \\
\rowrule
Local Transformer & 4 layers; hidden size 1,024; 16 attention heads \\
\rowrule
Codec representation & 8 codebooks; 4,096 entries per codebook \\
\rowrule
Duration prediction & Adaptive codec: final local position, after all 8 codes; durations 1--64 frames on the 50-Hz grid. Fixed-rate codec: known frame interval, without a duration head \\
\rowrule
Maximum target length & 1,024 speech events \\
\rowrule
Speaker conditioning & A different utterance from the same speaker, supplied with probability 0.9; at most 375 reference events \\
\rowrule
Distributed training & 4 GPUs; 32 examples/GPU; global batch 128; no gradient accumulation \\
\rowrule
Precision & bfloat16 mixed precision \\
\rowrule
Optimizers & Muon \cite{jordan2024muon,liu2025muonscalable} for hidden weight matrices; AdamW \cite{loshchilov2019adamw} for embeddings, output heads, and remaining parameters \\
\rowrule
Optimizer parameters & Muon momentum 0.95; AdamW \(\beta=(0.9,\allowbreak 0.95)\), \(\epsilon=10^{-8}\) \\
\rowrule
Peak learning rates & Pretrained backbone: \(2\times10^{-4}\); newly initialized modules: \(8\times10^{-4}\) \\
\rowrule
Learning-rate schedule & Linear warmup for 1,000 steps, then constant learning rate; linear decay over the final 20\% of training to 1\% of the peak rate \\
\rowrule
Weight decay & 0.1; excluded for normalization parameters, biases, and text, codebook, and duration embeddings \\
\rowrule
Gradient clipping & Global gradient norm 1.0 \\
\rowrule
Loss weights & Equal weights for all prediction heads \\
\rowrule
Training length & 3 epochs \\
\rowrule
Training seed & 42 \\
\bottomrule
\end{tabularx}
\end{table}

For evaluation (Sections~\ref{sec:tts-evaluation}--\ref{sec:generated-event-timing}), codebook tokens and durations are sampled with temperature 0.8, top-p 0.95, and top-k 50; a repetition penalty of 1.1 applies only to codebook tokens, using the history of the corresponding codebook. Continuation or stop is sampled from the binary distribution at temperature 1.0, without truncation or a repetition penalty.

\section{Code, Models, and Data}\label{sec:code-models-and-data}

This appendix lists public sources for the codec implementations, pretrained components, evaluation tools, and datasets used in this study. Experimental configurations and data selection are described in Sections~\ref{sec:training-data-and-optimization}--\ref{sec:duration-aware-text-to-speech-modeling} and Appendix~\ref{sec:training-settings}.

\subsection{Codec implementations and checkpoints}

\begin{table}[htbp]
\centering
\small
\caption{\textbf{Codec implementations and checkpoints.}}\label{tab:codec-resources}
\begin{tabularx}{\linewidth}{P{0.3\linewidth}P{0.3\linewidth}L}
\toprule
\textbf{Resource} & \textbf{Role in this study} & \textbf{Source links} \\
\midrule
RubatoCodec & Proposed codec & \href{https://github.com/third-intelligence/RubatoCodec}{Code}\newline \href{https://huggingface.co/third-intelligence/RubatoCodec}{Model} \\
\rowrule
EnCodec \cite{defossez2023encodec} & Codec baseline & \href{https://github.com/facebookresearch/encodec}{Code}\newline \href{https://huggingface.co/facebook/encodec_24khz}{Model: 24 kHz} \\
\rowrule
DAC \cite{kumar2023dac} & Codec baseline & \href{https://github.com/descriptinc/descript-audio-codec}{Code}\newline \href{https://huggingface.co/descript/dac_24khz}{Model: 24 kHz} \\
\rowrule
BigCodec \cite{xin2024bigcodec} & Codec baseline & \href{https://github.com/Aria-K-Alethia/BigCodec}{Code}\newline \href{https://huggingface.co/Alethia/BigCodec}{Model} \\
\rowrule
WavTokenizer \cite{ji2025wavtokenizer} & Codec baseline & \href{https://github.com/jishengpeng/WavTokenizer}{Code}\newline \href{https://huggingface.co/novateur/WavTokenizer-large-speech-75token}{Model: 75 Hz} \\
\rowrule
X-Codec-2 \cite{ye2025llasa} & Codec baseline & \href{https://github.com/zhenye234/X-Codec-2.0}{Code}\newline \href{https://huggingface.co/HKUSTAudio/xcodec2}{Model} \\
\rowrule
Mimi \cite{defossez2024moshi} & Codec baseline & \href{https://github.com/kyutai-labs/moshi}{Code}\newline \href{https://huggingface.co/kyutai/mimi}{Model} \\
\rowrule
MiMo-Audio-Tokenizer \cite{xiaomillmcoreteam2025mimoaudio} & Codec baseline & \href{https://github.com/XiaomiMiMo/MiMo-Audio}{Code}\newline \href{https://huggingface.co/XiaomiMiMo/MiMo-Audio-Tokenizer}{Model} \\
\rowrule
MOSS-Audio-Tokenizer \cite{gong2026mossaudiotokenizer} & Codec baseline & \href{https://huggingface.co/OpenMOSS-Team/MOSS-Audio-Tokenizer}{Model and inference code} \\
\rowrule
Qwen3-TTS-Tokenizer-12Hz \cite{hu2026qwen3tts} & Codec baseline & \href{https://github.com/QwenLM/Qwen3-TTS}{Code}\newline \href{https://huggingface.co/Qwen/Qwen3-TTS-Tokenizer-12Hz}{Model} \\
\rowrule
DTM-Codec \cite{sohn2026dtmcodec} & Codec baseline & \href{https://github.com/hoyso48/DTM-Codec}{Code}\newline \href{https://huggingface.co/hoyso48/DTM-Codec/tree/80hz}{Model: 80 Hz} \\
\rowrule
CodecSlime \cite{wang2026codecslime} & Codec baseline & \href{https://github.com/X-LANCE/codecslime}{Code}\newline \href{https://github.com/X-LANCE/codecslime/releases/tag/v0.1.0}{Model} \\
\rowrule
FlexiCodec \cite{li2026flexicodec} & Codec baseline & \href{https://github.com/AmphionTeam/FlexiCodec}{Code}\newline \href{https://huggingface.co/jiaqili3/flexicodec}{Model} \\
\rowrule
DyCAST \cite{dellalibera2026dycast} & Codec baseline & \href{https://github.com/lucadellalib/dycast}{Code}\newline \href{https://huggingface.co/lucadellalib/dycast}{Model} \\
\bottomrule
\end{tabularx}
\end{table}

X-Codec-2 is also the architecture backbone of RubatoCodec; the evaluated codec configurations are listed in Table~\ref{tab:baseline-codec-specifications}.

\FloatBarrier
\subsection{Datasets}

\begin{table}[htbp]
\centering
\small
\caption{\textbf{Datasets.}}\label{tab:dataset-resources}
\begin{tabularx}{\linewidth}{P{0.3\linewidth}P{0.3\linewidth}L}
\toprule
\textbf{Resource} & \textbf{Role in this study} & \textbf{Source links} \\
\midrule
HiFiTTS-2 \cite{langman2025hifitts2} & Codec and TTS training / validation & \href{https://huggingface.co/datasets/nvidia/hifitts-2}{Dataset} \\
\rowrule
Emilia-YODAS \cite{he2025emilia} & Multilingual codec training & \href{https://huggingface.co/datasets/amphion/Emilia-Dataset}{Dataset} \\
\rowrule
ReazonSpeech \cite{yin2023reazonspeech} & Multilingual codec training & \href{https://huggingface.co/datasets/reazon-research/reazonspeech}{Dataset} \\
\rowrule
The People's Speech \cite{galvez2021peoplesspeech} & Multilingual codec training & \href{https://huggingface.co/datasets/MLCommons/peoples_speech}{Dataset} \\
\rowrule
VoxPopuli \cite{wang2021voxpopuli} & Multilingual codec training & \href{https://huggingface.co/datasets/facebook/voxpopuli}{Dataset} \\
\rowrule
Multilingual LibriSpeech \cite{pratap2020mls} & Multilingual codec training & \href{https://huggingface.co/datasets/facebook/multilingual_librispeech}{Dataset} \\
\rowrule
Omnilingual ASR Corpus \cite{omnilingualasrteam2025omnilingualasr} & Multilingual codec training & \href{https://huggingface.co/datasets/facebook/omnilingual-asr-corpus}{Dataset} \\
\rowrule
IndicVoices \cite{javed2024indicvoices} & Multilingual codec training & \href{https://huggingface.co/datasets/ai4bharat/IndicVoices}{Dataset} \\
\rowrule
LibriSpeech \cite{panayotov2015librispeech} & English reconstruction evaluation & \href{https://www.openslr.org/12}{Dataset} \\
\rowrule
VCTK 0.92 \cite{yamagishi2019vctk} & English reconstruction evaluation & \href{https://doi.org/10.7488/ds/2645}{Dataset} \\
\rowrule
JVS \cite{takamichi2019jvs} & Japanese reconstruction evaluation & \href{https://sites.google.com/site/shinnosuketakamichi/research-topics/jvs_corpus}{Dataset} \\
\rowrule
JVNV \cite{xin2024jvnv} & Japanese reconstruction evaluation & \href{https://sites.google.com/site/shinnosuketakamichi/research-topics/jvnv_corpus}{Dataset} \\
\rowrule
seed-tts-eval \cite{anastassiou2024seedtts} & Mandarin reconstruction and English TTS evaluation & \href{https://github.com/BytedanceSpeech/seed-tts-eval}{Dataset} \\
\bottomrule
\end{tabularx}
\end{table}

The multilingual training mixture also includes internal data, as described in Section~\ref{sec:multilingual-model-training}.

\FloatBarrier
\subsection{Pretrained components and evaluation tools}

\begin{table}[htbp]
\centering
\small
\caption{\textbf{Pretrained components and evaluation tools.}}\label{tab:evaluation-resources}
\begin{tabularx}{\linewidth}{P{0.3\linewidth}P{0.3\linewidth}L}
\toprule
\textbf{Resource} & \textbf{Role in this study} & \textbf{Source links} \\
\midrule
w2v-BERT 2.0 \cite{seamlesscommunication2023seamless} & Semantic feature extraction and supervision & \href{https://huggingface.co/facebook/w2v-bert-2.0}{Model} \\
\rowrule
wav2vec2-XLS-R-300M \cite{babu2022xlsr} & Perceptual-loss feature extraction & \href{https://huggingface.co/facebook/wav2vec2-xls-r-300m}{Model} \\
\rowrule
Qwen3-0.6B \cite{yang2025qwen3} & TTS temporal backbone & \href{https://huggingface.co/Qwen/Qwen3-0.6B}{Model} \\
\rowrule
Whisper-large-v3 \cite{radford2023whisper} & English and Japanese ASR & \href{https://huggingface.co/openai/whisper-large-v3}{Model} \\
\rowrule
Paraformer-zh \cite{gao2022paraformer} & Mandarin ASR & \href{https://huggingface.co/funasr/paraformer-zh}{Model} \\
\rowrule
WavLM-Large speaker verification \cite{chen2022wavlm} & Speaker similarity evaluation & \href{https://github.com/microsoft/UniSpeech/tree/main/downstreams/speaker_verification}{UniSpeech implementation}\newline \href{https://drive.google.com/file/d/1-aE1NfzpRCLxA4GUxX9ITI3F9LlbtEGP/view}{SV checkpoint}\newline \href{https://github.com/BytedanceSpeech/seed-tts-eval}{Evaluation instructions} \\
\rowrule
UTMOSv2 \cite{baba2024utmosv2} & Speech naturalness evaluation & \href{https://github.com/sarulab-speech/UTMOSv2}{UTMOSv2 implementation} \\
\rowrule
UTMOS \cite{saeki2022utmos} & Speech naturalness evaluation & \href{https://github.com/tarepan/SpeechMOS/tree/v1.2.0}{Implementation and weights} \\
\rowrule
DNSMOS \cite{reddy2022dnsmos} & Speech quality evaluation & \href{https://github.com/Lightning-AI/torchmetrics/blob/master/src/torchmetrics/functional/audio/dnsmos.py}{TorchMetrics implementation} \\
\rowrule
Muon \cite{jordan2024muon} & TTS optimization & \href{https://github.com/KellerJordan/Muon}{Code} \\
\bottomrule
\end{tabularx}
\end{table}

For speaker similarity, we use the speaker-verification checkpoint \texttt{wavlm\_large\_finetune.pth} with the UniSpeech ECAPA-TDNN head and WavLM-Large features, following the checkpoint distribution linked by seed-tts-eval.

UTMOS uses the \texttt{utmos22\_strong} model from SpeechMOS v1.2.0.

\end{document}